\documentstyle[10pt,aaspp4,psfig,flushrt]{article}

\begin{document}
\pagestyle{myheadings}
%\markright{{\tt {Date: \today } }}

\def\spose#1{\hbox to 0pt{#1\hss}}
\newcommand\ginga{\it Ginga}

\newcommand{\fluxu}{{\rm erg \ cm^{-2} \ s^{-1}}}
\newcommand{\zsun}{z_{\odot}}
\newcommand{\cps}{\mbox{\rm c\ s$^{-1}$}}
\newcommand\lsim{\mathrel{\spose{\lower 3pt\hbox{$\mathchar"218$}}
     \raise 2.0pt\hbox{$\mathchar"13C$}}}
\newcommand\gsim{\mathrel{\spose{\lower 3pt\hbox{$\mathchar"218$}}
     \raise 2.0pt\hbox{$\mathchar"13E$}}}
\newcommand\simless{\mathrel{\spose{\lower 3pt\hbox{$\mathchar"218$}}
     \raise 2.0pt\hbox{$\mathchar"13C$}}}
\newcommand\simgreat{\mathrel{\spose{\lower 3pt\hbox{$\mathchar"218$}}
     \raise 2.0pt\hbox{$\mathchar"13E$}}}

\newcommand\powspec{(see Appendix~\ref{appb:powspec})}
\newcommand\efold{(see Appendix~\ref{appb:efold})}
\newcommand\phase{(see Appendix~\ref{appb:phase})}
\newcommand\flux{(see Appendix~\ref{appb:flux})}
\newcommand\spec{(see Appendix~\ref{appb:spec})}

\def\be{\begin{equation}}
\def\ee{\end{equation}}
\def\chk{$\surd$}
\def\huh{{\bf ??}}
\def\etal{et al.}

% These have to be hardwired because they exist in separate files
% which will be made photo-ready.
\newcommand\tabpulsars{1}
\newcommand\taborbits{3}
\newcommand\tabspectra{4}

\newcommand\hps{Hz~s$^{-1}$}
\newcommand\final{{}}
\newcommand\nopfrac{No pulsed fraction measurements have been made 
in the BATSE energy range.}

\newcommand{\rsun}{{R}_{\odot}}
\newcommand{\msun}{{M}_{\odot}}
\newcommand{\lacc}{L_{\rm a}}
\newcommand{\facc}{F_a}
\newcommand{\mdot}{\dot{M}}
\newcommand{\mdotg}{\dot{M}_{g}}
\newcommand{\mdotscale}{10^{-8} \ \msun \ {\rm yr^{-1}}}
\newcommand{\rwdscale}{5\times 10^8 {\rm cm}}
\newcommand{\racc}{r_{\rm acc}}
\newcommand{\rmag}{r_{\rm mag}}
\newcommand{\mwd}{M_{wd}}
\newcommand{\rwd}{R_{wd}}

\bibliographystyle{apj_noskip}

\title{OBSERVATIONS OF ACCRETING PULSARS}
\bigskip

\author{Lars~Bildsten\altaffilmark{1},
Deepto~Chakrabarty\altaffilmark{2}, John Chiu\altaffilmark{3}, 
Mark~H.~Finger\altaffilmark{4,5}, Danny~T.~Koh\altaffilmark{3}, \\
Robert~W.~Nelson\altaffilmark{3,6}, Thomas~A.~Prince\altaffilmark{3},
Bradley~C.~Rubin\altaffilmark{4,7}, D.~Matthew~Scott\altaffilmark{4,5}, \\
Mark~Stollberg\altaffilmark{4,8}, Brian~A.~Vaughan\altaffilmark{3}, 
Colleen~A.~Wilson\altaffilmark{4}, and Robert~B.~Wilson\altaffilmark{4}}

\smallskip

\altaffiltext{1}{Department of Physics and Department
  of Astronomy, University of California, Berkeley, CA 94720;
  bildsten@fire.berkeley.edu}
\altaffiltext{2}{Center for Space Research,
  Massachusetts Institute of Technology, Cambridge MA 02139;
  deepto@space.mit.edu}
\altaffiltext{3}{Space Radiation Laboratory, 
  California Institute of Technology, Pasadena, CA 91125;
  chiu@srl.caltech.edu, koh@srl.caltech.edu, nelson@tapir.caltech.edu, 
  prince@caltech.edu, brian@srl.caltech.edu}
\altaffiltext{4}{Space Science Laboratory, NASA/Marshall Space
  Flight Center, ES84, Huntsville, AL 35812; 
  finger@gibson.msfc.nasa.gov,
  rubin@crab.riken.go.jp, scott@gibson.msfc.nasa.gov, 
  stollberg@gibson.msfc.nasa.gov, 
  wilsonc@ssl.msfc.nasa.gov, wilson@gibson.msfc.nasa.gov}
\altaffiltext{5}{Universities Space Research Association}
\altaffiltext{6}{Theoretical Astrophysics 130-33, California Institute
  of Technology, Pasadena, CA 91125}
\altaffiltext{7}{Current address: Cosmic Radiation Laboratory,
  Institute for Physical and Chemical Research (RIKEN), Wako-shi,
  Saitama 351-01, Japan}
\altaffiltext{8}{Department of Physics, University of Alabama,
  Hunstville, AL 35899} 

\bigskip
\centerline{To Appear In {\sc Astrophysical Journal Supplements} 
1997, 113, \#2}
\bigskip
%\centerline{\today}

%%%%%%%%%%%%%%%%%%%%%%%%%%%%%%%%%%%%%%%%%%%%%%%%%%%%%%%%%%%%%%%%%%%%%%%%%%%%%%%

%\input abstract.tex

We summarize five years of continuous monitoring of accretion-powered
pulsars with the Burst and Transient Source Experiment (BATSE) on the
{\em Compton Gamma Ray Observatory}. Our 20--70 keV observations have
determined or refined the orbital parameters of 13 binaries,
discovered 5 new transient accreting pulsars, measured the pulsed flux
history during outbursts of 12 transients
(GRO~J1744--28, 4U~0115+634, GRO~J1750--27, GS~0834--430, 2S~1417--624,
GRO~J1948+32, EXO~2030+375, GRO~J1008--57, A~0535+26, GRO~J2058+42,
4U~1145--619 and A~1118--616),
and also measured the
accretion torque history of during outbursts of 6 of those transients
whose orbital parameters were also known.  We have also continuously
measured the pulsed flux and spin frequency for eight persistently
accreting pulsars (Her X-1, Cen X-3, Vela X-1, OAO 1657--415, GX
301--2, 4U 1626--67, 4U 1538--52, and GX 1+4). Because of their
continuity and uniformity over a long baseline, BATSE observations
have provided new insights into the long-term behavior of accreting
magnetic stars.  We have found that all accreting pulsars show
stochastic variations in their spin frequencies and luminosities,
including those displaying secular spin-up or spin-down on long time
scales, blurring the conventional distinction between disk-fed and
wind-fed binaries. Pulsed flux and accretion torque are strongly
correlated in outbursts of transient accreting pulsars, but
uncorrelated, or even anticorrelated, in persistent sources. We
describe daily folded pulse profiles, frequency, and flux measurements
that are available through the {\em Compton Observatory} Science
Support Center at NASA-Goddard Space Flight Center. 

%%%%%%%%%%%%%%%%%%%%%%%%%%%%%%%%%%%%%%%%%%%%%%%%%%%%%%%%%%%%%%%%%%%%%%%%%%%%%%%

%\input keywords.tex

\keywords{Accretion, Accretion Disks --- 
      Stars: Binaries: General --- Stars: Pulsars: General --- 
      X-Rays: Stars  --- Stars: Neutron}

%%%%%%%%%%%%%%%%%%%%%%%%%%%%%%%%%%%%%%%%%%%%%%%%%%%%%%%%%%%%%%%%%%%%%%%%%%%%%%%

%\input intro_new.tex

\newpage
\tableofcontents

\newpage
\section{INTRODUCTION}

  Accreting X-ray pulsars were discovered over 25 years ago (Giacconi
\etal 1971), and a qualitative description of both the accretion
process and the origin of the pulsed emission was understood almost
immediately (Pringle \& Rees 1972; Davidson \& Ostriker 1973; Lamb,
Pethick \& Pines 1973). X-ray pulsars are rotating and strongly
magnetized ($B \simgreat 10^{11}$ G) neutron stars which accrete gas
from a stellar companion.  As the accreting material approaches the
neutron star, the plasma is channeled to the magnetic polar caps,
where it releases its gravitational energy as X and
$\gamma$-radiation; these rotating hotspots are the sources of the
pulsed emission.  Despite more than two decades of study, however,
many details of this scenario remain poorly understood.

\nocite{Pringle72,Davidson73,Lamb73} 

  The accreting pulsars are also important evolutionary links to other
binary systems containing neutron stars. Some young neutron stars with
high mass companions may begin their lives as rotation-powered radio
pulsars (Johnston et al. 1992, Kaspi et al. 1994) and become X-ray
sources only during episodes of significant mass transfer later in
life. On the other hand, there is some evidence that extended episodes
of accretion onto neutron stars with low mass companions can cause
their magnetic fields to decay (Bhattacharya \& Srinivasan 1995).
If the inner accretion disk can then extend nearly down to the
stellar surface, these neutron stars should spin up to millisecond
rotation periods (Alpar \etal 1982); there is mounting evidence
that the low mass X-ray binaries contain rapidly rotating neutron
stars (see for example Strohmayer et. al. 1996) and are the birthplace of
the millisecond radio pulsars.  Perhaps most importantly, the
qualitative picture developed in the early 1970s to explain the
behavior of X-ray pulsars has become the paradigm for accretion
onto other types of magnetic stars, such as magnetic CVs and T Tauri
stars (\cite{Warner90,Konigl91}).  It is thus becoming increasingly
important that the standard picture of X-ray pulsars developed more
than 20 years ago be tested critically.

\nocite{Johnston92,Bhattacharya95,Kaspi94,Konigl91,Alpar82,
        Strohmayer96,Warner90}

Much of our understanding of accretion-powered pulsars originates from
accurate timing of the pulsed emission. Just as in binary radio
pulsars, the orbital motion causes a modulation in the observed pulse
frequency, which allows the determination of the binary
orbital parameters. The small moment of inertia of a neutron star
makes it possible to  measure directly the intrinsic changes in the pulsar
spin frequency caused by angular momentum gained (or lost) during the
accretion process on $\sim$days timescales. 
This can potentially reveal the nature of the accretion flow --
a persistent trend in the spin frequency indicates the presence of an
accretion disk, while short term changes with no persistent trend
are usually indicative of a wind-fed system.  As of this writing, 
there are 44 known accreting pulsars in our Galaxy and the
Magellanic Clouds, with spin periods ranging from 0.069\,s through
1413\,s. Approximately half of these objects are observed only
during episodes of transient accretion.

 The physics and astronomy of accretion-powered pulsars have been
reviewed previously. White, Nagase \& Parmar (1995) reviewed accreting
X-ray binaries in general. Nagase (1989) reviewed observations of
accreting pulsars.  Hayakawa (1985) provided a theoretical overview of
accretion physics and spectral formation in strong magnetic fields.
Joss \& Rappaport (1984) reviewed neutron stars in binaries.
White, Swank \& Holt (1983) presented energy spectra and pulse
profiles.  Rappaport \& Joss (1977a,b) reviewed the ``standard
model'' for accretion torques and pulse profiles.

\nocite{White95}
\nocite{White83}
\nocite{RappaportJoss77a}
\nocite{RappaportJoss77b}
\nocite{Hayakawa85}

 In this paper we summarize over five years of observations of
accreting binary pulsars with the all-sky BATSE instrument on the {\it
Compton Gamma Ray Observatory}.  BATSE's principal advantage over
previous instruments for studying accreting pulsars is its continuous
monitoring capability.  The timing data we present here represent a
$\sim$100-fold increase in the time resolution of spin frequency histories
of persistent pulsars, and the first long-term, spatially-uniform
monitoring program for the detection of new pulsars and recurrent
transients. We have thus detected and studied nearly half of the known
accreting pulsars and determined accurate orbital parameters for 13 of
these systems. Table {\tabpulsars} 
lists all known accreting pulsars along with their
positions in galactic coordinates, spin and orbital periods, 
and companion type (where known).

  BATSE's continuous timing of X-ray pulsars gives the neutron star
spin period history over timescales of days to years and is ideally
suited for detailed studies of the accretion torque.  Our observations
give a qualitatively different picture of the spin behavior of
disk-fed pulsars on long timescales ($\sim$ years) than understood
from pre-BATSE measurements (see Nagase 1989 and references therein).
Moreover, BATSE has been able to test theories of accretion torque on
short timescales ($\sim$ days). This has led to several unexpected
discoveries in disk-fed pulsars: (1) the transition between spin-up
and spin-down in 4U 1626--67 (\cite{Chakrabarty97a}) and Cen~X-3, (2)
anticorrelated behavior of torque and luminosity in GX 1+4
(\cite{Chakrabarty97b}) and (3)  evidence that transient
accretion disks sometimes form in GX 301--2 (\cite{Koh97}). By
monitoring these changes along with variations in the pulsed
luminosity, we may be able to learn about the complex interaction
between the magnetosphere and the accretion flow -- physics which is
at work in a broad variety of accreting systems but can only be
measured dynamically in accretion-powered pulsars.

Five of the 13 transient systems detected by BATSE are new binaries.
Combining these discoveries with the ``recovery'' of previously known
transients yields new information on the population and
typical distance of these sources.  In addition to the recent
discovery that the bursting transient GRO J1744--28 is a 2 Hz pulsar
(\cite{Finger96a}), the discovery of quasi-periodic oscillations in
the accreting transient A 0536+26 (\cite{Finger96c}) gave us the best
evidence yet for periodic phenomena originating from the
magnetospheric radius.

  Section 2 outlines how we take maximum advantage of the BATSE
instrument by actively processing the standard data sets (DISCLA and
CONT). We also summarize our data analysis methods and give our
typical flux sensitivity as a function of spin period. Appendices A
and B contain additional details about our data analysis
technique. A summary and brief review of the science that can be done
by pulse timing of accreting pulsars is provided in \S 3.
Section 4 presents a synopsis of BATSE observations with
frequency and flux histories for each accreting pulsar we detected.
We also provide pointers to the literature where more details can be
found.  Section 5 is a discussion of how the BATSE
observations have changed our understanding of the long-term spin
evolution of accreting pulsars and the nature of transient sources.
We conclude in \S 6 with a brief summary of our primary discoveries.

\centerline{
\psfig{file=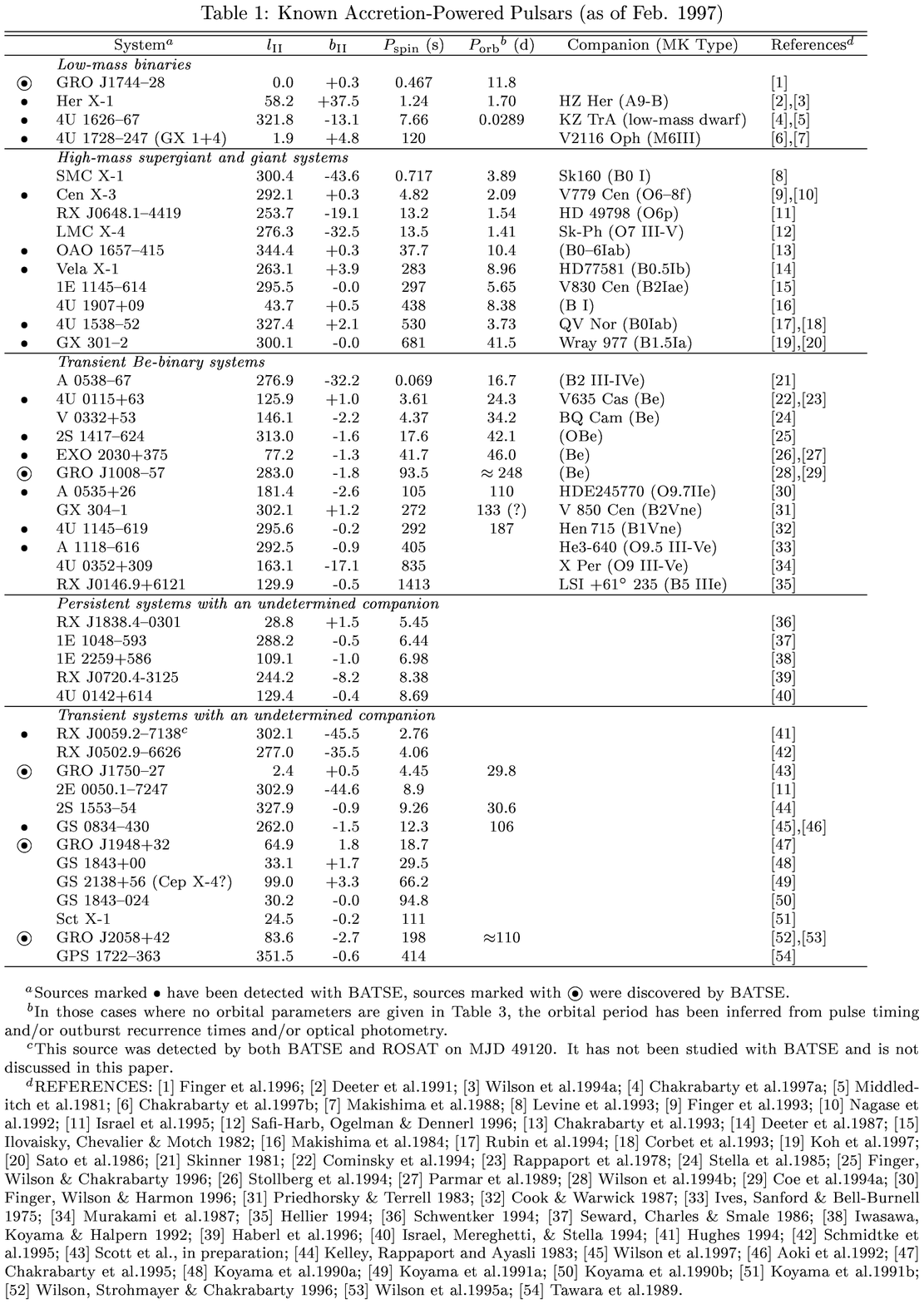,height=10in}
}

%%%%%%%%%%%%%%%%%%%%%%%%%%%%%%%%%%%%%%%%%%%%%%%%%%%%%%%%%%%%%%%%%%%%%%%%%%%%%%%

%\input batse_analysis.tex

\section{PULSAR DETECTION AND STUDY WITH BATSE}

 The BATSE detectors have provided unprecedented continuous all-sky
monitoring for both pulsed and unpulsed sources above 20 keV since 1991.
This section briefly summarizes our methods and
describes the resulting flux sensitivity as a function of pulse
frequency. This is the most crucial quantity for determining the detection
sensitivity for new sources and shapes our discussion in \S 5 of what
BATSE has learned about the populations of X-ray transients.

%\newpage
%\begin{figure}
\centerline{\psfig{file=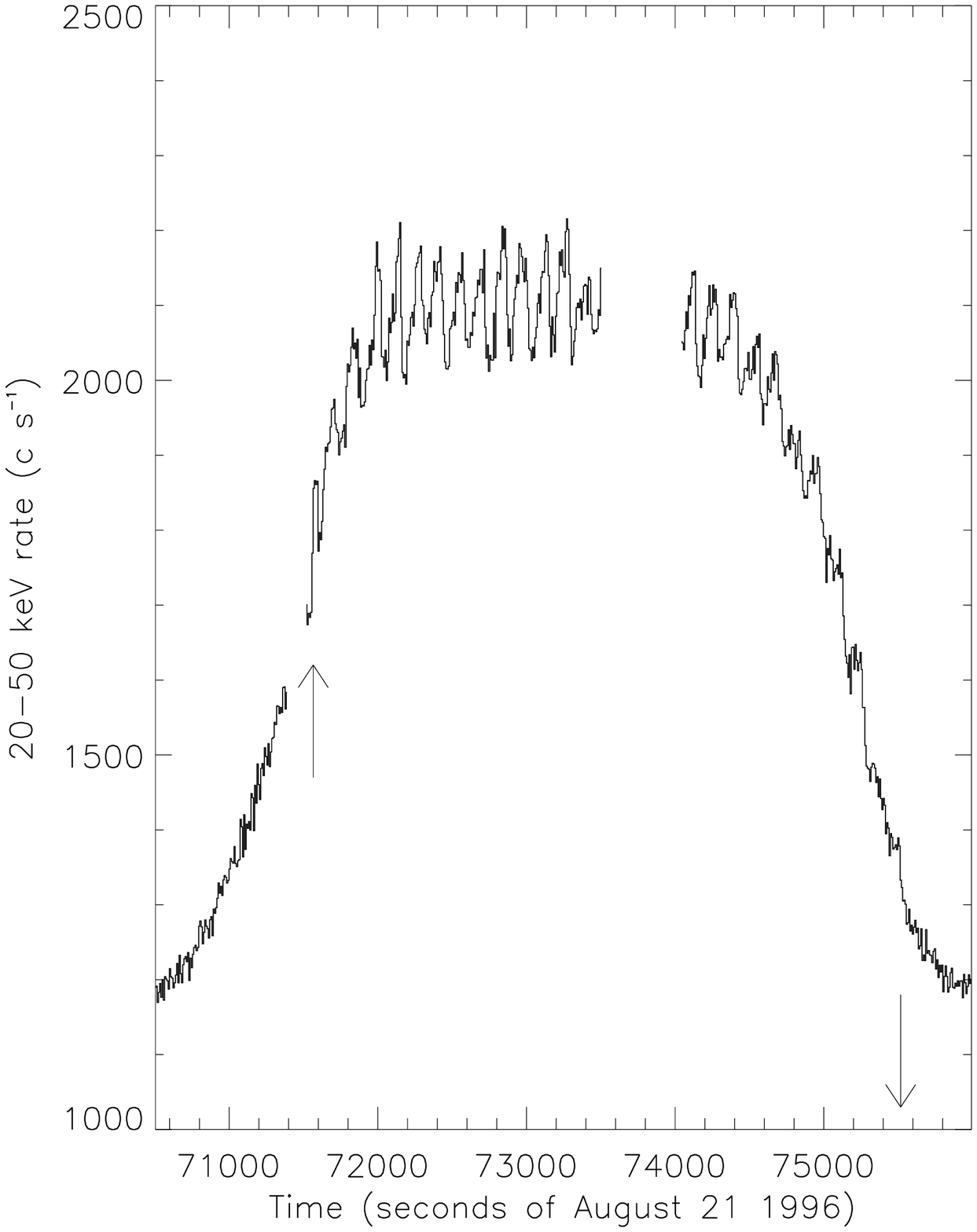,height=5.5in}}
\figcaption{DISCLA data from a single BATSE detector for a full
spacecraft orbit. The rates are from the 20-50 keV channel of Large
Area Detector 2, on August 21 1996 when Vela X-1 was in a high
state. Shown are 8.2 second average rates, with errors $<16\  \cps$.
The 800 $\cps$ modulation is due to the difference between the cosmic
diffuse background and the diffuse background from the Earth's
atmosphere. The lowest rates are when the detector mainly sees the
Earth, while at the highest rates it sees the sky. The data gaps
are due to a loss of telemetry. Vela X-1 rises above the horizon at
71566 s and sets at 75517 s, as noted by the arrows. 
\label{fig:velax1}
}
%\end{figure}

  BATSE consists of eight detector modules facing outward from the
corners of the CGRO spacecraft. Each module contains a large area
detector (LAD) (with geometric area of 2025\,cm$^2$ and an energy
range of 20--1800\,keV) and a smaller spectroscopy detector. The LADs
are non-imaging NaI(Tl) scintillators with $2\pi$ steradian fields of
view.  Our pulsar studies primarily use the background data from these
detectors, which are folded on-board or available continuously at
1.024 second time resolution with 4 energy channels (DISCLA data) and
at 2.048 second time resolution with 16 energy channels (CONT
data). The large field of view allows for multiple contributions to
the background, which varies by factors of two during each $\approx$93
min satellite orbit. Figure \ref{fig:velax1} shows a selected orbit of
DISCLA data from a single BATSE LAD detector. The large pulses every
141.5 seconds are from the bright accreting pulsar Vela X-1
[$P_{\rm spin}=283.2185(18)$ s on this day] which has a double peaked
pulse profile.  Very few sources are persistently bright enough to
observe directly in this $\approx 2000 \ \cps$ background.

Accreting pulsars are typically detected in the lowest DISCLA channel,
covering 20--50\,keV, and in CONT channels 1--4, typically covering
20--70\,keV, with detections sometimes extending to energies as
high as 160\,keV. In this sense, BATSE is only measuring the
high energy spectral tails of accreting pulsars.  Often most of the
flux from these objects is at lower energies, so that our flux
measurements are subject to a large (and often unknown) bolometric
correction. The pulsed flux histories we provide thus reflect the
history of the bolometric luminosity only when the overall spectral shape, and
the pulsed fraction in the BATSE energy band are independent of time.

  The signal $C_{\rm S}$ (in $\cps$) in the BATSE detectors from most
accreting pulsars is $10^2$--$10^3$ times smaller than the background
count rate $C_{\rm B}$, so that a Fourier or epoch-folding analysis
proves to be the best way to detect them. For pulse periods shorter
than a few minutes the signal-to-noise for an observation of length
$t$ is just governed by the Poisson variations of the background, $S/N
\propto C_{\rm S} t/(C_{\rm B} t)^{1/2}$. The background variations on
timescales longer than a few minutes are mostly due to the satellite
orbit and other effects (see Appendix A) and hence exceed the Poisson
variations. Our sensitivity for detecting long period pulsars would be
substantially degraded if nothing was done to remedy this.  Our
solution (presented in Appendix A) is to fit a phenomenological model
to these background variations and subtract it prior to scientific
analysis. Though this does not bring the sensitivity down to the
Poisson level at all frequencies (see Figure \ref{fig:powerback} in
Appendix A) it is a great improvement relative to the raw count rates.

  The resulting 1--day pulsed-intensity sensitivities of the CONT data as a
function of energy are shown in Figure \ref{fig:pulsens} for three
representative pulse frequencies. The upper panel shows the minimum
count rate needed so that the count rate of the pulse is found to
within 20\% accuracy.  This is a more stringent criterion than that
for detecting the pulse in a narrow frequency range. As is evident
from the figure, the excess noise at low frequencies strongly reduces
our sensitivity to long period pulsars. The lower panel is the
resulting pulsed flux assuming that the pulsar has a power law energy
spectrum with index $\alpha=4$ (see the figure caption). Most of our
searches are carried out with combinations of various channels. Figure
\ref{fig:threshold} shows the one-day sensitivity in a single detector
for the sum of CONT channels 1-4. When Poisson-limited, the resulting
sensitivity to a pulsed source of high frequencies is
\begin{equation}\label{eq:countlimit}
C_{\rm S}\approx 1.1 \ \cps 
  \left( \frac{C_{\rm B}}{2000\ \cps}\right)^{1/2}
  \left( \frac{42000 \ {\rm s}}{t} \right)^{1/2}. 
\end{equation} 
Depending on how steep the spectrum is (see Table~\ref{tab:flux2cts}
for the conversion
from BATSE LAD $\cps$ to flux units) this is a flux of $\approx
10^{-10} \fluxu$ in the 20-60 keV band. The corresponding luminosity
at the Galactic center (8.5 kpc) is $L(20-60 \ {\rm keV}) \approx 8\times
10^{35} \ {\rm erg \ s^{-1}}$, allowing detection of the majority 
of the known accretion pulsars.
For the purposes of comparison, the one day sensitivity
for detection using occultation steps 
is about $10~\cps$ (\cite{Harmon92}), a factor of
ten worse than the pulsed sensitivity. This is basically due to the
difference in effective exposure time in one day for both methods.

We have performed a uniform, standard analysis on all pulsars viewable
with BATSE that have spin periods longer than about 4 seconds, the
Nyquist frequency of the CONT data.  Power spectra are computed using
the fast Fourier transform for a daily estimate of the spin frequency
$\nu$ of each system, followed by epoch folding at $\nu$ for a daily
pulse profile.  The profiles are then used to determine the pulsed
count rate in each CONT energy channel, which are then fit with
standard models to determine the spectral shape and pulsed flux (see
Appendix B). These folded profiles form the basis of detailed timing
studies.  Frequency, pulsed flux and folded pulse profile histories
generated by this analysis are available from the BATSE pulsed source
database at the {\em Compton Observatory} Science Support Center.
This database forms the basis of many of the figures displayed in this
paper.  For systems with spin periods comparable to or shorter than 4
seconds [GRO J1744--28 (0.467\,s), Her X-1 (1.24\,s), 4U0115+63
(3.6\,s), GRO J1750--27 (4.4\,s), and Cen X-3 (4.8\,s)] we utilize a
combination of DISCLA and folded-on-board data, also described in
Appendix B.

\newpage
\setcounter{table}{1}
\begin{deluxetable}{cccc}
\tablecolumns{4}
\tablewidth{0pt}
\tablecaption{Conversion to Pulsed Flux from Counts
\label{tab:flux2cts}}

\tablehead{\multicolumn{4}{c}{Power Law} \\ \cline{1-4} 
           \colhead{$\alpha^a$} &
           \colhead{DISCLA$^b$} &
           \colhead{CONT$^c$} & 
           \colhead{\% Error$^d$}
        }
\startdata
2.0     &       0.84    &       0.79    & 5 \nl
3.0     &       0.94    &       0.87    & 10 \nl
4.0     &       1.05    &       0.95    & 15 \nl
5.0     &       1.17    &       1.02    & 20 \nl
\hline 
\multicolumn{4}{c}{OTTB} \\ \cline{1-4} 
\colhead{$kT$ (keV)$^e$} &
\colhead{DISCLA$^b$} &
\colhead{CONT}$^c$ &
\colhead{\% Error$^d$} \\ \cline{1-4}
10      &       1.18    &       1.03    & 15    \nl
25      &       0.94    &       0.88    & 8     \nl
40      &       0.89    &       0.84    & 6     \nl
55      &       0.87    &       0.82    & 5     \nl
\enddata
\tablecomments{These conversions are for normal incidence. 
For other incident photon angles, see Appendix A and
Figure \ref{fig:angresp}}
\tablenotetext{a}{Spectral model used is $dN/dE \propto E^{-\alpha}$}
\tablenotetext{b}{Energy flux in units 
of $10^{-10}$ ergs~cm$^{-2}$~s$^{-1}$ corresponding 
to a pulsed intensity of 1~$\cps$
in DISCLA Channel 1 (20--50~keV).}
\tablenotetext{c}{Energy flux in units 
of $10^{-10}$ ergs~cm$^{-2}$~s$^{-1}$ corresponding 
to a pulsed intensity of 1~$\cps$
in CONT Channels 1--4 (20--70~keV) summed .}
\tablenotetext{d}{Percentage error in energy flux due
to variations in energy edges for different detectors.}
\tablenotetext{e}{Spectral model used is $dN/dE \propto 
(1/E) \exp(-E/kT)g_{ff}(E,kT)$}
\end{deluxetable}

%\newpage
%\begin{figure}
\psfig{file=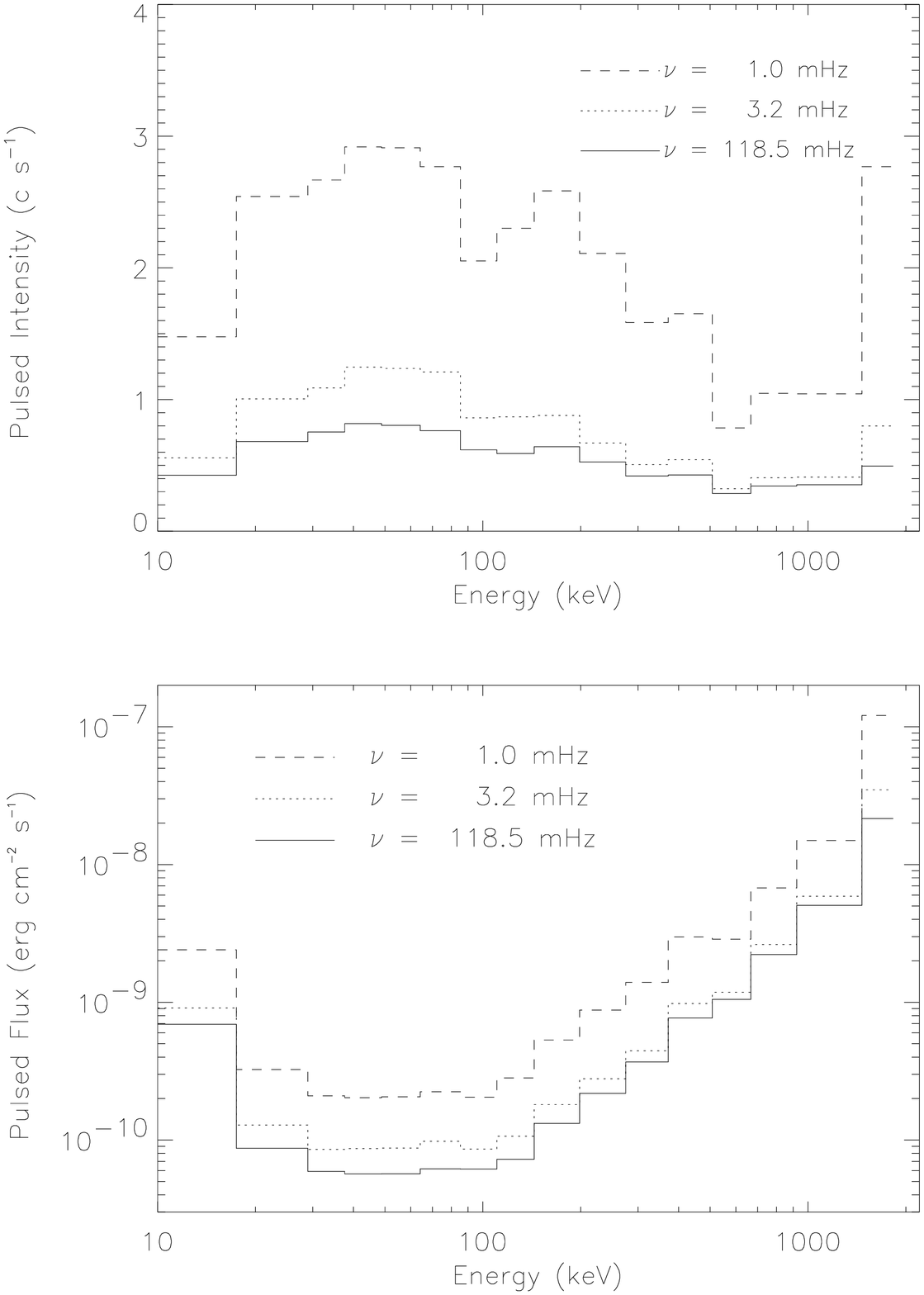,height=7in}
\figcaption{The 1--day BATSE sensitivity 
for a single large-area detector to an unocculted 
pulsed source at three pulse frequencies. We define the threshold
pulsed intensity (upper panel) as the count rate where the error in
the measured rate is 20 \%. The degradation in sensitivity at low
frequencies is due to residual background not fully accounted for by
the background subtraction process. The lower panel shows the energy
flux corresponding to these threshold count rates, assuming that the
source has a power--law spectrum with photon number index $\alpha =
4.0$ and is viewed at normal incidence.  A typical live time of 
$\sim$ 42000 s is obtained per day.
\label{fig:pulsens}
}
%\end{figure}

%\newpage
%\begin{figure}
\psfig{file=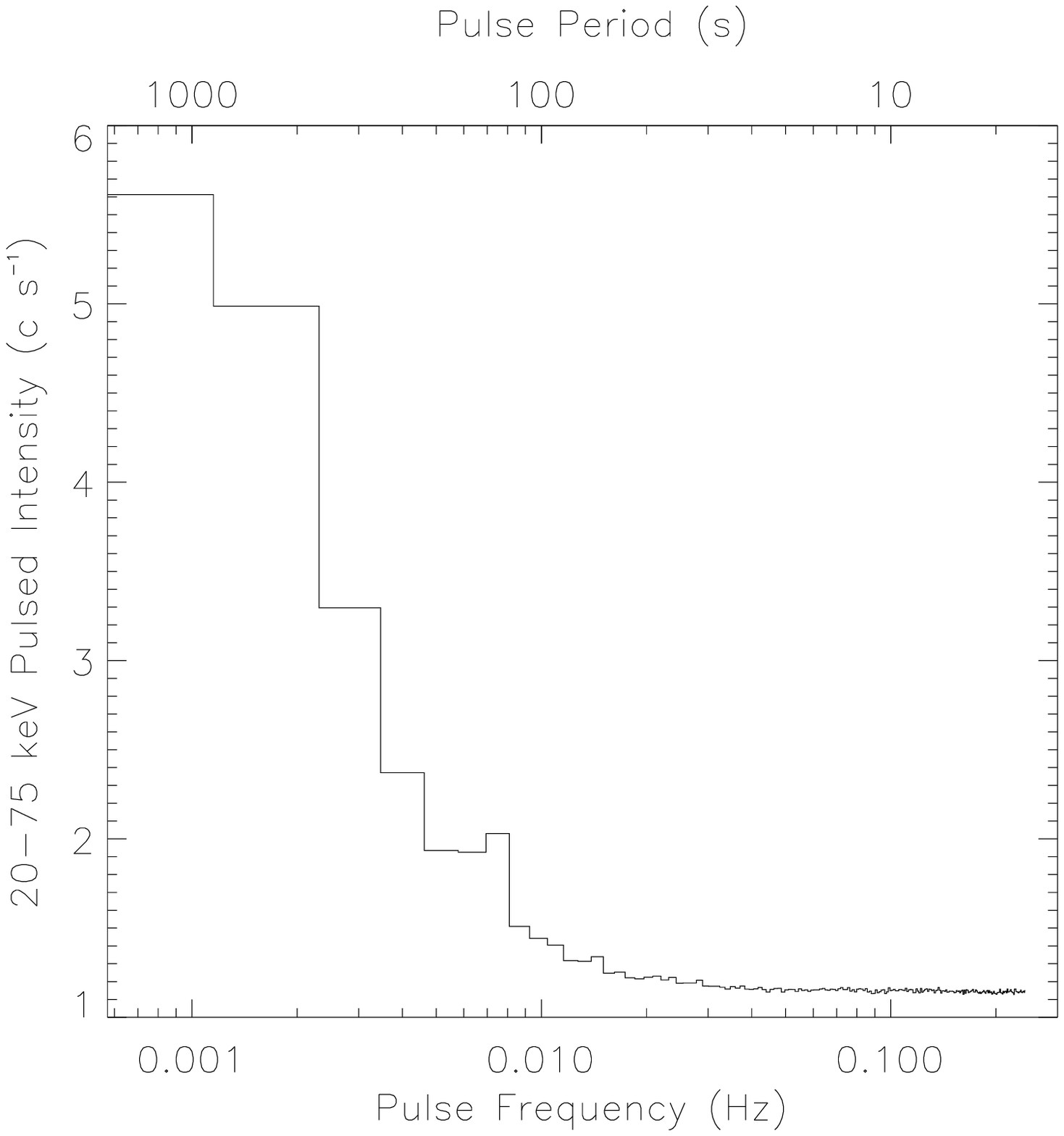,height=5in}
\figcaption{The 1-day BATSE 5$\sigma$ detection threshold for an
unocculted source (CONT channels 1 to 4 summed) following background
subtraction.  The thresholds plotted were obtained by averaging the
thresholds for each of the 8 BATSE detectors from MJD 49081 to
49093.  A typical day is characterized by a mean background rate
(CONT channels 1 to 4 summed) of $\approx 2000\ \cps$ and $\approx
42000$ seconds of useful data.  At frequencies $\nu \gsim$ 0.02~Hz,
the background noise is essentially Poisson and the threshold is
simply that given by equation (\protect\ref{eq:countlimit}). The thresholds at
lower frequencies are found by assuming local Gaussian statistics for
the measured noise strength.  The anomalous rise in the thresholds at
$\nu \approx 0.007$~Hz is attributable to the second harmonic of Vela
X-1, which was not accounted for in the background subtraction model.
\label{fig:threshold} }
%\end{figure} 

%%%%%%%%%%%%%%%%%%%%%%%%%%%%%%%%%%%%%%%%%%%%%%%%%%%%%%%%%%%%%%%%%%%%%%%%%%%%%%%

%\input science.tex

\section{OVERVIEW OF ACCRETION-POWERED PULSARS}

The discovery of orbitally-modulated, periodic X-ray pulsations from
Cen X-3 by {\em Uhuru} (Giacconi et al. 1971; Schreier et al. 1972)
\nocite{Giacconi71,Schreier72}) quickly led to a qualitative
understanding of X-ray pulsars as rotating magnetized neutron stars
accreting matter from a binary companion (Pringle \& Rees 1972;
Davidson \& Ostriker 1973; Lamb et al. 1973).  The neutron star accretes
matter either by capturing material from the stellar wind of the
companion or through Roche lobe overflow of the mass-donating
star. The strong surface magnetic field (typically $B \sim 10^{12}$ G)
controls the accretion flow close to the neutron star, where, in the
simplest picture, the ionized matter follows the field lines onto the
magnetic poles. The resulting accretion luminosity from the polar
regions is
\begin{equation}
L_{\rm acc} = \frac{GM_{\rm x}\dot M}{R_{\rm x}}\simeq 1.2 \times 10^{36}
{\mbox{\rm erg} \ \mbox{\rm s}^{-1}}
\left(\frac{\dot M}{10^{-10} M_\odot \mbox{\rm\ yr}^{-1}}\right)
\left(\frac{M_{\rm x}}{1.4 M_\odot}\right)
\left(\frac{10\ \mbox{\rm km}}{R_{\rm x}}\right)
\end{equation}
where $\dot M$ is the instantaneous mass accretion rate, and $M_{\rm
x}$ and $R_{\rm x}$ are the neutron star mass and
radius. Both the misalignment of the rotation axis with the dipolar
field and asymmetric emission from the accreting polar cap leads to
pulsed emission at the neutron star spin period. Many of these
accreting pulsars were known and studied prior to the launch of the
BATSE instrument and this section is mostly an overview of the
``standard'' picture of these objects developed with pre-BATSE
observations. We discuss, in \S 5, how our understanding of accreting
pulsars has been re-shaped by BATSE observations.

\nocite{White83}
\nocite{Mihara95}
  
The simple blackbody temperature estimate (i.e. $L_{\rm acc}=A_{\rm
cap} \sigma_{\rm SB} T_{\rm eff}^4$, where $A_{\rm cap}\sim {\rm
km^2}$ is the typical polar cap area) gives $kT_{\rm eff} \sim 3 \
{\rm keV}$, comparable to where the peak in $\nu F_\nu$ usually
appears in the X-ray spectrum. The observed 2--100 keV X-ray spectra
are much harder than a blackbody and have been represented by a
variety of models, most commonly a power law with an exponential
cutoff or a broken power law (e.g., White et al. 1983; Mihara
1995).  The exponential cutoff energy falls in the 5--25 keV
range, while the power-law photon index is typically $\alpha\lesssim
1.5$ below the cutoff energy.  It is the hard power-law tail that we
typically detect with the BATSE instrument.  The pulse profiles of
accreting pulsars are relatively smooth and simple (i.e. single or
double peaked) above 20 keV. The pulsed fraction (see appendix B.2)
is typically greater than 50\% and normally increases 
with photon energy (see Figure \ref{fig:pulsemosaic} 
for examples of pulse profiles in the BATSE energy
range). The pulse profiles at lower energies are generally more
complex (see White et al. 1983 for examples).

 Table {\tabpulsars} shows the presently known accreting pulsars, which are
generally classified according to the mass of the donor star as either
{\em low-mass} ($M_{\rm c} \lesssim 2.5\,M_\odot$) or {\em high-mass}
($M_{\rm c}\gtrsim 6\,M_\odot$) systems (\cite{Shore94}).  Systems
which have been detected by BATSE are marked with a bullet and those
discovered by BATSE are marked with a circled bullet.  
Table {\taborbits} shows
the presently known orbital parameters, with BATSE measurements and
discoveries marked as in Table {\tabpulsars}. 
There are only four known low-mass
binaries with accreting pulsars: Her X-1, 4U 1626-67, GX 1+4, and GRO
J1744--28. The overwhelming majority of low-mass X-ray binaries are
not pulsars and thus evidently have fields too weak to strongly affect the
accretion flow ($B \lesssim 10^{9} \ {\rm G}$). The high mass binaries
may be divided into those with main sequence Be star companions and
those with evolved OB supergiant companions. The Be systems, which
account for more than half of the known accreting pulsars, are
generally observed during transient outbursts. The mass donor in these
systems is an O or B star still on the main sequence and lying well
inside its Roche surface. The episodic outbursts are often correlated
with periastron passage of the neutron star in its eccentric
orbit. These systems are thought to undergo a ``propeller'' phase
during X-ray quiescence (\cite{Stella86}).  

The supergiant binaries may be further subdivided into two groups
according to the dominant mode of mass transfer: Roche-lobe overflow
or capture from the stellar wind. In some systems, both types of mass
transfer may be taking place (\cite{Blondin91}). 
Most OB supergiants have stellar winds driven by the radiation
pressure from resonance lines of highly-ionized atoms, with mass loss
rates of $\dot M \sim 10^{-6} M_\odot \ {\rm yr^{-1}}$ being quite
typical. Although capture from a high velocity wind is inefficient,
the large mass loss rate in the wind can result in an appreciable
mass accretion rate onto the neutron star. Vela X-1 is the best known
example of a wind-fed supergiant pulsar binary.

\psfig{file=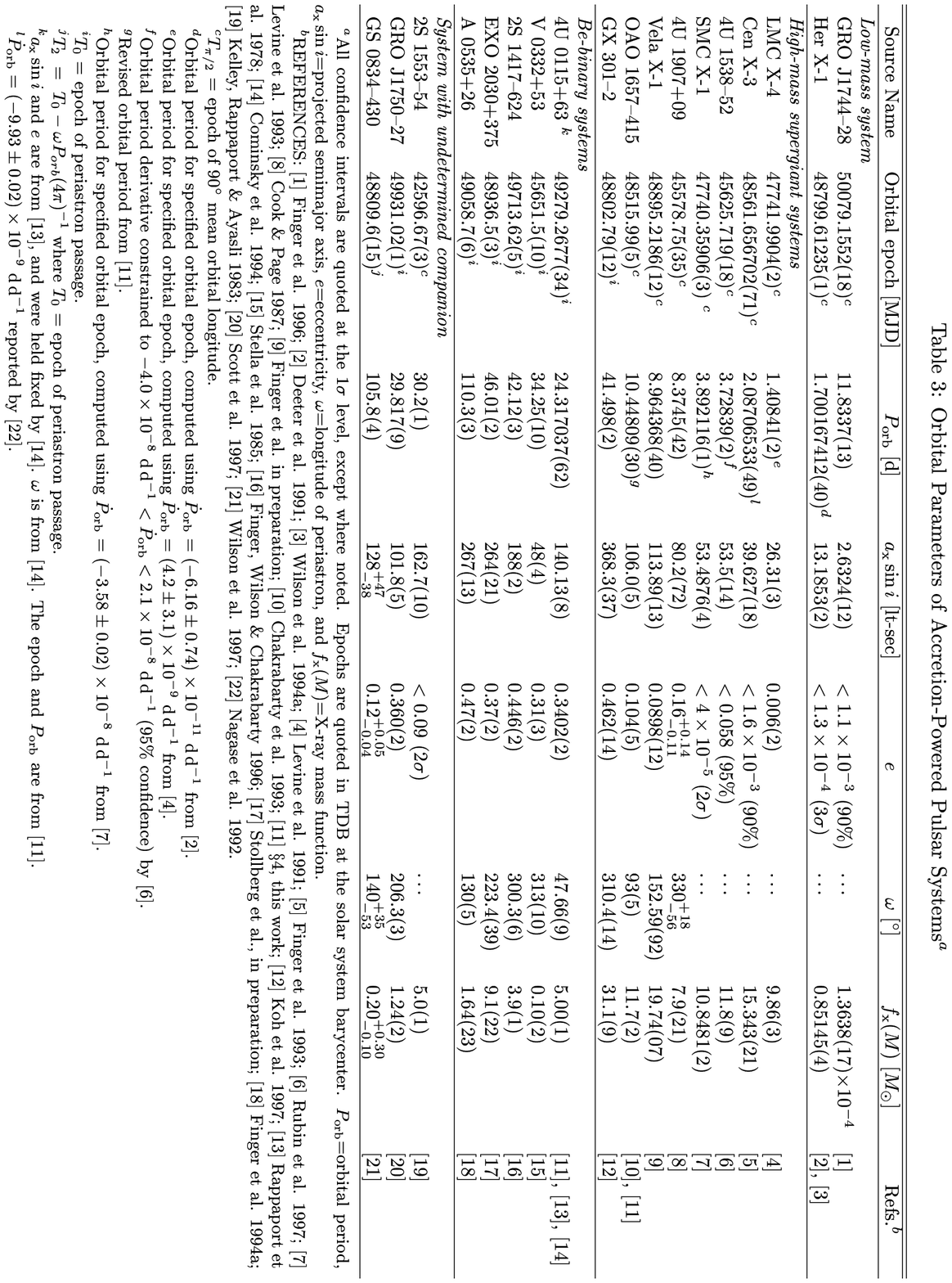,height=9in}

 If the mass donor fills its Roche lobe, material from the companion
flows with high specific angular momentum through the first Lagrange
point and forms an accretion disk around the neutron star. This is a
very efficient form of accretion, and results in a mass transfer rate
much larger than by capture of the wind alone. The large
persistent accretion rates in SMC X-1, Cen X-3, and LMC X-4 make them
prime candidates for disk-fed (via Roche lobe overflow) supergiant
pulsar binaries. Optical photometric observations of these systems
show both ellipsoidal variations consistent with a tidally distorted
companion and excess light due to the presence of an accretion disk
(\cite{vanParadijs95} and references therein). As we
discuss later, the accretion torque magnitude and temporal behavior is
also indicative of accretion from a Keplerian disk.

 A convenient organization of the high-mass systems emerges by
plotting those with known orbital periods on a $P_{\rm spin}$-$P_{\rm
orb}$ diagram (\cite{Corbet86,Waters89}), where
$P_{\rm spin}$ is the neutron star spin period. As is evident in
Figure \ref{fig:corbet}, the neutron stars orbiting Roche lobe filling
supergiants (asterisks) have short spin periods ($P_{\rm spin} \lesssim
10$~s) and short orbital periods ($P_{\rm orb} \lesssim 4$~d).  They are
quite luminous ($L_{\rm x} \gtrsim 10^{37}$
erg~s$^{-1}$) and tend to show long episodes ($\gtrsim P_{\rm orb}$)
of relatively steady torques.  On the other
hand, the wind-fed supergiant binaries (squares) have longer orbital
periods (as required to avoid Roche lobe overflow), longer spin
periods, and are less luminous ($L_{\rm x}\sim 10^{35}$--$10^{37}$
erg~s$^{-1}$). In addition, the observed accretion torque on these
wind-fed objects often fluctuates (even between spin-up and spin-down)
on timescales much shorter than the orbital period. Finally, the Be
transients (circles) populate a third region of the diagram,
displaying a marked correlation between their spin and orbital
periods.  It has been suggested that this correlation arises
from the fact that, 
given identical companion masses and mass loss rates, 
neutron stars in systems with longer orbital periods are further away 
from their companions, thus leading to lower mass accretion rates
and higher equilibrium periods.  In 
addition, Waters \& van~Kerkwijk (1989) argue
that selection effects favor the detection of Be systems which are
in equilibrium with the slow equatorial wind of the companion.
The observed anti-correlation of spin and orbital periods for 
the Roche-lobe filling supergiants is not understood
As Figure \ref{fig:galdistrib} shows, most of the high mass systems are found 
in the Galactic plane, consistent with the short lifetime of the massive
companions.

\nocite{Waters89}

The torque exerted on an accreting star depends on the nature of the
angular momentum transfer during the accretion of matter. Accreting
pulsars are the only objects where such measurements have been made
repeatedly. The much larger moment of inertia of an accreting magnetic
white dwarf (in particular the DQ Her systems; \cite{Patterson94})
requires a decade-long baseline to measure the change in spin period
and so only one torque measurement has typically been made for each
object. Patterson's (1994) Table 1 shows five which are spinning up
and one (V1223 Sgr) which is spinning down.  BATSE's ability to
repeatedly measure the spin frequency of accreting pulsars has allowed
us to {\it monitor} the torque exerted during accretion. We
have found that spin-up and spin-down are nearly equally prevalent in
these systems, contrary to the picture in the 1970s, when most
accreting pulsars were then observed to be spinning up steadily (see
Figure 5 in \cite{Joss84}).

 Assuming the gas deposits its angular momentum at the magnetospheric
boundary and that field lines transport all of this angular momentum
to the star (\cite{Pringle72,RappaportJoss77b}),
the accreting pulsar will experience a spin-up torque
\begin{equation}
N \approx \dot M\sqrt{GM_{\rm x} r_{\rm m}},
\label{eq:torque}
\end{equation} 
where $r_{\rm m}=\xi r_{\rm A}$ is the magnetospheric radius with 
the Alf\'ven radius
\begin{equation}
r_{\rm A} = \left( \frac{\mu^4}{2GM_{\rm x}\dot M^2}\right)^{1/7}\simeq
6.8\times 10^8 \ {\rm cm}
\left({\mu\over 10^{30} \ {\rm G\ cm^3}}\right)^{4/7}
\left({{10^{-10} \ M_\odot \ {\rm yr^{-1}}}
\over {\dot M}}\right)^{2/7}
\left( \frac{1.4M_\odot}{M_{\rm x}}\right)^{1/7}
\label{eq:alfven}
\end{equation}
being a characteristic length found by equating magnetic and fluid
stresses for a neutron star with magnetic moment $\mu$. Estimates for
the model dependent dimensionless number $\xi$ range from 0.52
(\cite{GhoshLamb79}) to $\approx 1$ (Arons 1993; Ostriker \& Shu 1995;
Wang 1996) \nocite{Arons93,OstrikerShu95,Wang96} for the case at hand,
where $r_m\sim 10^8 \ {\rm cm}$.  The detailed physics by which
material at this magnetospheric boundary loses its orbital angular
momentum, becomes entrained on the magnetic field lines, and makes its
way to the magnetic polar caps is thought to involve
magnetohydrodynamical versions of Rayleigh-Taylor and Kelvin-Helmholtz
instabilities (Lamb et al. 1973; Arons \& Lea 1976, 1980; Elsner \&
Lamb 1977). \nocite{Lamb73,AronsLea76,AronsLea80,ElsnerLamb77}

Accretion will be inhibited by a centrifugal barrier if the pulsar
magnetosphere rotates faster than the Kepler frequency at the inner
disk boundary. For accretion to occur, the magnetospheric radius
should thus lie inside the corotation radius
\begin{equation}
r_{\rm co} = \left( \frac{GM_{\rm x} P_{\rm spin}^2}{4\pi^2} \right)^{1/3}
\simeq 1.7 \times 10^8 \mbox{\rm\ cm} 
\left( \frac{P_{\rm spin}}{1\ \mbox{\rm s}} \right)^{2/3}
\left( \frac{M_{\rm x}}{1.4 M_\odot} \right)^{1/3}.
\label{r_co}
\end{equation}
For the case that $r_{\rm m}< r_{\rm co}$ while accreting, there is 
a characteristic torque,
\begin{equation}
N_0\equiv \dot M\sqrt{GM_{\rm x} r_{\rm co}},
\label{eq:maxtorque}
\end{equation}
which is a convenient fiducial as it only depends on the observable
spin period of the pulsar and the inferred accretion rate.
The fiducial torque sets a scale, and the actual torque may
by significantly smaller.
A pulsar subject to the torque in (\ref{eq:torque}) will spin up
at a rate 
\begin{equation}
\dot \nu = {N\over 2\pi I}\simeq 1.6\times 10^{-13} {\rm s^{-2}}
\left({\dot M}\over {10^{-10} M_\odot \ {\rm yr^{-1}}}\right)
\left(P_{\rm spin}\over {\rm s}\right)^{1/3}
\left(\frac{r_{\rm m}}{r_{\rm co}}\right)^{1/2}
\end{equation}
where $I\simeq0.4 M_{\rm x} R_{\rm x}^2$ 
is the neutron star's moment of inertia
(\cite{Ravenhall94}). The timescale for spinning up the neutron
star is then
\begin{equation}
t_{\rm spinup}\equiv {\nu\over \dot \nu} \simeq 2\times 10^5 \ {\rm yr}
\left({10^{-10} M_\odot \  {\rm yr^{-1}}}\over \dot M\right)
\left( \frac{1\ \mbox{\rm s}}{P_{\rm spin}}\right)^{4/3}
\left({r_{\rm co}\over r_{\rm m}}\right)^{1/2},
\label{eq:tspin}
\end{equation}
\nocite{Elsner80}
much shorter than the ages of most X-ray binaries (Elsner, Ghosh and
Lamb 1980).  Hence, in this simple picture, the neutron star spins up
until the spin frequency matches the Kepler frequency at the
magnetosphere (or where $r_{\rm m}\approx r_{\rm co}$)
\begin{equation}
P_{\rm spin,eq} \approx 8 \ {\rm s}
\left({10^{-10} M_\odot \ {\rm yr^{-1}}}\over 
\langle {\dot M} \rangle \right)^{3/7}
\left({\mu\over 10^{30} \ {\rm G\ cm^{3}}}\right)^{6/7}.
\label{eq:peq}
\end{equation}
Here $\langle \dot M \rangle$ is an appropriately averaged mass
accretion rate. Presumably, neutron stars with shorter periods than
$P_{\rm spin,eq}$ cannot accrete easily, and may experience a strong
spin-down torque --- the so-called ``propeller effect'' (Illarionov and
Sunyaev 1975). If the observed accreting pulsars are near their
equilibrium spin periods, then one infers magnetic field strengths
from eq. (\ref{eq:peq}) in the range $10^{11} -10^{14}$\,G. The
instantaneous accretion rates and observed torques can be much
different than their long-term averages, however, so one only obtains
a rough measure of the pulsar magnetic field in this way.

A more complex picture of accretion torques emerged as more
systems were discovered and spin histories were extended.  Her
X-1 and LMC X-4 were found to be spinning up on a much longer
timescale than predicted by equation (\ref{eq:tspin}). Some pulsars
show secular spin-down behavior while continuing to accrete (
4U~1626-67, GX~1+4, 1E~1048.1--5937, 1E~2259+586, 
E1145.1--614, 4U~1538--52), while others show
more erratic variations in spin period (Cen X--3, Vela X--1, X Per). 
More sophisticated theories of accretion torque were developed
to explain these observations, which take into account the magnetic 
interaction between the inner accretion disk and the neutron star 
magnetosphere (\cite{GhoshLamb79,Anzer80,Arons84}).  However, as we show 
in \S 4, the continuous 
monitoring of accreting pulsars by the BATSE instrument has 
now substantially changed our view of many of these systems. 
We discuss these new observations and their implication for 
theories of accretion torque in \S 5.

%\begin{figure}
\psfig{file=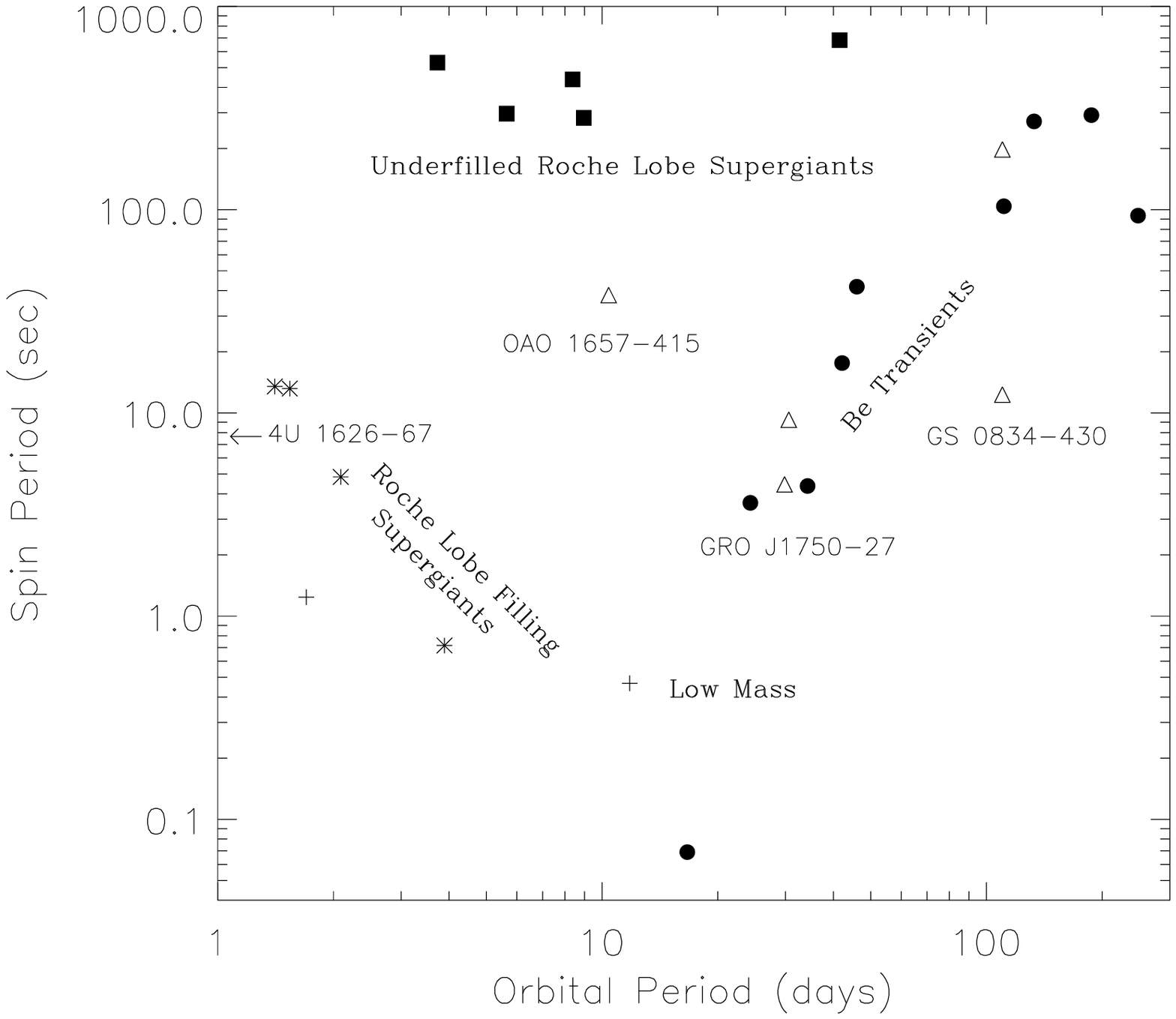,height=5in}
\figcaption{ The spin period of the accreting neutron stars
versus the binary period (the Corbet diagram). The different symbols
refer to the type of binary the neutron star resides in. Asterisks are
supergiant companions which are Roche-Lobe filling, squares are
supergiant companions that underfill their Roche lobe, circles are
confirmed Be transient binaries, and crosses are those with low
mass ($\lesssim 2 M_\odot$) companions (Her X-1, GRO J1744--28, 4U
1626--67). Triangles refer to sources for which there are no optical
companions yet identified, though orbits have been measured (GS
0834--430, OAO 1657--415 and GRO J1750--27).
\label{fig:corbet}}
%\end{figure}

%\begin{figure}
%\psfig{file=pulsars_1.ps,height=8in,angle=90}

\centerline{
\psfig{file=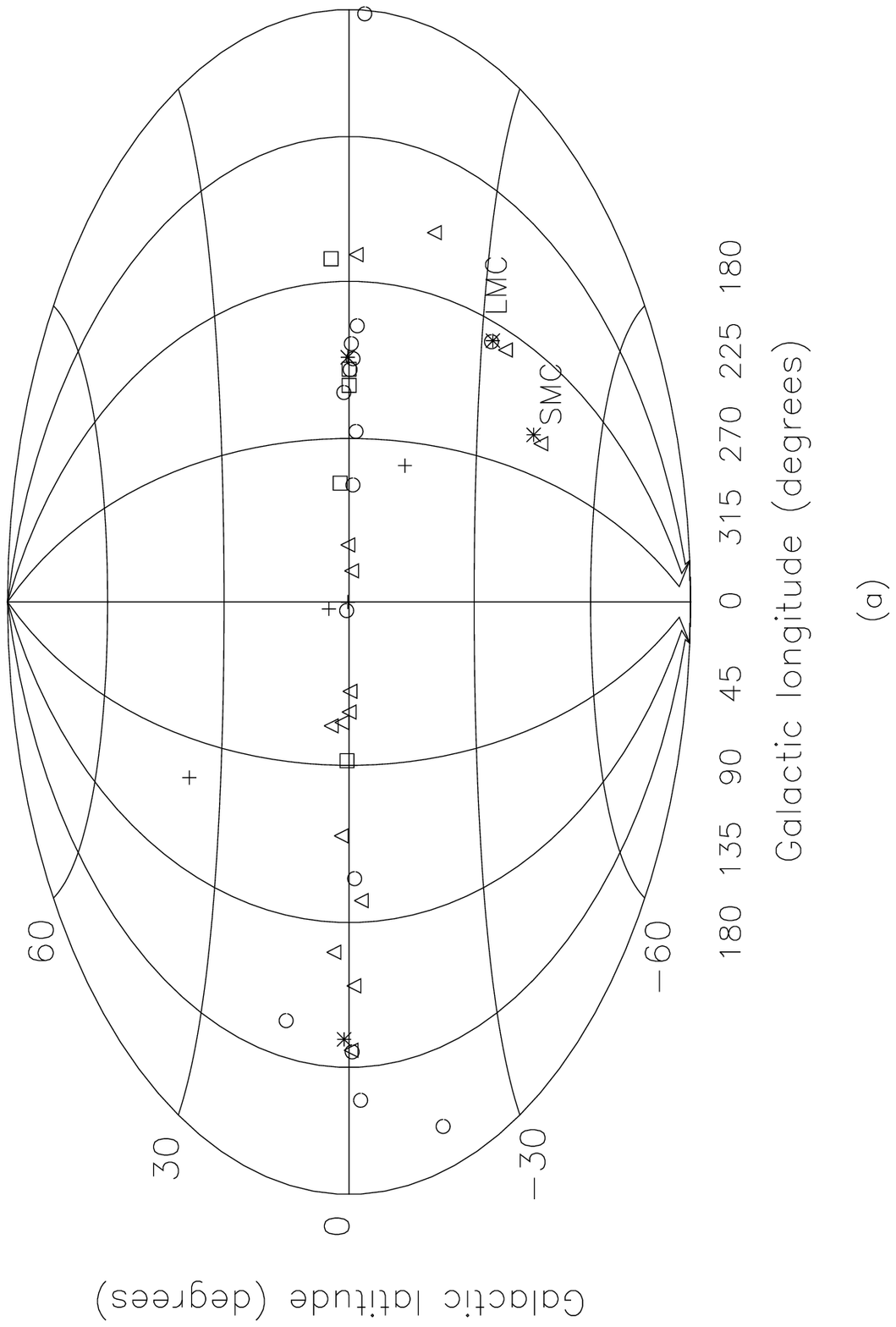,height=8in}
}
\centerline{
\psfig{file=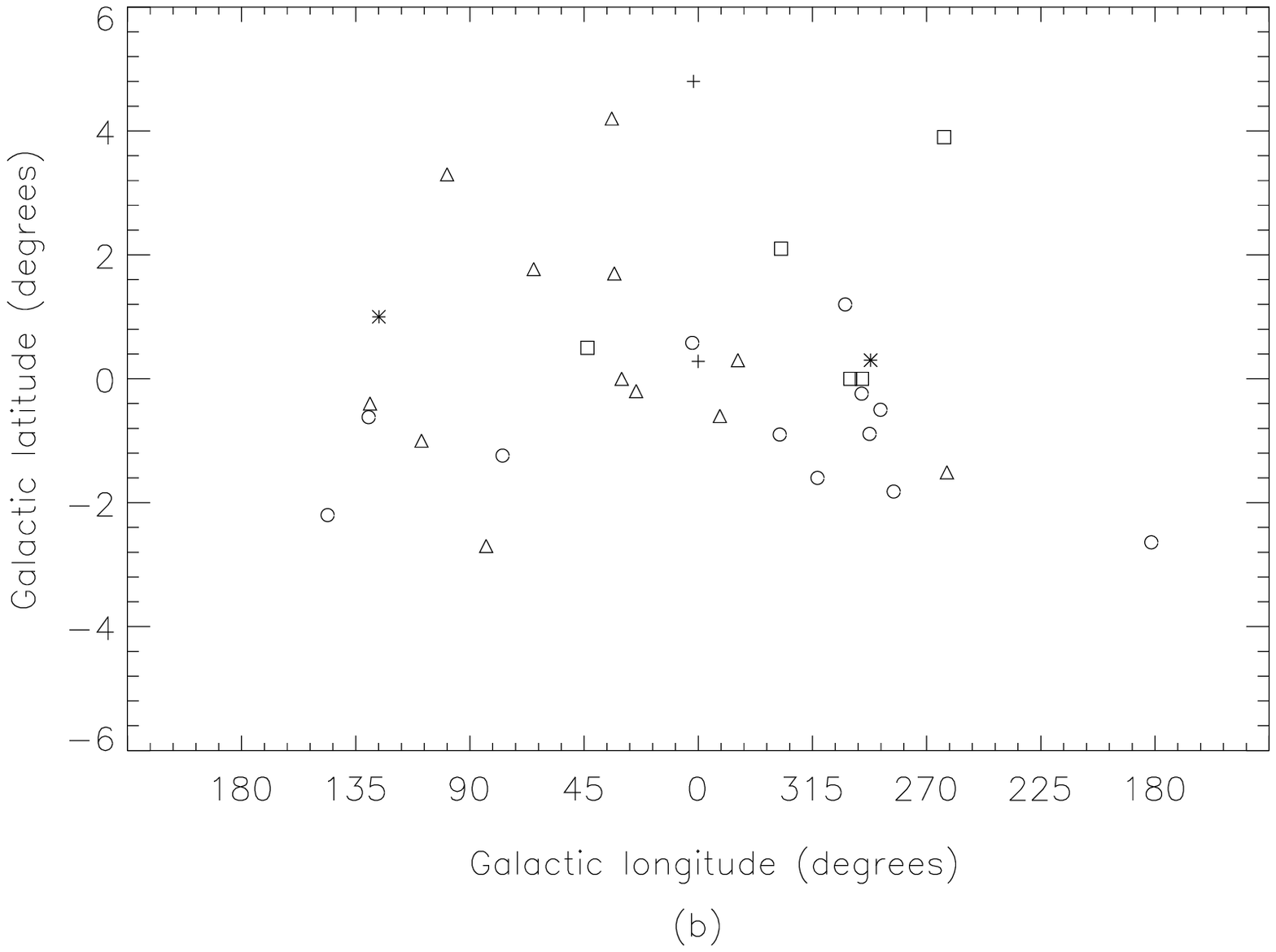,height=7in}
}
\vskip -8cm
\figcaption{The Galactic distribution of accreting pulsars. We plot
the known Be transients as open circles, the low-mass objects as
pluses, the disk-fed OB supergiants with asterisks and the wind-fed OB
supergiants with squares.  This contains all pulsars listed in Table
\protect{\tabpulsars}. Triangles denote sources 
with unknown companions and orbit. This
symbol convention is the same as in Figure \protect\ref{fig:corbet}, except
here the symbols are not filled.
\label{fig:galdistrib}}
%\end{figure}

\begin{figure}
\psfig{file=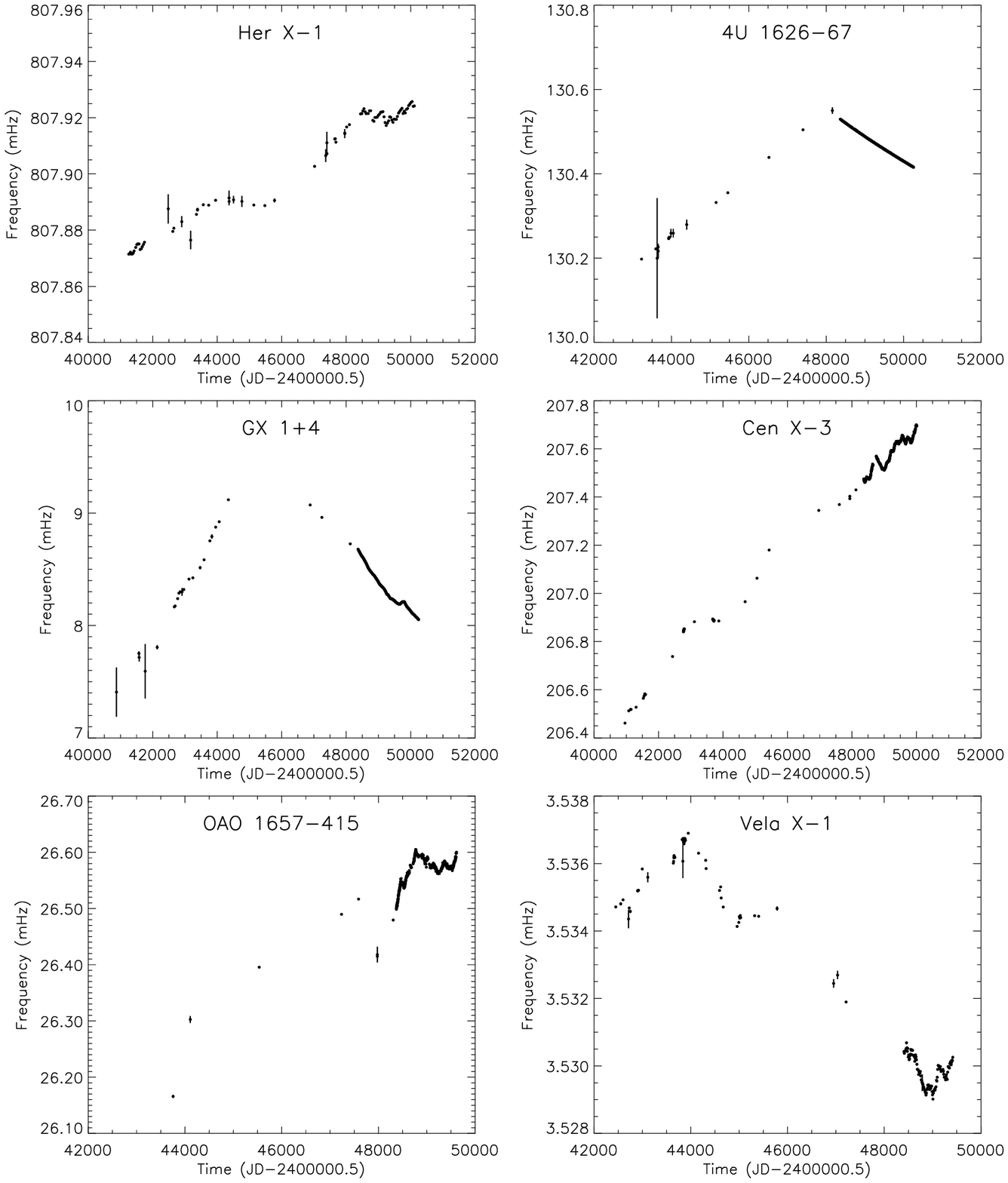,height=7in}
\end{figure}

\begin{figure}
\psfig{file=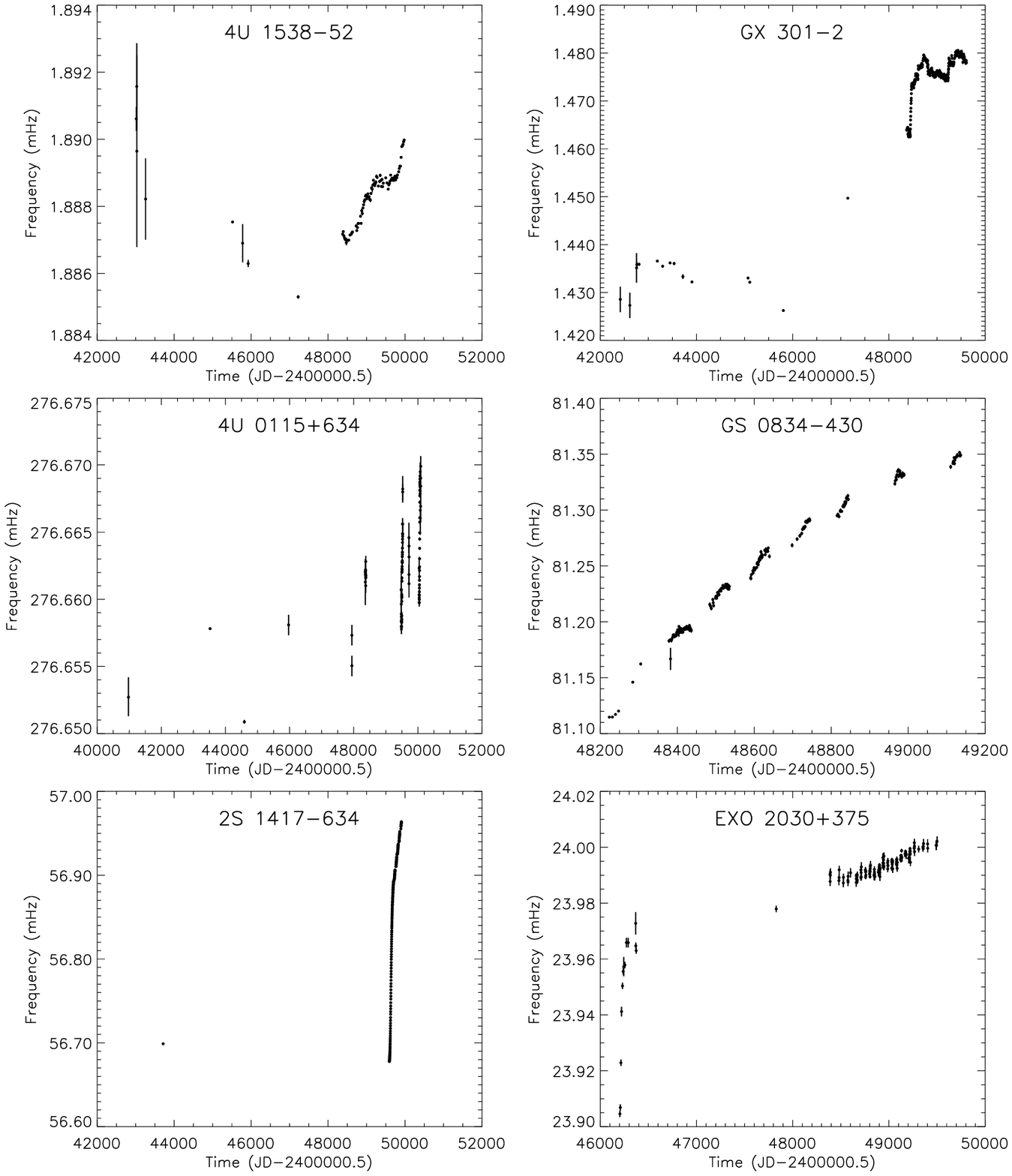,height=7in}
\end{figure}

%\addtocounter{figure}{-2}
\begin{figure}
\psfig{file=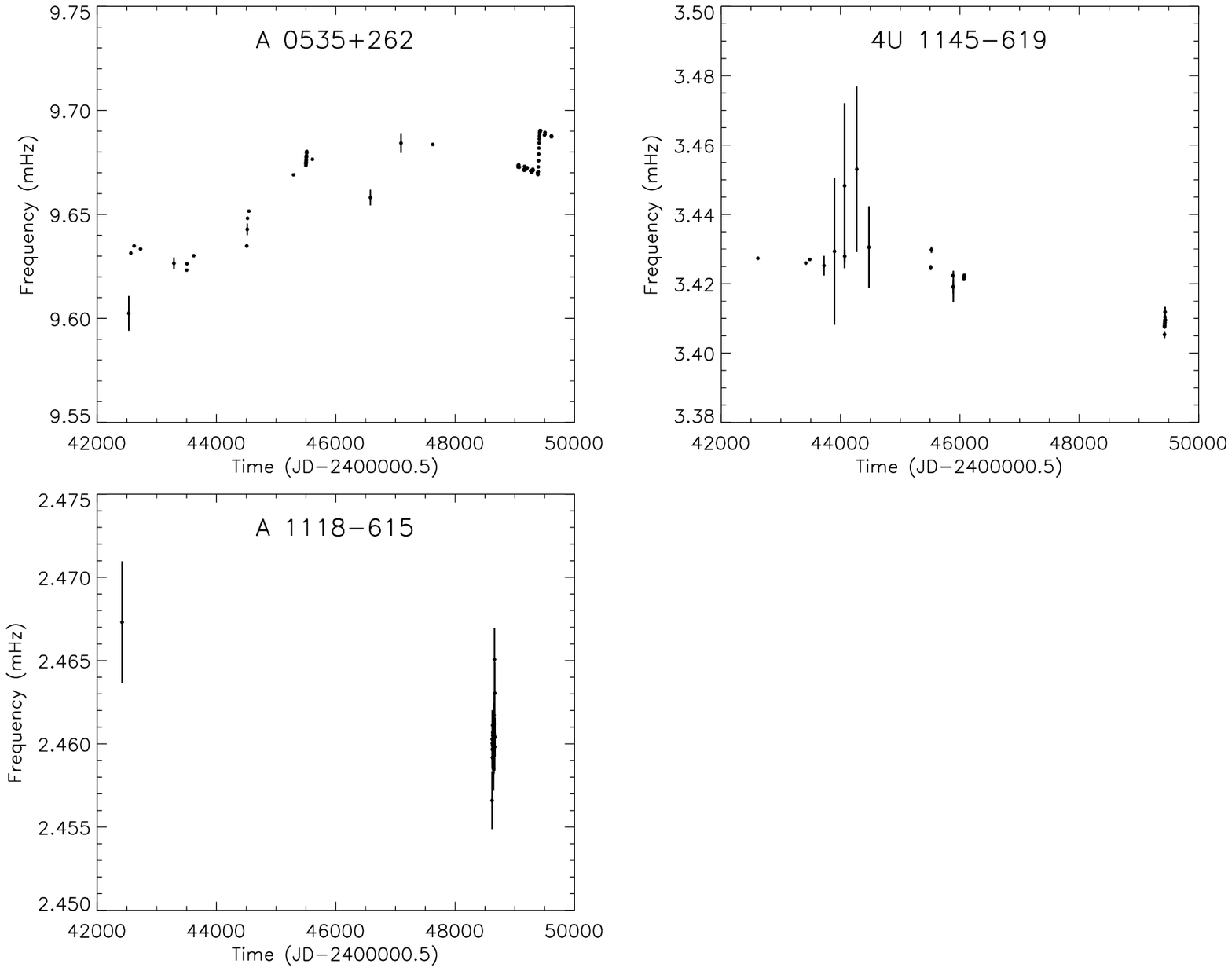,height=7in}
\vspace{-5cm}
\figcaption{
The long-term frequency history for all pulsars detected by BATSE that
were previously known. The squares show the pre-BATSE data taken from
Nagase (1989) and additional references.  The line is the BATSE data,
which we discuss later in great detail.  The long term frequency
history for X-ray pulsars observed by BATSE that were known prior to
the {\em Compton Observatory} launch commences April 1991.  For Her X-1, Cen
X-3, Vela X-1, 4U 1538--52, GX 301--2, 4U 0115+634, and EXO 2030+375
all frequencies have been orbitally corrected. For OAO 1657--415, GS
0834--430, 2S 1417--62, and A~0535+262 orbital corrections have been
applied only to the BATSE observations. No orbital corrections have
been applied for 4U 1626--67, GX 1+4, 4U 1145--619, or A1118--615,
which have unknown, or incompletely known, orbital elements. The BATSE
frequencies for OAO 1657--415, GS 0834--430, EXO 2030+375, 4U
1145--619 and A 1118--615 are from the daily frequency history files
we have deposited at the {\em Compton Observatory} archive. For the
remainder of the objects, the frequencies are from source specific
studies: Her X-1 (Wilson, et al., in prep.)  4U 1626--67 (Chakrabarty
et al. 1997), GX 1+4 (Chakrabarty et al., in prep.), Cen X-3 (Finger
et al. in preparation); Vela X-1 (Finger et al. in preparation); 4U
1538--52, (Rubin et al. 1997); GX 301--2 (Koh et al. 1997);
4U 0115+634 (Cominsky et al. in preparation); 2S 1417--624 (Finger et
al.(1996); and A~0535+262 (Finger et al. 1996).
\label{fig:freqmosaic} }
\end{figure}

\psfig{file=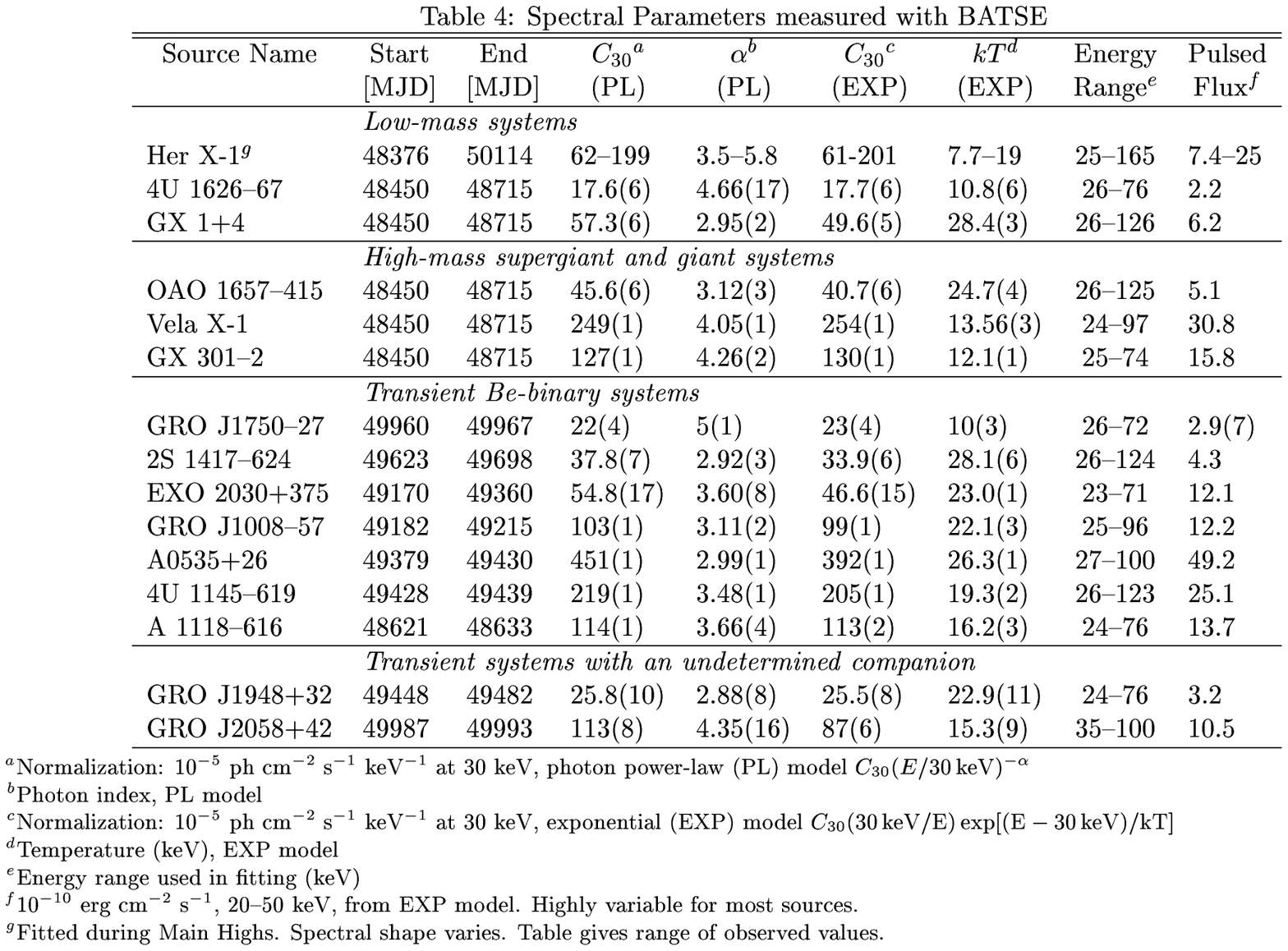,height=9in}

%%%%%%%%%%%%%%%%%%%%%%%%%%%%%%%%%%%%%%%%%%%%%%%%%%%%%%%%%%%%%%%%%%%%%%%%%%%%%%%

%\input results_lm.tex		% low mass sources

\section{BATSE OBSERVATIONS OF INDIVIDUAL SOURCES}

BATSE continuously monitors the spin frequency and pulsed flux of 3
low-mass systems (Her X-1, 4U~1626--67, and GX 1+4) and 5 high-mass
systems (Cen~X-3, OAO~1657--415, Vela~X-1, 4U~1538--52, and
GX~301--2). BATSE has also observed one or more outbursts from 7 known
transient systems (4U~0115+63, GS~0834--430, 2S~1417--624,
EXO~2030+375, A~0535+26, 4U~1145--619, and A~1118--616). In addition,
it has discovered 5 new transients (GRO~J1744--28, GRO~J1750--27,
GRO~J1948+32, GRO~J1008--57, and GRO~J2058+42).  In 
Figure~\ref{fig:freqmosaic} we display long-term frequency histories of
all sources seen with BATSE that were known prior to BATSE, including
archival data.

In this section, intrinsic spin
frequency and flux histories for the persistently accreting binaries
are presented for the first four years of BATSE monitoring, 1993 April
23--1995 Feb 11 (MJD 48370--49760).  We also show frequency and flux
histories of the outbursts for the transient sources. For those
transients where we have yet to measure the orbital parameters, we
display the observed frequencies, whereas we display the intrinsic
spin frequencies for neutron stars with measured orbital parameters.
Up-to-date results on these sources are being made available through
the public archive at the {\em Compton Observatory} Science Support
Center (http://cossc.gsfc.nasa.gov/cossc/COSSC\_HOME.html).

  In the following subsections, we categorize the accreting pulsars by
the type of star they are accreting from: low mass stars ($M\lesssim 2
M_\odot$), OB supergiants, and main sequence Be stars. We then briefly
summarize the BATSE observations of each individual binary, focusing
on those results obtained from continuous timing and pulsed flux
monitoring. 

Spectral fits to the pulsed flux
are tabulated in Table~{\tabspectra} for most sources. 
These are typical spectra.  Exposure varies from source to source.  The
interval used for spectral fitting is not always the brightest the source
displayed.  For some sources, spectra could not be determined
because the spin frequency was too high, the pulse profile varied with 
energy and/or luminosity, or the source was too weak.

Pulse profiles for all sources seen with BATSE are displayed in
Figure~\ref{fig:pulsemosaic}.  
High energy pulse profiles in accretion-powered pulsars are
generally simpler and smoother than those $\lesssim$10\,keV, as they
are less affected by circumstellar scattering and absorption. Thus,
they may be more indicative of the instrinsic radiation pattern from
the neutron star.  The profiles in Figure~\ref{fig:pulsemosaic} can be
classified broadly as single or double peaked.
Aside from 4U~1145--619
and Her~X-1 the profiles are essentially featureless, although they are
visibly asymmetic, as can be seen clearly in OAO~1657--415 and Vela~X-1.
Cases where the pulse shape changes dramatically with energy or
flux are discussed individually.  Fluxes measured with BATSE have been
found using both power-law and simple exponential models. 

%\begin{figure}[t]
%\psfig{file=all_profs.ps,width=7.0in,height=7.0in}

\centerline{
\psfig{file=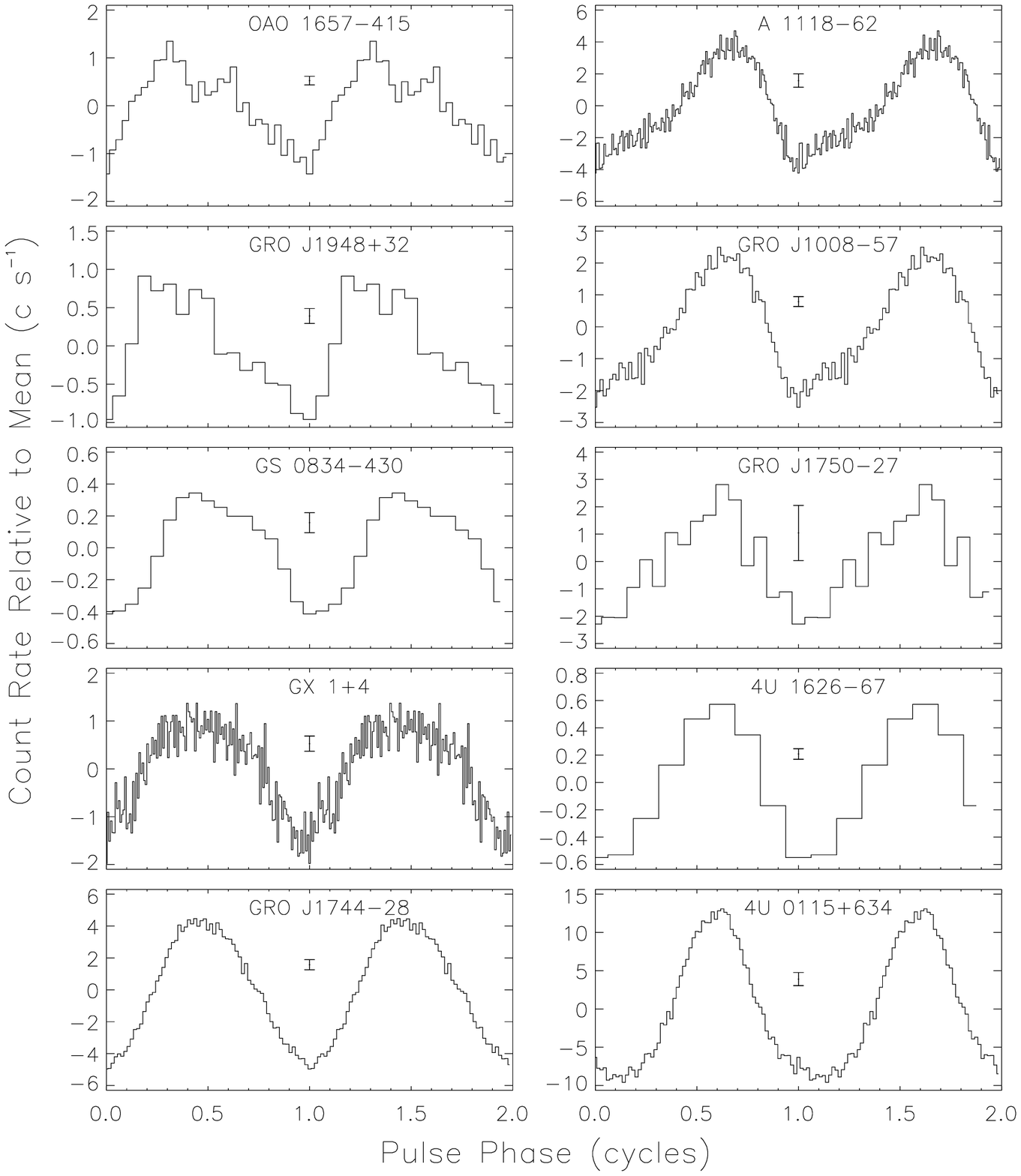,width=6.0in}%,height=7.0in}
}
\centerline{
\psfig{file=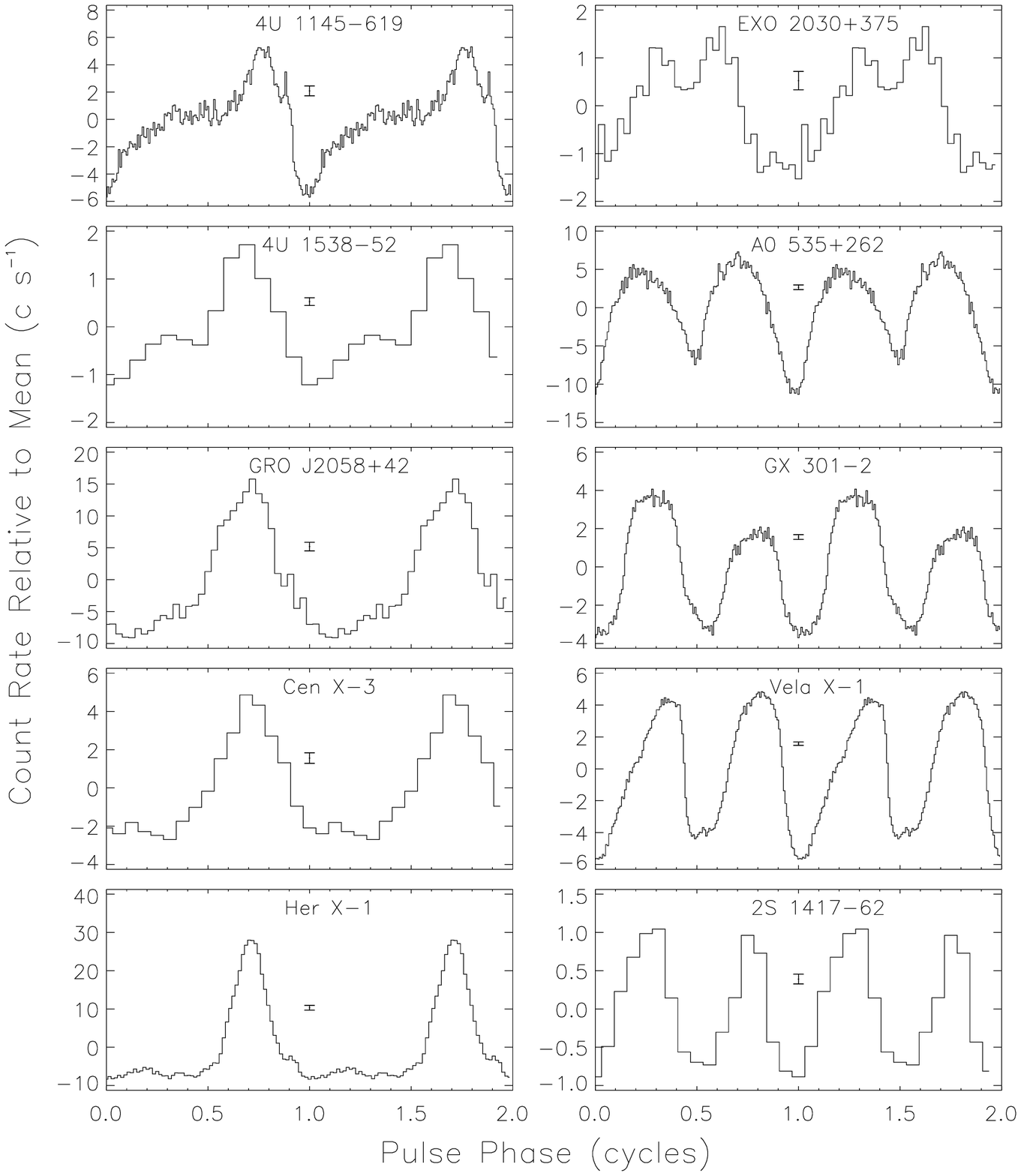,width=5.5in}%,height=7.0in}
}
%\vspace{-2.0cm}
\figcaption{
Pulse profiles of accreting pulsars from BATSE, in
20--35 keV except as noted. The $1\sigma$ 
error bar in each phase bin is also shown. The 
profile for GS 0834-430 was constructed using bin splitting rather than whole 
binning \spec. %(See Appendix B.2.2).
The range of days (in MJD) summed to construct 
the profiles for each of the sources (and energy range if different from above)
are: 
a. OAO 1657--415 (48450--48715), 
A 1118--62 (48621--48633), 
GRO J1948+32 (49448--49482), 
GRO J1008--57 (49182--49215), 
GS 0834--430 (48518--48522), 
GRO J1750--27 (49960--49967; 20--70 keV), 
GX 1+4 (48450--48715), 
4U 1626--67 (48450--48715),
GRO J1744--28 (50092--50098), 
4U 0115+634 (50049--50057; 20--40 keV) 
b. 4U 1145--619 (49428--49439), 
EXO 2030+375 (48450--48715), 
4U 1538--52 (49350--49421; 20--50 keV), 
A0 535+262 (49379--49430), 
GRO J2058+42 (49987--49993; 20--40 keV), 
GX 301--2 (48450--48715), 
Cen X-3 (48985--48992), 
Vela X-1 (48450--48715), 
Her X-1 (49586--49593; 20--40 keV), 
2S 1417--62 (49623--49698)
\label{fig:pulsemosaic}}
%\end{figure}

%\newpage
\subsection{Low-Mass Systems}

Only four accreting pulsars are definitely known to be orbiting
low mass ($M\lesssim 2 M_\odot$) stars: Her X-1 and GRO J1744--28 on the
basis of timing-based measurements of their companions' mass functions
(\cite{Tananbaum72,Finger96a}), 4U 1626--67 from optical
photometry (\cite{Middleditch81,Chakrabarty97c}), and GX 1+4 from spectroscopy
(\cite{Davidsen77,ChakrabartyRoche97}).  This is a very
heterogeneous class of objects: the mass donors are a main sequence A
star (Her X-1), a $< 0.1 M_\odot$ helium or carbon-oxygen degenerate
dwarf (4U 1626--67), and two red giants (GX 1+4, GRO J1744--28). 
The absence of an observable companion in very deep optical and IR
searches and the lack of orbital detections also suggest low mass
companions for 4U 0142+61, 1E 1048.1--5937, RX J1838.4--0301 and 1E
2259+589 (see Mereghetti \& Stella 1995 and references therein).  We
now discuss the BATSE observations of Her X-1, 4U 1626--67, GX 1+4,
and GRO J1744--28. 

\nocite{Mereghetti95}

%\newpage
%\begin{figure}
\psfig{file=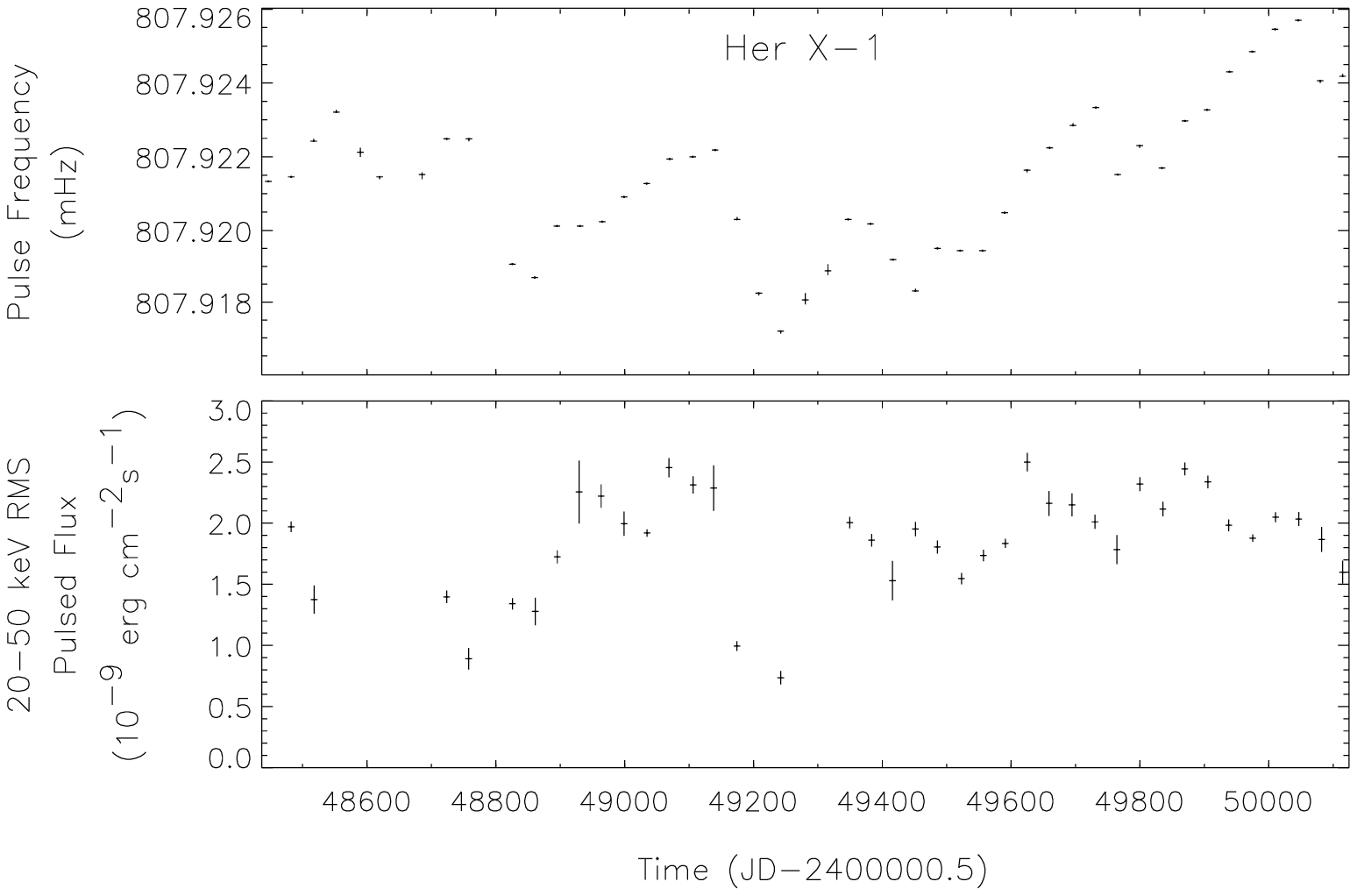}
\figcaption{
Her X-1 frequency and pulsed flux measurements from BATSE. A
mean frequency has been determined for each Main High state for which
adequate scheduled folded-on-board data was available. The 
frequencies, which have been orbitally corrected using parameters
from Wilson \etal (1994a), were obtained using linear fits 
to pulse phases of 20--70~keV data \phase.  Pulsed fluxes
are the pulse phase-averaged flux obtained during one orbital
period (averaged from eclipse egress to eclipse ingress) during the
brightest portion of each available Main High state.  An exponential model,
with the e--folding energy allowed to vary, was used to obtain the fluxes
\flux.
\label{herx1freqflux}}
%\end{figure}

\nocite{Wilson94a}

{\em Hercules X-1.} ---
{\it Uhuru} discovered 1.2~s pulsations from Her X-1 (4U1656+354)
in 1971 (\cite{Tananbaum72}) and the pulsar was subsequently found to be in 
an eclipsing, circular, 1.7~d orbit (\cite{DBP81}) around 
the low--mass companion HZ Her (\cite{Doxsey73,Gottwald91}). 
This disk-fed system 
exhibits ``super-cycles''  of intensity modulated with 
a period of $\approx $35~d (\cite{Giacconi73,Soong90}). At
energies of 1-10 keV, the source is observed during both a 
``Main High'' and ``Short High'' interval of the 35 day cycle.
Detailed discussion of BATSE observations have appeared elsewhere
(\cite{Wilson94b,Wilson94a}). 

BATSE detects pulsations for 5--10 days during each Main 
High interval.  We report an average frequency for each Main High interval
since reliable measurements of $\dot \nu$ 
within a Main High are hampered by the low signal-to-noise ratio
of the BATSE data as well as pulse shape variations.
Analyses of 20 Her X-1 Main High Observations through July 1993 found that
the pulsed 20--70 keV luminosity varied by a factor of $\sim$4, and that the
pulsar was spinning down during 35d cycles when the immediately preceding
Main High interval luminosity (averaged over the peak days of the interval),
and presumably the mass accretion rate,
was low (Wilson et al. 1994d). Subsequent observations do not
universally show a significant correlation of luminosity and $\dot \nu$,
with some episodes of spindown following intervals with high flux levels. 
The neutron star is usually spinning up 
with $\dot\nu$ between zero and a maximum of $5\times10^{-13}$~{\hps} 
between 32 of the 48 observed intervals.  As is evident
in Figure \ref{herx1freqflux}, the 
maximum spindown rates are larger in magnitude, reaching 
$7\times10^{-13}$~{\hps}. 

\psfig{file=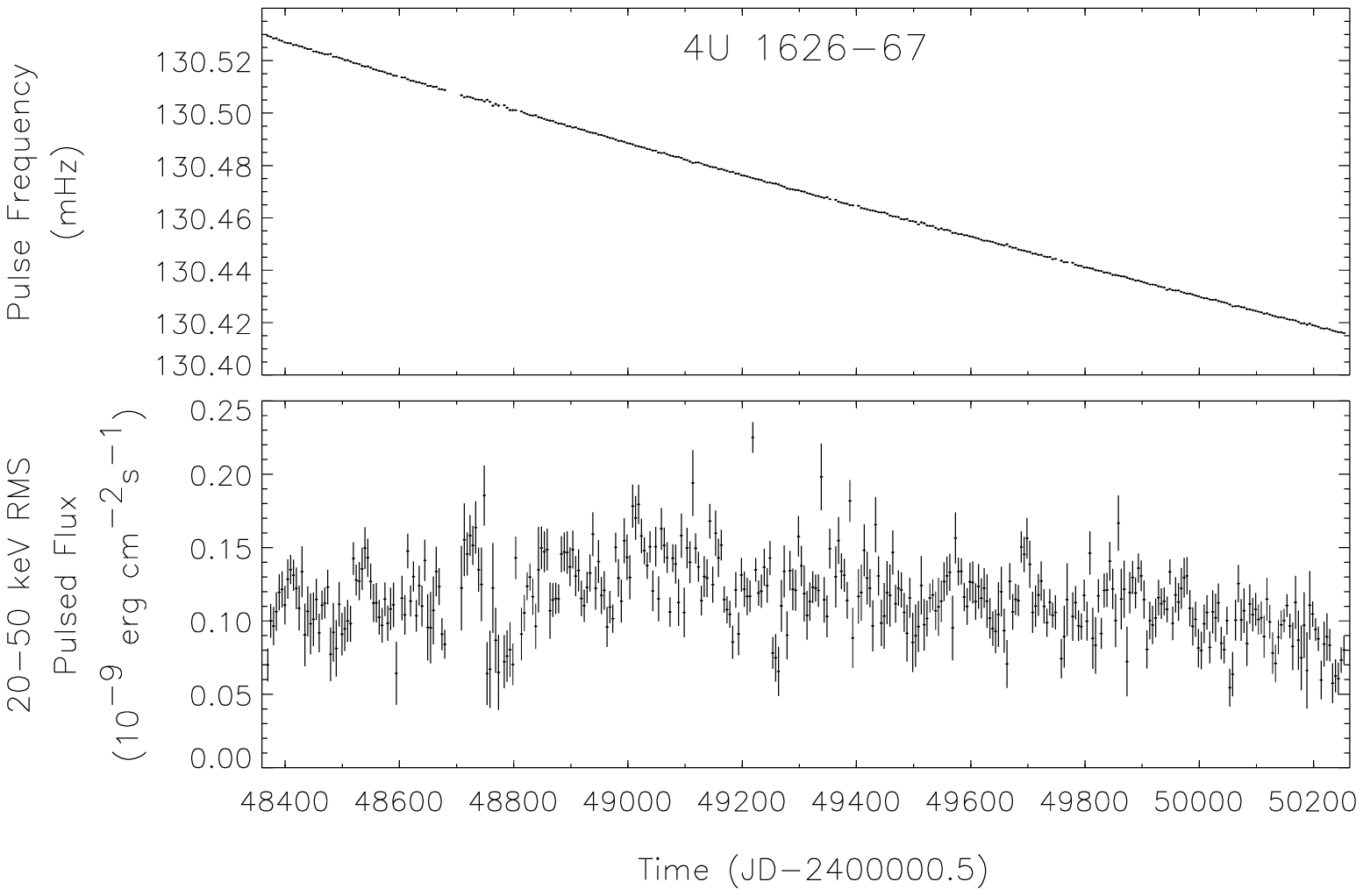}
\figcaption{{\final} 
4U~1626-67 frequency and pulsed flux measurements from BATSE. 
The spin frequencies, which have not been orbitally corrected
since the orbit is unknown, were determined at 5--day intervals by searching
the Fourier power spectrum of the 20-50 keV DISCLA data for the strongest
signal in a small range around 7.7~s \powspec. 
The pulsed fluxes were obtained 
by assuming a power law spectral model
with a photon number index of 4.9  \flux.}

{\em 4U 1626--67.} ---
{\it SAS-3} discovered 7.68~s pulsations from 4U 1626-67 in 1977
(\cite{Rappaport77}) and the optical counterpart was later
identified to be KZ TrA (\cite{McClintock77}).
X-ray timing limits imply an ultracompact binary with an extremely low-mass
companion (\cite{Levine88,Chakrabarty97a}).  There is optical
photometric evidence for a 42-min orbital period
(\cite{Middleditch81,Chakrabarty97c}), 
suggesting a mass of $<0.1 M_\odot$ for the mass donor. 

Detailed discussion of BATSE observations have
appeared elsewhere (\cite{Bildsten94,Chakrabarty97a}). The
neutron star was observed to spin up steadily during 1977-1991 
and made a transition to steady spin-down at nearly the same
rate of $|\nu/\dot{\nu}|\approx$\,5000\,yr by the start of 
BATSE observations in 1991. Despite this torque
reversal, there is no evidence for a large change in
the bolometric flux from the source. The torque exerted on the neutron
star is quiet, in the sense that the torque-fluctuation power
measured by BATSE is the lowest measured for any X-ray
pulsar and is comparable to the timing noise observed in
young rotation-powered radio pulsars.  %{\nopfrac}

%\newpage

\psfig{file=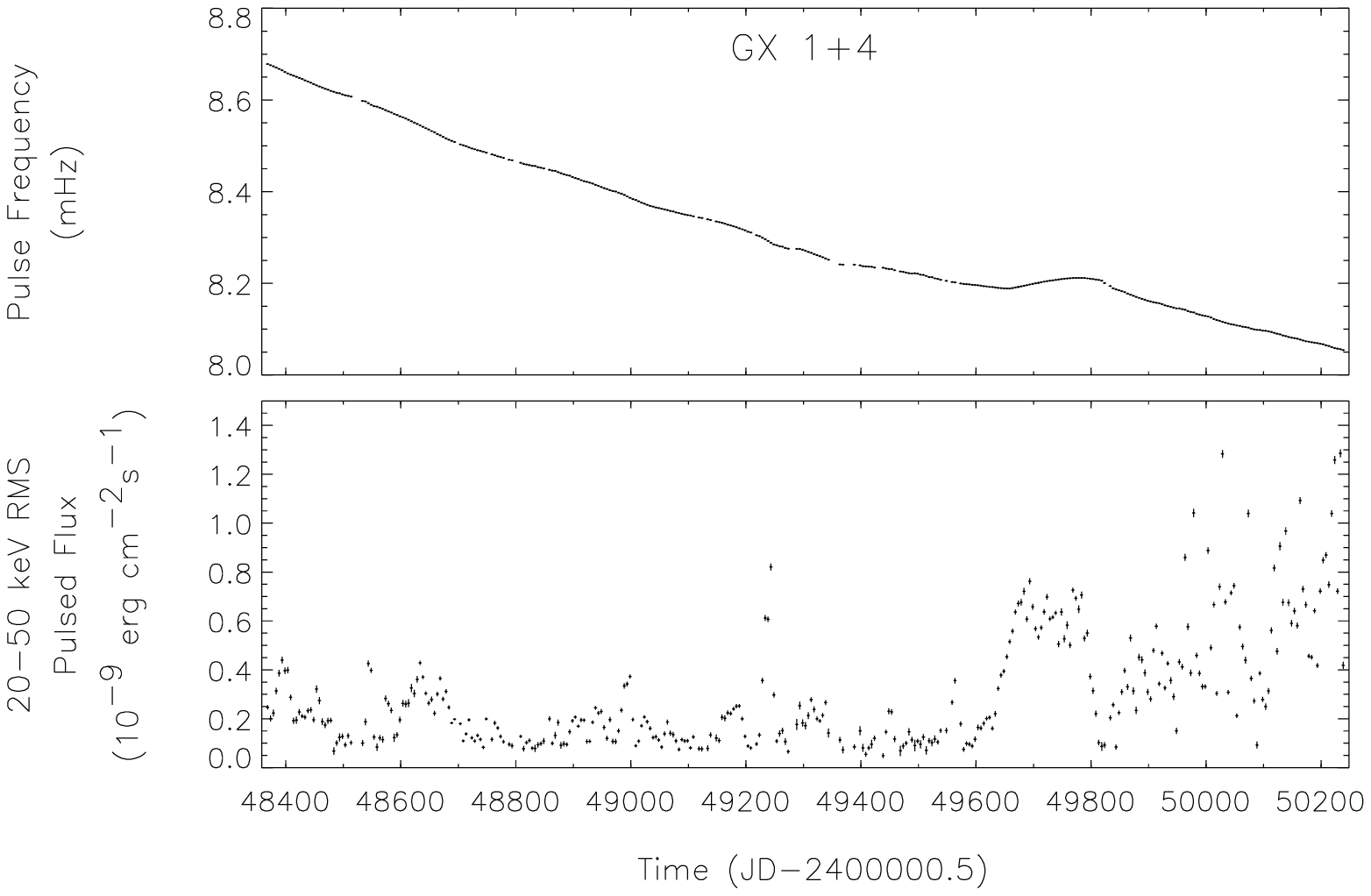}
\figcaption{{\final}
GX~1+4 frequency and pulsed flux measurements from BATSE.
The spin frequencies, which have not been orbitally corrected since
the orbit is unknown, were determined at 5--day intervals by
searching the Fourier power spectrum of the 20--50 keV DISCLA data 
for the strongest signal in the pulse period range 
110~s $\lesssim P_{\rm spin} \lesssim $ 130~s  \powspec.
The pulsed fluxes were determined at 5--day intervals by assuming a 
power law spectral model with a photon number index of 2.5  \flux.}

{\em GX 1+4.} ---
An 18--50 keV X-ray balloon experiment discovered $\approx$ 2~min
pulsations from GX~1+4 in 1970 (\cite{Lewin71}). It is now known to be
orbiting an M5 III giant ({\cite{Davidsen77,ChakrabartyRoche97}), 
making it the only verified accreting pulsar with a red giant donor. 
The binary period is unknown but believed to be of order 
years ({\cite{ChakrabartyRoche97}}).  Throughout the 1970s, GX 1+4 was
persistently bright and was spinning up on a time scale
$|\nu/\dot{\nu}|\sim$\,40\,yr, increasing in frequency from
$\sim$7.5\,mHz to $\sim$9\,mHz between 1970 and 1980 (\cite{Nagase89}).  
After decreasing in flux by at least two orders of magnitude in the
early 1980s, GX 1+4 was found by Ginga to be rapidly spinning down
(\cite{Makishima88}).

Detailed discussions of the BATSE observations of GX 1+4 have appeared
elsewhere (\cite{Chakrabarty94b,Chakrabarty97b})
These observations found GX~1+4 to have the hardest spectrum of any
accretion-powered pulsar, with pulsations clearly detected up to
energies of 160 keV (\cite{Chakrabarty97b}). GX 1+4 is spinning down on
average, on a time scale $|\nu/\dot{\nu}|\approx$\,40\,yr.  
During 1991--1994, BATSE observed a number of bright flares in the
hard X-ray (20--100 keV) band which were accompanied by episodes of
enhanced spin-down. A smooth torque reversal to spin-up accompanied an
extended bright state during late 1994 and early 1995 (\cite{Chakrabarty94}),
followed by a return to spin-down and a lower average hard
X-ray flux (\cite{Chakrabarty95a}).  During spin-down, the torque
fluctuations exhibit a $1/f$ power density spectrum, similar to that seen in 
Cen X-3.  %{\nopfrac}

%\newpage
%\begin{figure}
\psfig{file=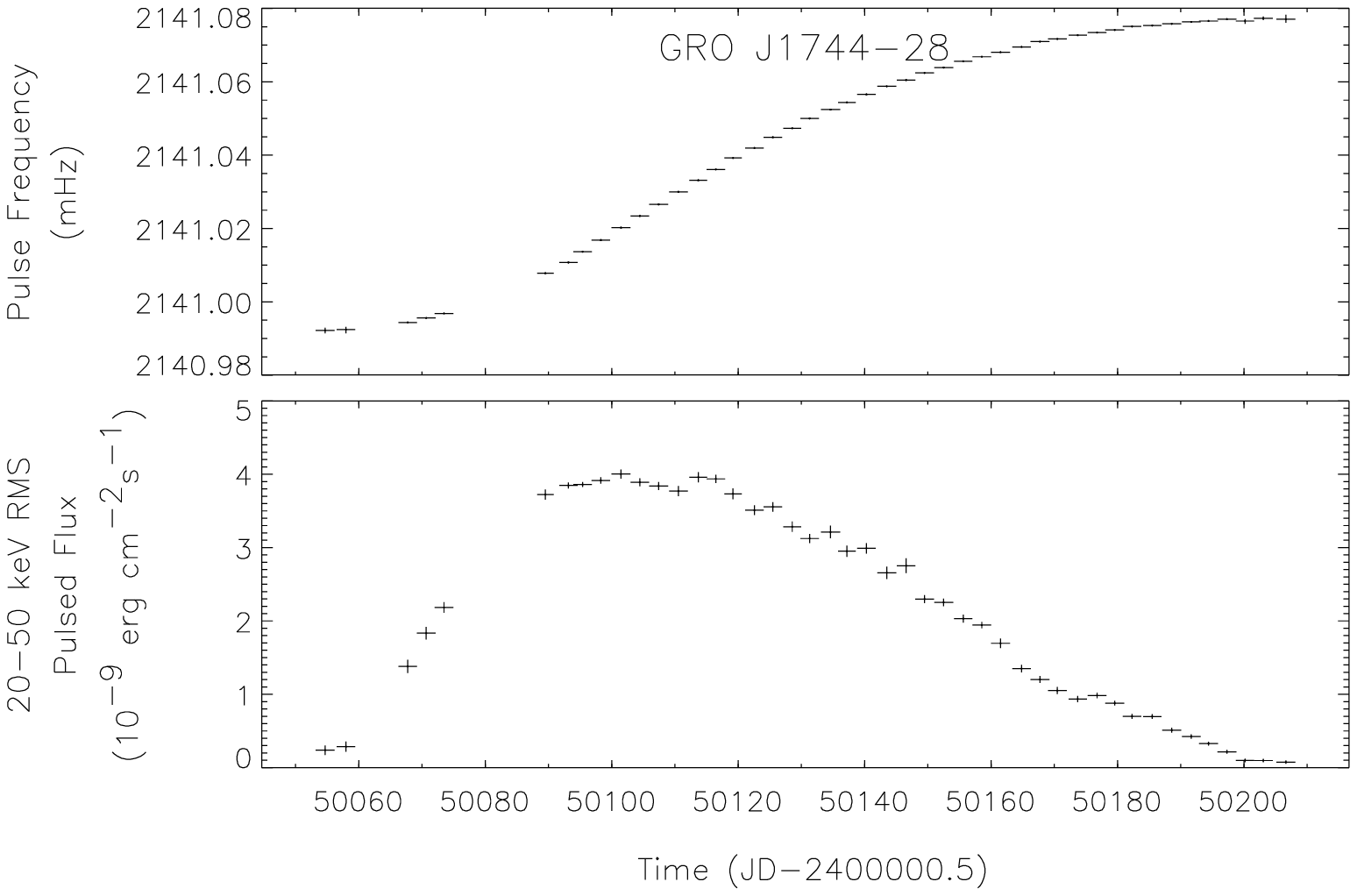}
\figcaption{{\final} GRO~J1744-28 frequency and pulsed flux measurements 
from BATSE. The spin frequencies, which have been orbitally
corrected using parameters from Finger et al. (1996), were determined
at 2--day intervals by fitting pulse phases derived from pulsar
folded-on-board data in the 20-40 keV band  \phase.  The
pulsed fluxes were obtained by assuming a spectra of the form $F(E) =
A E^{-\lambda} \exp(-E/kT)$ with $\lambda = 2.0$ and $kT=15$ keV, as
determined by OSSE measurements (Strickman et al. 1996) 
\flux
\label{fig:j1744freqflux}
}

\nocite{Strickman96}

{\em GRO J1744-28.} ---
GRO~J1744-28 was initially discovered by BATSE as an unusual bursting source 
in the direction of the Galactic Center (\cite{Kouveliotou96}).  
%Its unique bursting behavior, unprecedented among transient
%X-ray and $\gamma$-ray sources, was suggested to arise
%from spasmodic accretion onto a neutron star (Kouveliotou \etal 1996).
The discovery of coherent 467~ms pulsations by BATSE, and subsequent
pulse timing unambiguously established GRO J1744-28 to be a neutron
star in a circular 11.8--day orbit around a low--mass companion and
indicated that the neutron star was spun--up by an accretion disk
during the outburst (\cite{Finger96a}).  These are the first
persistent pulsations seen in a bursting X-ray source. 

Detailed discussions of the BATSE observations of GRO~J1744--28 have appeared
elsewhere (\cite{Finger96a}).
One major outburst has been observed to date, spanning $\approx$ MJD
50053--50223.  Another outburst which began on $\approx$ MJD 50253
lasted for only $\approx$ 1 week.   The initial outburst showed
enough dynamic range that the relation between accretion torque and 
pulsed flux could be tested directly (see \S \ref{sect:torque-lum}).
The 20--40~keV pulse profile is nearly sinusoidal, 
in stark contrast to the more
complicated pulse shapes seen in other accretion powered pulsars (see
Figure \ref{fig:pulsemosaic}).  Simultaneous 20--40 keV pulsed and
Earth occultation DC flux measurements on 10--16 January 1996 
(MJD 50092--50098) yielded
a peak-to-peak pulsed fraction of $\approx$ 25\% (\cite{Finger96a}).

%%%%%%%%%%%%%%%%%%%%%%%%%%%%%%%%%%%%%%%%%%%%%%%%%%%%%%%%%%%%%%%%%%%%%%%%%%%%%%%

%\input results_hm.tex		% high mass sources

\subsection{High-Mass Supergiant Systems}

   BATSE continuously monitors 5 pulsars which accrete from high-mass
evolved supergiants: Cen~X-3, OAO~1657--415, Vela~X-1, 4U~1538--52 and
GX~301--2. The long-term spin frequency evolution of these pulsars has
revealed several surprises which challenge the standard model of such
systems, as we discuss in \S 5. For example, Cen~X-3 (the only
Roche-Lobe filling high-mass supergiant system observed by BATSE)
exhibits short term ($\sim$ 50~d) spin--up and spin--down
episodes. Moreover, the underfilled Roche-Lobe system GX~301-2
exhibits transient spin--up episodes, also of $\sim$ 50~d
durations.

%\newpage

%\begin{figure}
\psfig{file=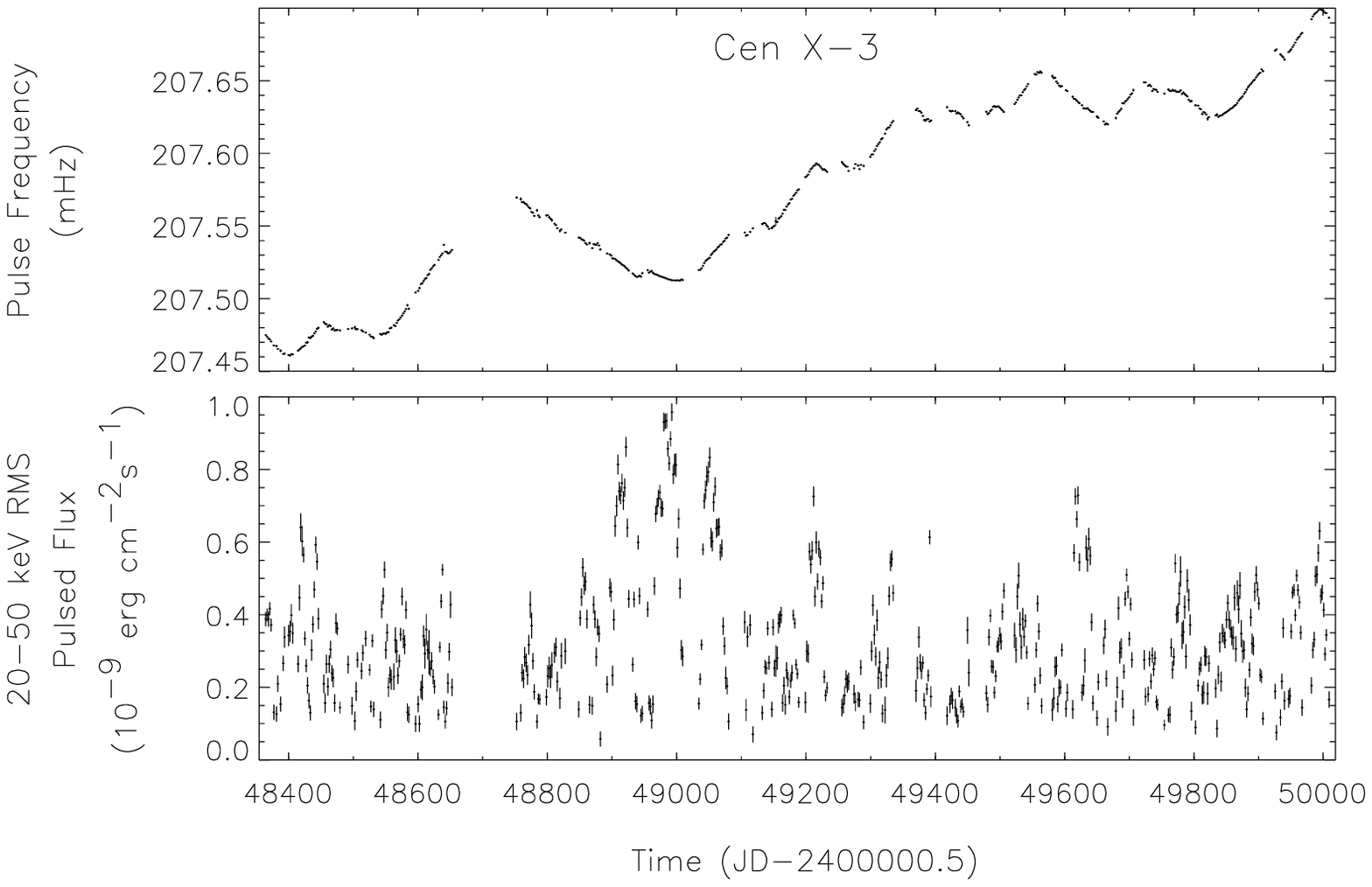}
\figcaption{{\final}
Cen X--3 frequency and pulsed flux measurements from BATSE.  The
intrinsic spin frequencies, which have been orbitally
corrected using parameters from Finger \etal (1993), were determined
at 2.1--day intervals by epoch-folding the 20-50 keV DISCLA data \efold.
The pulsed fluxes were determined at 2.1--day intervals
by assuming an exponential spectrum with a e--folding energy of 12 keV 
\flux. \label{cenx3freqflux}}

{\em Centaurus X-3.} ---
{\it Uhuru} discovered 4.8~s pulsations from Cen~X-3 in 1971, 
the first observation of an accreting pulsar (\cite{Giacconi71}).  
This bright, persistent, eclipsing pulsar is in a 2.1\,d 
orbit (\cite{Kelley83})
around the O6--8 supergiant V779 Cen
(\cite{Krzeminski74,Rickard74,Hutchings79}).  An accretion disk is
apparent from the optical lightcurve (\cite{Tjemkes86}).  Pre-BATSE
observations by numerous pointed instruments found the neutron star to
be gradually spinning up, 
although episodes of spin-down have been observed (\cite{Nagase89}).

  Detailed discussions of the BATSE observations of Cen~X-3 have
appeared elsewhere (\cite{Finger92b,Finger94d}).  These continuous observations
found that the long term ($\sim$years) spin-up trend is
actually the average effect of alternating 10$\sim$100\,d intervals of
spin-up and spin-down at a constant rate (\cite{Finger94d}).
Large excursions in the X-ray intensity occur on timescales of days to
weeks, including bright flares lasting 10--40 days.  A comparison
of the orbital measurements made over the last 20 years reveals
that the orbital period is decreasing
(\cite{Kelley83,Nagase92}).  BATSE
confirms this orbital decay (\cite{Finger93}), which is thought to be due
to the tidal interaction of the neutron star with its companion. 
%{\nopfrac}

%\newpage
%\begin{figure}
\psfig{file=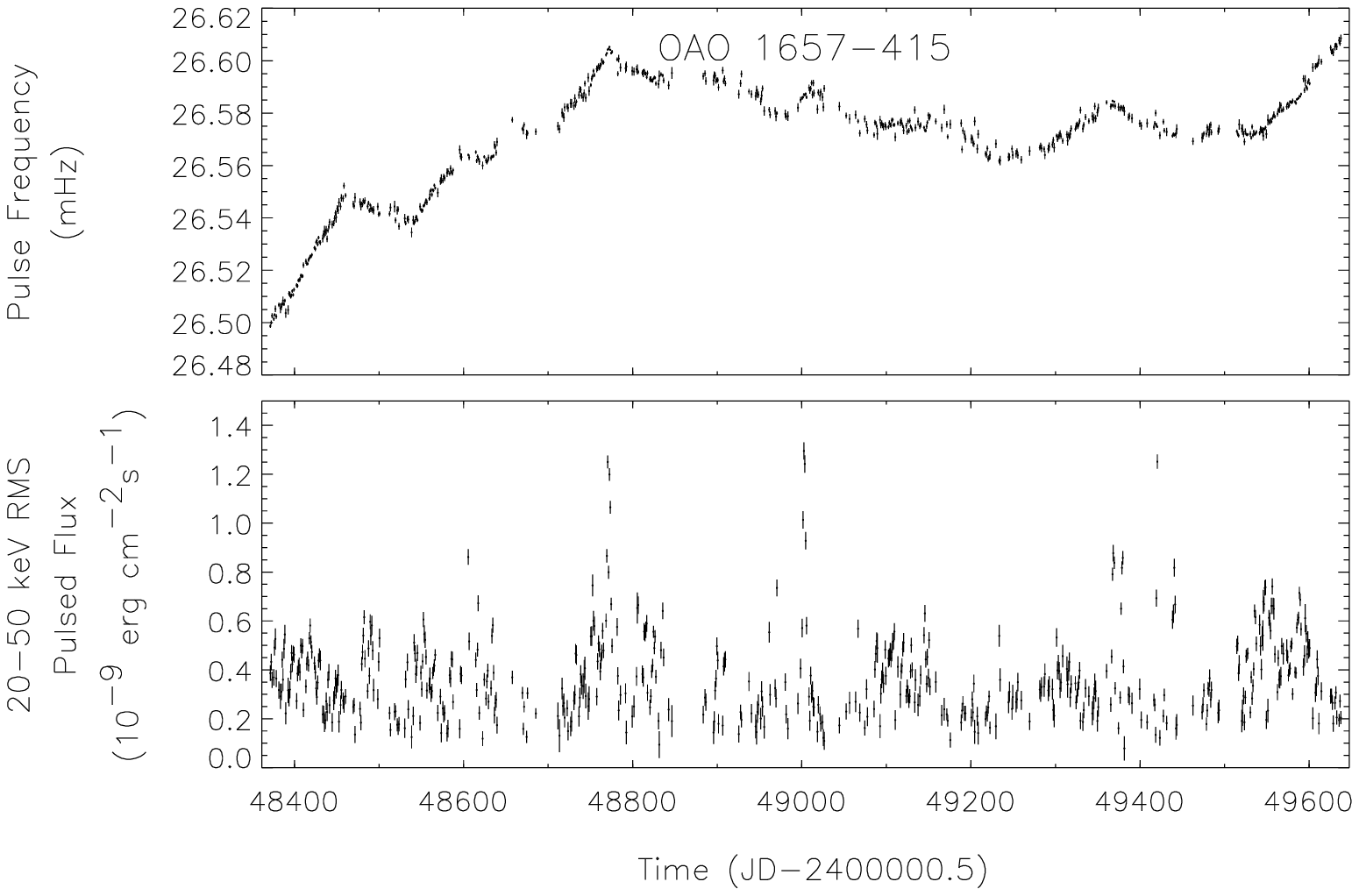}
\figcaption{{\final}
OAO~1657--415 frequency and pulsed-flux measurements from BATSE.
The intrinsic spin frequencies, which have been orbitally corrected
using parameters from Chakrabarty \etal (1993) and a refined orbital
period of $P_{\rm orb} = 10.44809(30)$\,d (see text below for
details), were measured at 1--day intervals from the power spectra
of the 15--55 keV CONT data  \powspec.  The pulsed fluxes were
measured at 1--day intervals by assuming an exponential spectrum
with a e--folding energy of 20~keV  \flux.  
\label{OAO1657freqflux}
}

{\em OAO 1657--415.} ---
 {\it HEAO 1} discovered 38.22~s pulsations from OAO~1657-415 in 1978
(\cite{White79}).  BATSE observations revealed a 10.4~d binary
orbit with a 1.7~d eclipse by the stellar companion
(\cite{Chakrabarty93}), making it the seventh eclipsing X-ray pulsar
discovered.
The intrinsic spin frequency history reveals strong,
stochastic variability and alternating episodes of steady spin-up
and spin-down lasting 10--200\,d, similar to what is
seen in Cen X-3.  Although the companion remains unidentified, it is
inferred to be an OB supergiant from the neutron-star orbit
(Table \taborbits) and eclipse duration (\cite{Chakrabarty93}).

Chakrabarty et al. (1993) measured the binary orbital parameters using
BATSE data spanning 1991 April 24 to 1992
July 23 (MJD 48370--48460). For the  spin frequency history
presented here, which extends far beyond the data used in the original
orbital analysis, it was necessary to refine the orbital period. 
For this purpose the pulse frequencies obtained between
1991 April 24 and 1994 September 20 (MJD 48370--49615) were fitted
using the Chakrabarty et al. (1993) orbital elements, with $P_{\rm
orb}$ as a free parameter.  The contribution to the uncertainty in
$P_{\rm orb}$ from stochastic variations in accretion torque was estimated 
by assuming that $\nu$ performed
a random walk with a strength of $2.5 \times 10^{-17}$
Hz$^2~$s$^{-1}$, as estimated from
the power spectrum of the frequency
derivative measured at $P_{\rm orb}$ (see Figure~\ref{fig:powerspectra}).
The revised orbital period is $P_{\rm orb} = 10.44809(30)$\,d,
consistent with the value measured by Chakrabarty et al. (1993), but
of improved accuracy.  %{\nopfrac}

%\newpage
%\begin{figure}

\psfig{file=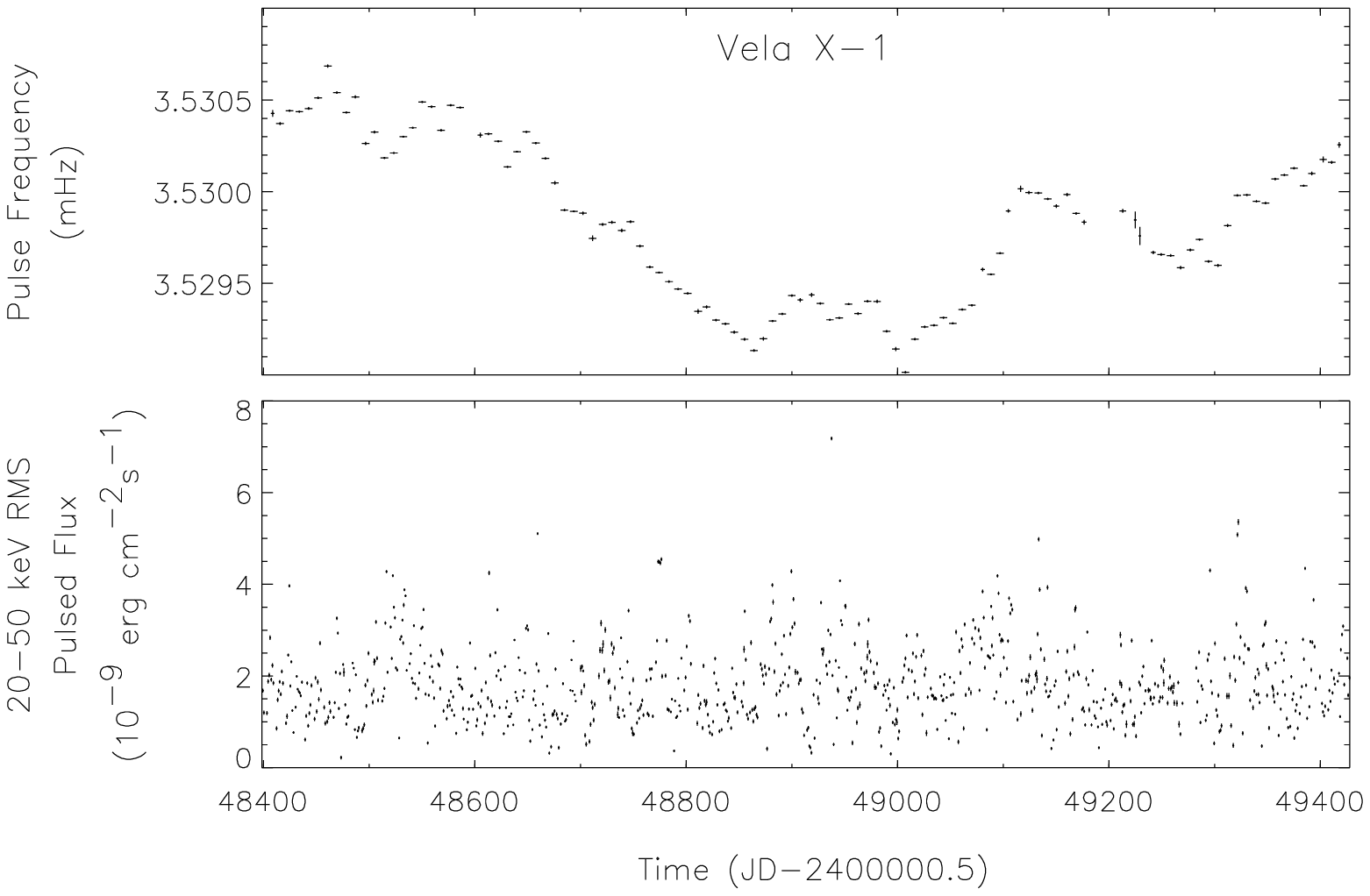}
\figcaption{{\final} 
Vela X-1 spin frequency and pulsed flux measurements from BATSE.
The intrinsic spin frequencies, which have been orbitally corrected
using parameters from Finger \etal\, (1997), in preparation, 
were determined by fitting pulse phase measurements
\phase. Each point uses the uneclipsed data of the
8.96~d binary orbit. The pulsed fluxes were determined at 1--day
intervals by assuming an exponential spectrum with a e--folding energy of
20~keV  \flux.
\label{velax1freqflux}}

{\em Vela X-1.} ---
  {\it SAS-3} discovered 283~s pulsations from the eclipsing binary
Vela X-1 in 1975 (\cite{McClintock76}) and pulse timing revealed this
pulsar to be in an 8.96~d eccentric orbit
(\cite{Rappaport76}, Finger \etal (1997), in preparation) around 
the B0.5Ib supergiant 
HD77581 (\cite{Hiltner72,VanKerkwijk95}).  The optical lightcurve of
the companion shows ellipsoidal variations, 
indicating that the star is substantially
distorted by the tidal field from the neutron star
(\cite{Tjemkes86}).

Vela X-1 is the brightest persistent accretion-powered pulsar in the
20--50 keV energy band. Individual pulses are often visible in the raw
data (Figure 1). BATSE observations showed Vela X-1 alternating
between spin-up and spin-down, with no long-term trend in spin
frequency, consistent with pre-BATSE 
observations of a random-walk in
spin frequency (\cite{Deeter89}).  Pulse profiles in the BATSE energy
range are double peaked and vary slightly with both energy and time,
with some evidence for a correlation between luminosity and pulse
shape.  At lower energies the pulse profiles are more complex, showing
dramatic changes with energy (\cite{Raubenheimer90}).  

%\newpage

%\begin{figure}
\psfig{file=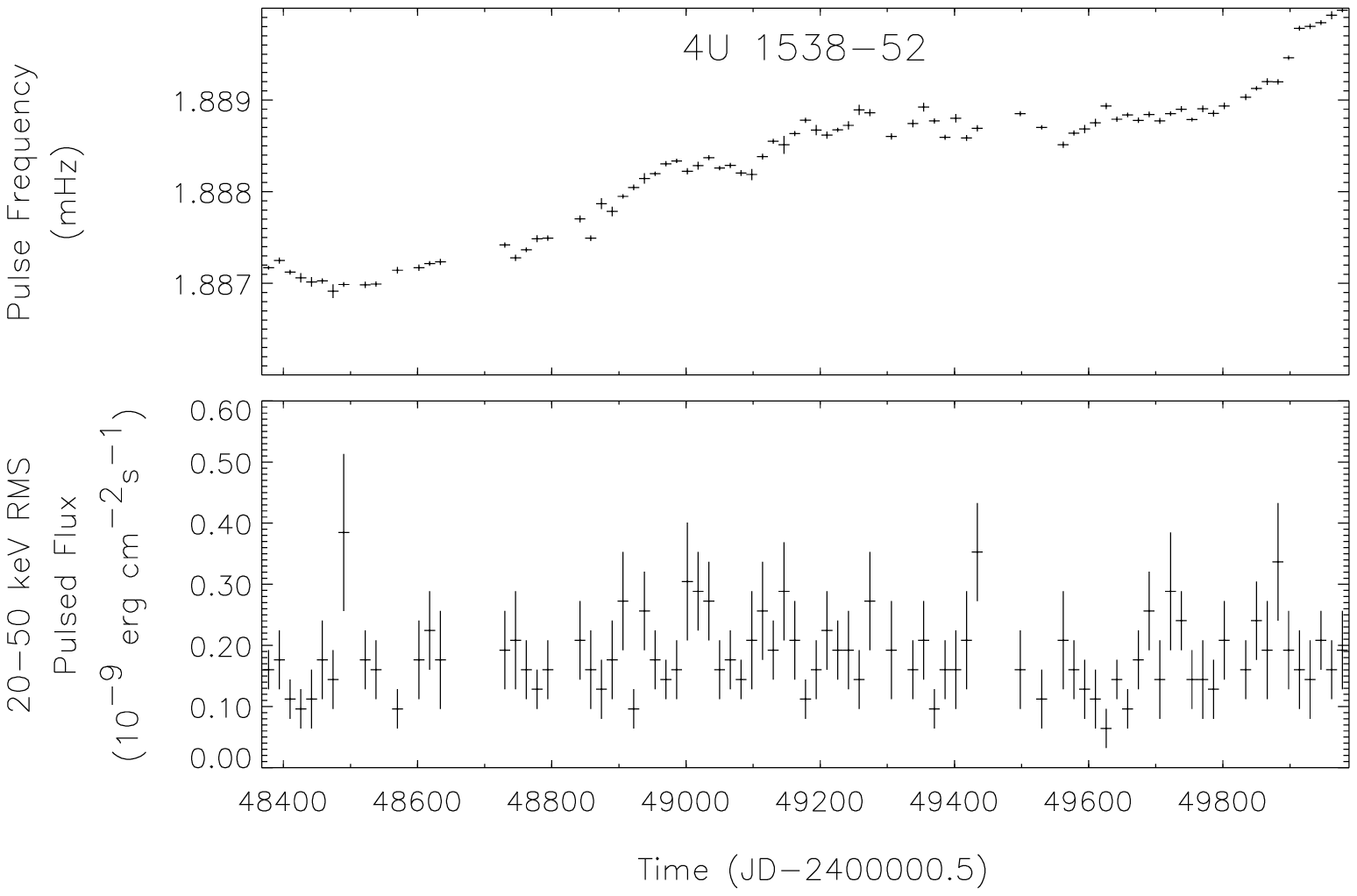}

\nocite{Rubin97}
\figcaption{{\final}
4U~1538-52 spin frequency and pulsed flux measurements from BATSE.
The intrinsic spin frequencies, orbitally corrected using
parameters from Rubin \etal (1997), were determined at 16\,d
intervals by epoch folding the 20--50\,keV DISCLA data at a range
of trial frequencies after subtracting
the background model of Rubin \etal (1996) \efold.
Measurements are obtained only
once every 16 days due to the low flux from the object and BATSE's
poor sensitivity at long periods.}
\label{4u1538freqflux}
%\end{figure}

{\em 4U 1538--52.} ---
{\it Ariel 5} discovered 530~s pulsations from 4U 1538-52 in 1976
(\cite{Davidson77}) and pulse timing revealed this pulsar to be in a
3.7~d circular orbit (\cite{Davidson77,Corbet93}) around the B0
supergiant companion, QV Nor (\cite{Parkes78}), which most likely
underfills its Roche lobe (\cite{Crampton78}).  Pre-BATSE data shows a
long term spin down trend with random pulse period variations on
shorter time scales (\cite{Nagase89}).

\nocite{Rubin94}
  Detailed BATSE observations of 4U~1538--52 have appeared in
Rubin \etal 1994.  These observations revealed a reversal of the secular
torque to long-term spin-up at an average rate of $\dot \nu =
1.8 \times 10^{-14}$~{\hps}. However, the change in $\nu$ is 
comparable in
magnitude to what one would predict from the observed torque-noise strength of
$\sim10^{-20}$\,Hz$^2$\,s$^{-2}$\,Hz$^{-1}$ (Figure~\ref{fig:powerspectra}), 
and
is thus consistent with being the result of a random walk in
frequency.  Combining orbital epochs measured with BATSE with those
determined from previous experiments has led to an improved value for
the orbital period (see footnote to Table 3)
and a 95\% confidence limit on the rate of change of $\dot P_{\rm orb}$ of
$-3.9\times 10^{-6} < \dot P_{\rm orb}/P_{\rm orb} <
2.1\times 10^{-6}$ yr$^{-1}$ (\cite{Rubin97}).

%\newpage

\psfig{file=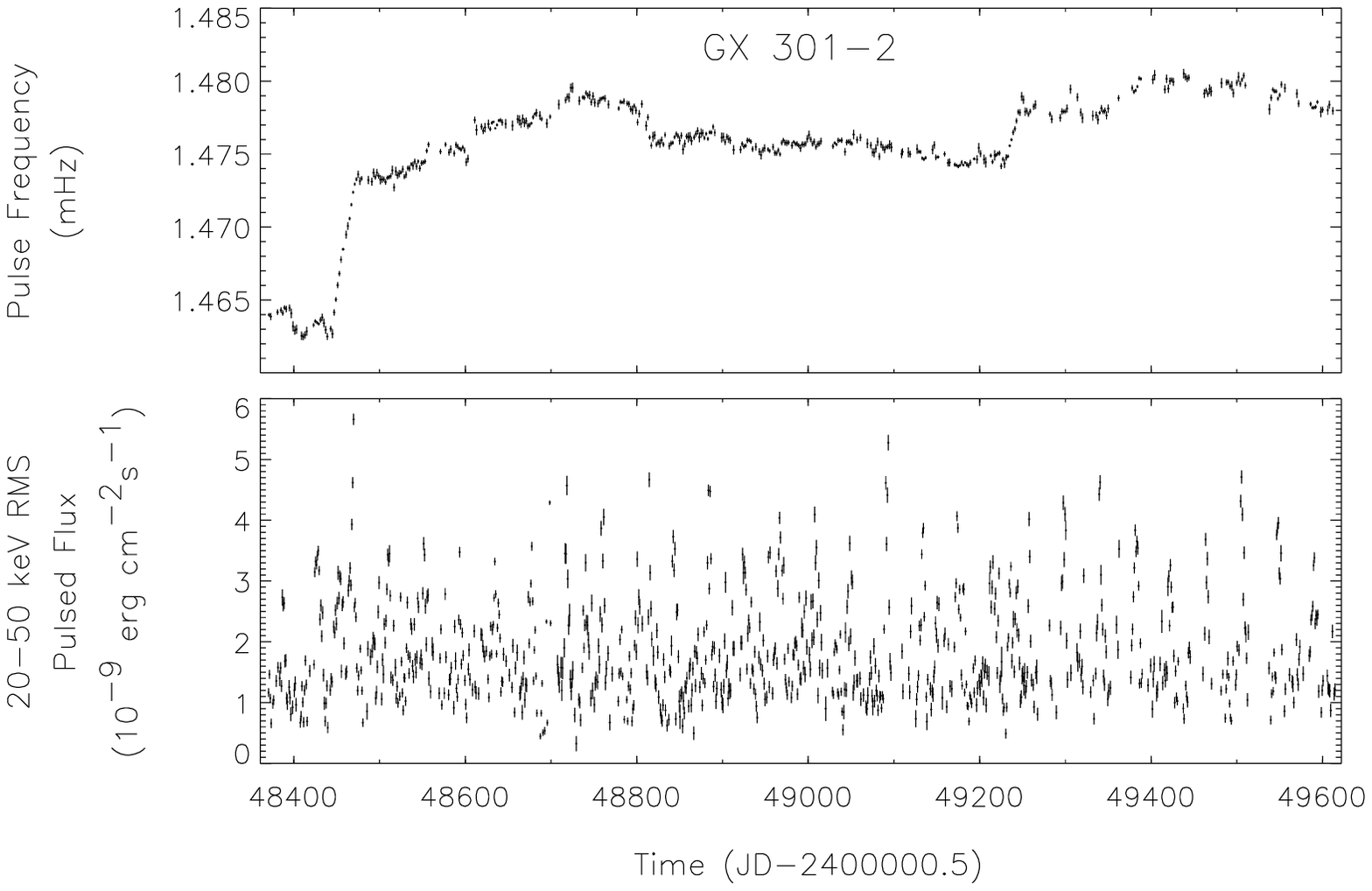}
\figcaption{{\final}
GX301--2 frequency and pulsed flux measurements from BATSE. The
intrinsic spin frequencies, which have been orbitally corrected using
parameters from Koh \etal (1997), were measured at 2--day intervals by
searching the Fourier power spectrum of the 15--55 keV CONT data for
the strongest signal in a range around the previously observed pulse
frequency \powspec.  The pulsed fluxes were measured at 2--day
intervals by assuming an exponential spectrum with a e--folding energy 
of 10\,keV \flux.
\label{fig:gx301freqflux}
}
\nocite{Koh97}
{\em GX 301--2.} ---
 {\it Ariel 5} discovered 700~s pulsations from GX~301-2 (4U 1223-62)
(\cite{White76}) and subsequent observations revealed the neutron star
to be in a 41.5 day eccentric ($e$ = 0.47) orbit (\cite{Sato86})
around the supergiant Wray 977 (\cite{Parkes80,Kaper95}).  Between
1975 and 1985, the neutron star was, on average, neither spinning up
nor down, indicative of wind accretion.  However, a prolonged period
of spin--up at $\dot \nu \approx 2 \times 10^{-13}$ began in 1985
(\cite{Nagase89}).

Detailed discussion of BATSE observations have appeared elsewhere
(\cite{Koh97}). %The most striking features found by BATSE are 
BATSE observed 
two rapid spin-up episodes with $\dot \nu \approx $ (3--5) $\times
10^{-12}$~{\hps}, each lasting $\sim$30 days, probably
indicating the formation of a transient accretion disk. Except for
these spin-up episodes, there are virtually no net changes in $\nu$
on long time scales, suggesting that the
long-term spin-up trend observed since 1985 may be due entirely to
brief ($\approx 30$~d) spin-up episodes similar to those we have
discovered.  In addition to confirming the previously known flare
which occurs $\approx$ 1.4~d before periastron, BATSE occultation and
pulsed-flux measurements folded at the orbital period reveal a smaller
flare near apastron (\cite{Pravdo95,Koh97}).  Orbital parameters
measured with BATSE are consistent with previous measurements, with
improved accuracy in the orbital epoch (\cite{Koh97}, Table~\taborbits).  
Simultaneous pulsed and occultation fluxes measured near periastron yield
a 20--55 keV peak-to-peak pulsed fraction of $\approx$ 0.5 (\cite{Koh97}).

%%%%%%%%%%%%%%%%%%%%%%%%%%%%%%%%%%%%%%%%%%%%%%%%%%%%%%%%%%%%%%%%%%%%%%%%%%%%%%%

%\input results_tr.tex		% transient sources

\subsection{High-Mass Transient Systems}

  BATSE has discovered four new high-mass, transient accreting pulsars (GRO
J1008--57, GRO J1948+32, GRO J2058+42, and GRO J1750--27),
two of which (GRO J2058+42 and GRO J1008--57) have
repeated. In addition, BATSE has observed multiple outbursts of the
previously known transient accreting pulsars 4U~0115+63, GS~0834--43,
2S~1417--62, A~0535+26, 4U~1145-619 and EXO~2030+375, and a single
outburst from A~1118--615.  Outburst times and durations are shown in
Figure \ref{fig:outburst_times}.

%\begin{figure}
\centerline{
\psfig{file=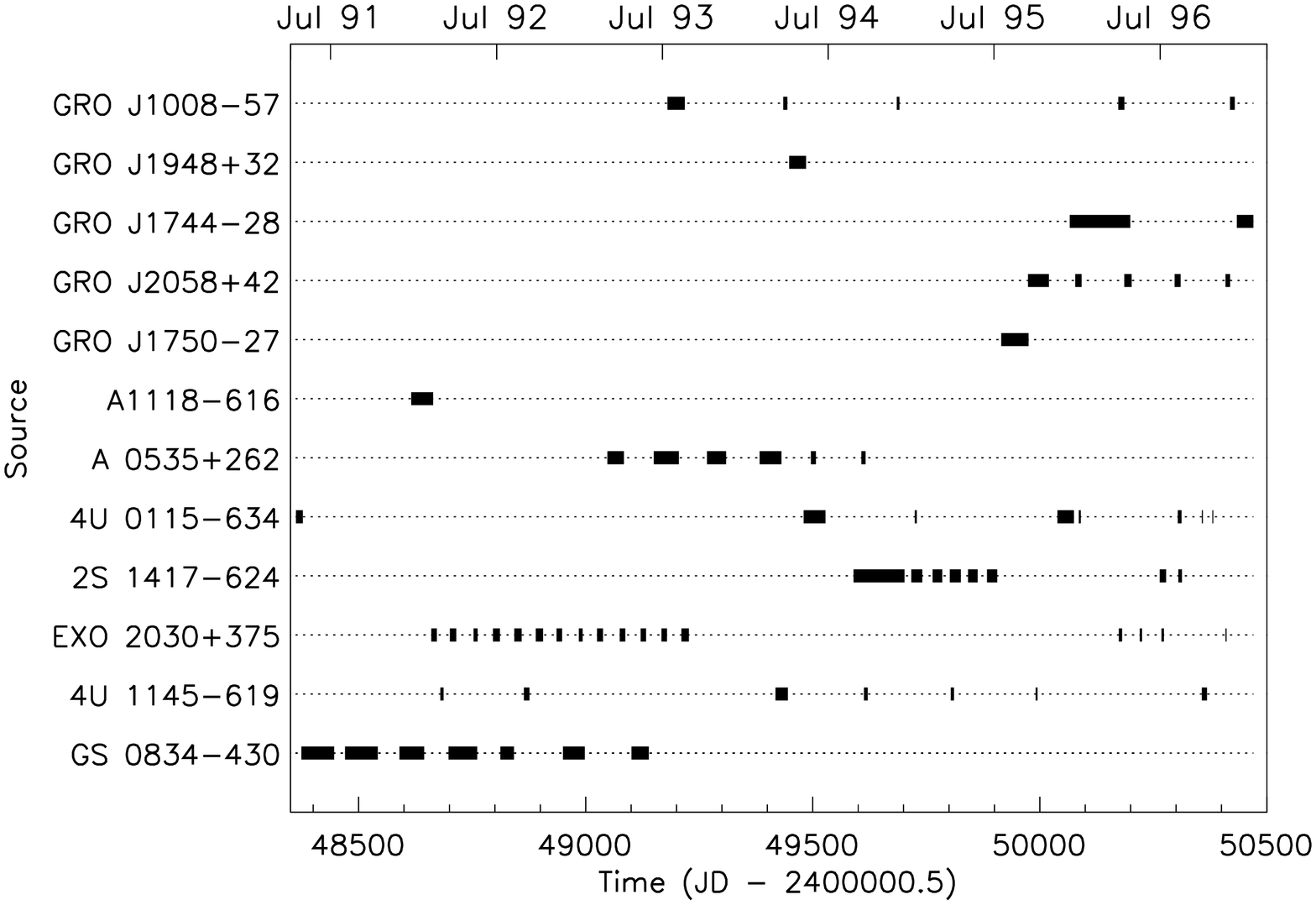,width=4.5in,angle=90}
}
\figcaption{Outburst times for all transient accreting pulsars
observed by BATSE.
\label{fig:outburst_times}}
%\end{figure}

BATSE has observed a series of regularly-spaced outbursts from several
transient pulsars. This was the case for GRO~J1008--57 (5 outbursts),
GRO~J2058+42 (5 outbursts), A~0535+26 (6 outbursts), 4U~0115--634 (4
outbursts), 2S~1417--62 (8 outbursts), EXO~2030+375 (17 outbursts),
4U~1145--619 (7 outbursts) and GS~0834--430 (7 outbursts). In some
cases one or more outbursts were missing from the sequence. In the
case of GS 0834--43 the spacing of the final two outbursts was
irregular. 

%\newpage
%\begin{figure}
\psfig{file=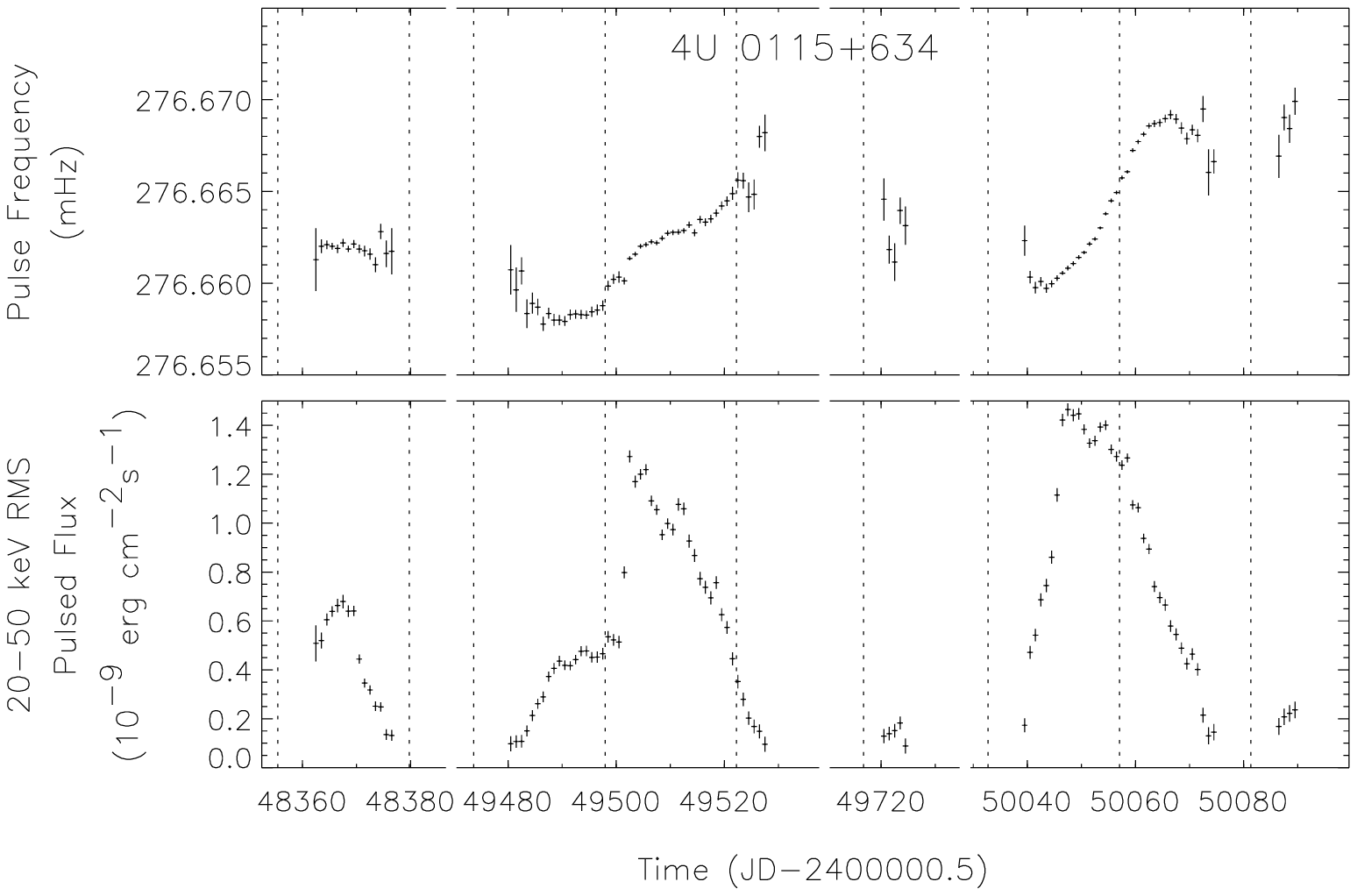}
\figcaption{{\final}
4U~0115+634 frequency and flux measurements from BATSE.  The
intrinsic spin frequencies, which have been orbitally corrected using
the orbital parameters discussed in the text, were determined at
1--day intervals by epoch folding the 20--50 keV DISCLA data 
at a range of trial frequencies \efold.
The pulsed fluxes were determined at 1--day intervals by
assuming an exponential spectrum with an e--folding energy of 15 keV
\flux.
The gaps in Figure~\ref{4u0115freqflux} are
extended intervals when the source was undetectable with BATSE.
\label{4u0115freqflux}}

{\em 4U 0115+634.} --- 
{\em SAS-3} discovered 3.6~s pulsations from 4U~0115+634 in 1978
(\cite{Cominsky78}) and subsequent pulse-timing revealed the pulsar to
be in a 24~d eccentric orbit (\cite{Rappaport78}) around the heavily
reddened Be Star, V635 Cas (\cite{Johns78}).  To date, BATSE has
observed 5 outbursts from 4U 0115+634.  
A 48 day outburst from 1994
May 7 -- June 24 (MJD 49480--49528) (\cite{IAUC5990,IAUC5999}) showed 
a sudden rise in
pulsed flux at the middle of the outburst, shortly following
periastron passage (MJD 49498.1). 
A 36 day outburst from 1995 November 17 --December 27 (MJD 50039--50075)
(\cite{IAUC6266}) was also seen by Granat/Watch
(\cite{IAUC6272}).  This was immediately followed in 1996 January by a
short weak outburst.  Not shown in Figure \ref{4u0115freqflux} is a 10
day outburst in August 1996 (\cite{IAUC6450}).

We estimated the epoch of periastron for the outbursts in
1991 April, 1994 May--June, and 1995 November--December by fitting the
phase measurements for each data set with a polynomial in pulse
emission time using  the orbital elements from Rappaport
\etal (1978), but allowing the epoch of periastron to vary.
This resulted in periastron epochs
of MJD 48355.44(7), 49498.1232(15) and 50057.4015(32), which are
plotted in Figure \ref{4U0115epochs} along with previous
determinations.  The Ginga result ({\cite{Tamura92}) deviates from the
trend of the other points. This may be due to an incorrect phase
connection in that poorly sampled data set. Discarding this point, we
find a best fit linear ephemeris of the periastron epoch $T_p =
$MJD49279.2677(34)$+n\times $24.317037(62).  
The frequencies in Figure
\ref{4u0115freqflux} are orbitally corrected using this ephemeris in
combination with the remaining Rappaport et al. (1978) elements.
%{\nopfrac}

%\begin{figure}

\centerline{
\psfig{file=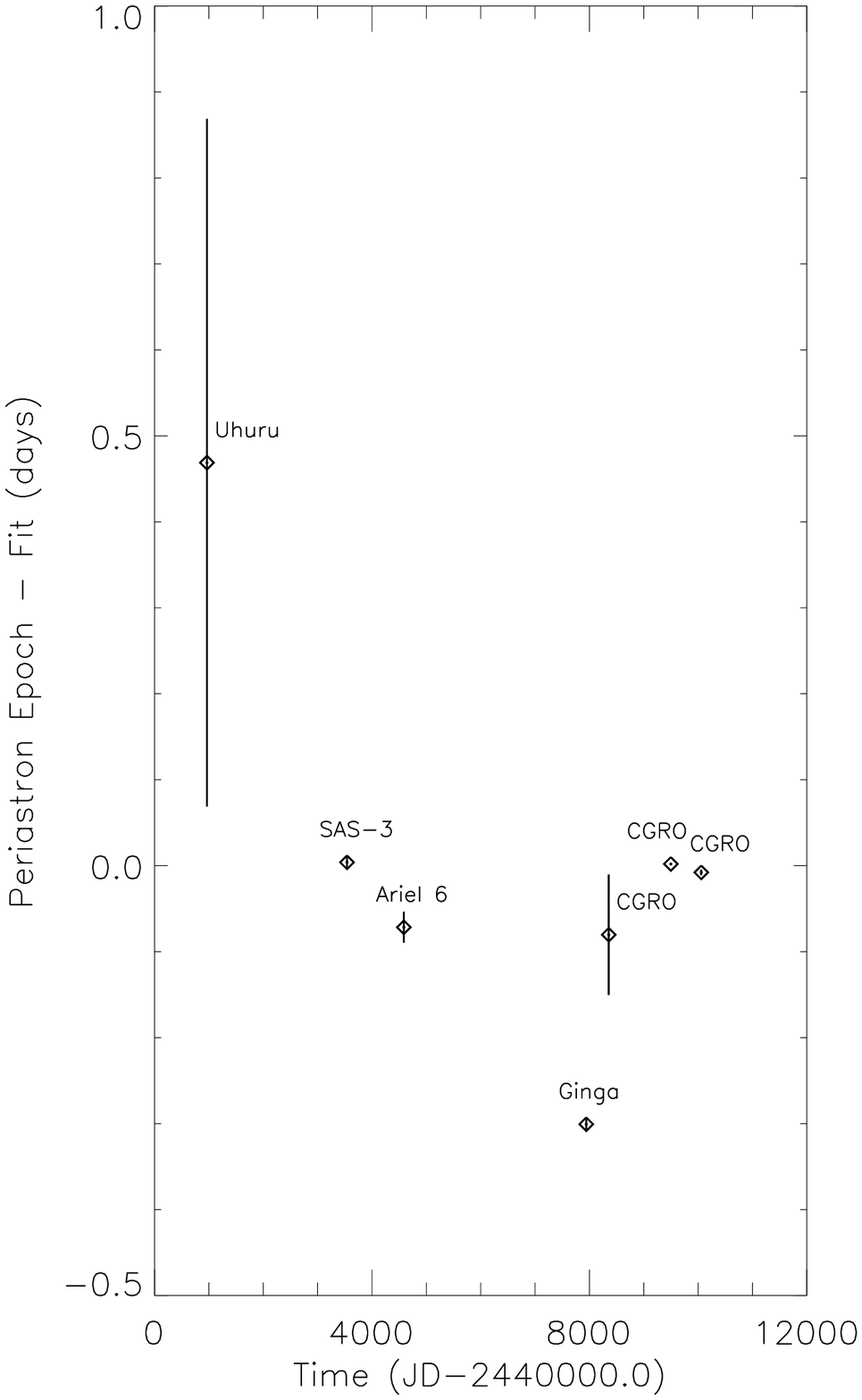,height=5in}
}
\figcaption{
Measurements of the 4U 0115+63 periastron epoch. The plot shows the periastron
epochs minus the linear ephemeris $MJD 49279.2677+ 24.317037*n$
where $n$ is an integral number of orbits. This ephemeris is discussed in the
text. The epoch measurements have been determined from 
Uhuru (Kelley et al. 1981), SAS-3 (Rappaport et al. 1978), 
Ariel 6 (Rickets et al. 1981), Ginga (Tamura et al. 1992) and this work. 
Excluding the Ginga measurement, the
observations are consistent with a constant orbital period.
\label{4U0115epochs}}
%\end{figure}

\nocite{Kelley81_0115,Rappaport78,Ricketts81,Tamura92}

%\newpage
%\begin{figure}
\psfig{file=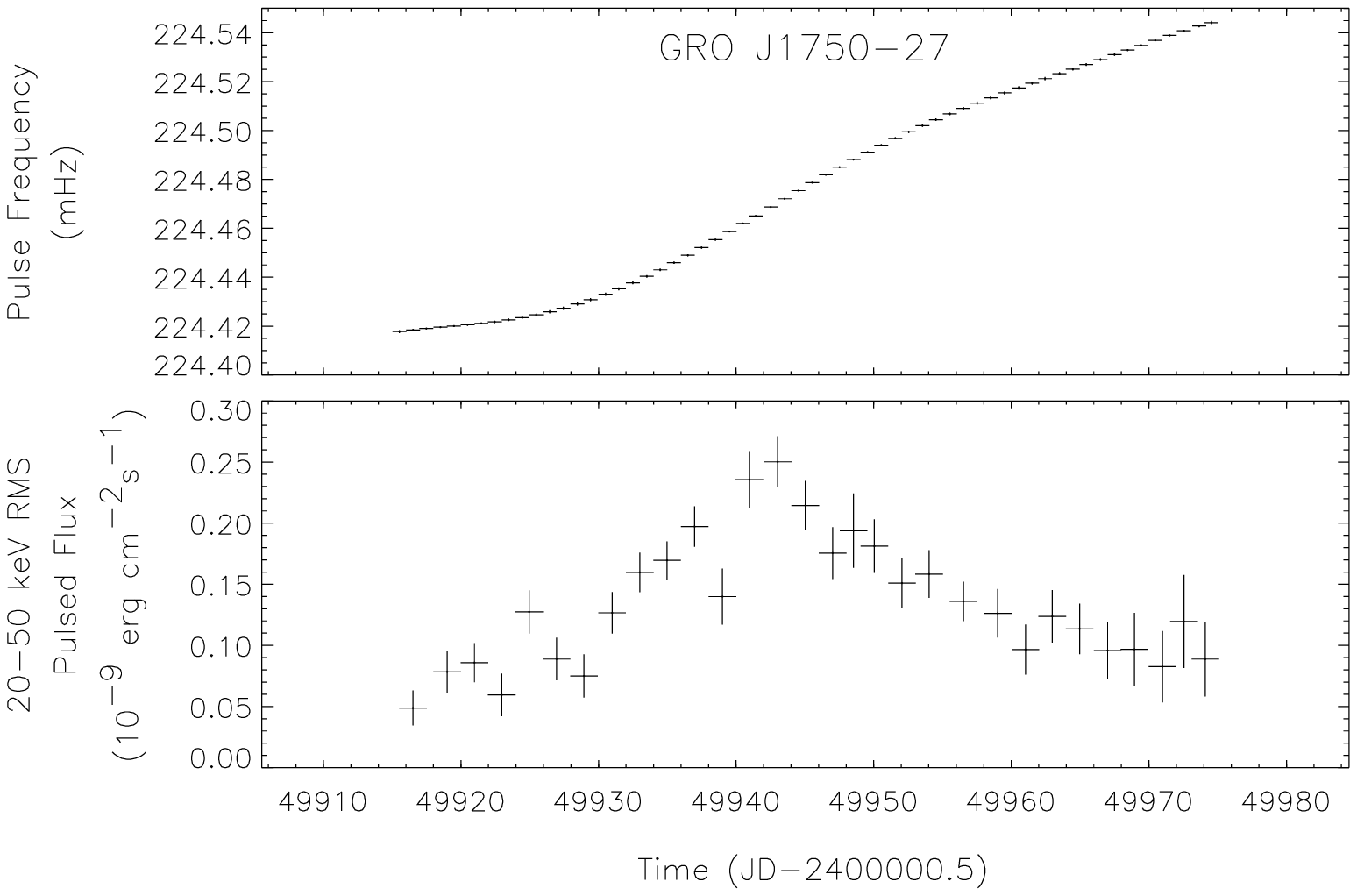}
\figcaption{{\final}
GRO J1750-27 spin frequency and pulsed flux measurements from BATSE.
The intrinsic spin frequencies, which have been orbitally
corrected using parameters from Scott \etal (1997), 
were determined at
1--day intervals from fits of phase measurements 
of the 20--50 keV DISCLA data \phase.  
The pulsed fluxes were determined at 1--day intervals by
assuming an exponential spectrum with an e--folding energy of 20~keV \flux.}

\nocite{Scott97}

{\em GRO J1750-27.} ---
  BATSE discovered and observed a single 60\,d 
outburst from the 4.4~s accreting
pulsar GRO J1750-27 from 1995 July 7 to September 18 (MJD
49915--49978) (\cite{IAUC6207,Scott97}).  Pulse timing revealed an
eccentric 29.82~d orbit.  
A 0.5 deg localization
with BATSE (\cite{Koh95}) motivated an ASCA TOO which successfully
localized the object to $\approx 2'$ (\cite{Dotani95}).  Although no
optical counterpart has been reported, the orbital period and
pulse period of GRO~J1750-27 place it squarely in the Be transient
region of the Corbet Diagram (Figure \ref{fig:corbet}). Steady spin--up
with a peak value of $3.8\times 10^{-11}$\,Hz~s$^{-1}$
coupled with a correlation between the spin--up rate and the pulsed
flux strongly suggests accretion from a disk (\cite{Scott97}).

%\newpage
%\begin{figure}
\psfig{file=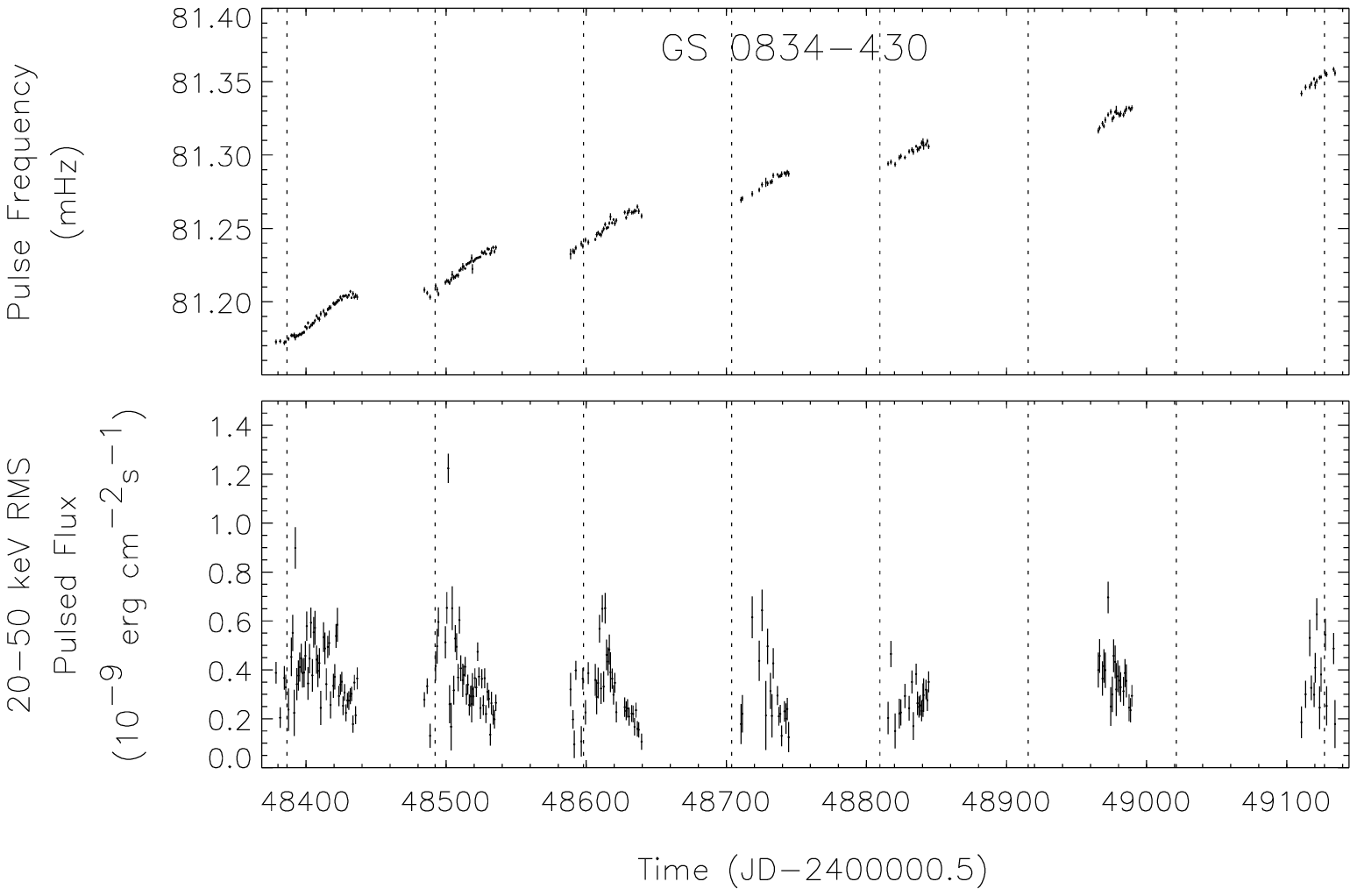}
\figcaption{{\final} 
GS 0834-430 frequency and flux measurements from BATSE.  The
intrinsic spin frequencies, which have been orbitally corrected using
parameters from Wilson \etal (1997), were determined at 1-day
intervals from the power spectra of the 20--70 keV CONT data \powspec.
The pulsed fluxes were determined at 1--day intervals by
assuming an exponential spectrum with an e--folding energy of 14~keV \flux.
\label{gs0834freqflux}}
%\end{figure}

{\em GS 0834-430.} ---
GS 0834-430 was first dectected in 1990 February by Granat/{\it WATCH}, 
but confusion with the X-ray burster MX 0836-42 
made unambiguous identification difficult (\cite{Lapshov92}).
Subsequent observations with Ginga revealed 12.3 s 
pulsations ({\cite{Aoki92}).
The optical counterpart is still unknown.  A
detailed discussion of BATSE observations has appeared elsewhere
(\cite{CWilson97}). To date, BATSE has observed 7 outbursts with
durations of 30--70 days, the first 5 of which were spaced at
105--107\,d intervals and the last 2 of which were unevenly spaced
(\cite{CWilson97}).  The eccentricity
$e$ and semi-major axis $a_x \sin i$, given in Table {\taborbits}, 
are individually poorly
determined due to large spin-up torques during the outbursts,
but $e\times a_x \sin i~=~15^{+6}_{-1}$~lt-s is well
constrained, thus establishing that the orbit is eccentric. This and
the recurrent outburst behavior is strongly reminiscent of the Be
transients, although GS~0834-430 falls below the Be-binary trend on the
Corbet diagram (Figure~\ref{fig:corbet}).
Pulsations are seen in the energy range 20--70 keV, and simultaneous
20--70 keV pulsed and Earth occultation DC flux measurements on 1991 
June 15--28 1991 (MJD 48422--48435), 
September 19--October 3 (MJD 48518--48532), 
December 15--27, (MJD 48605--48617) and 
1992 July 16--29 (MJD 48819--48832)
yielded consistent peak-to-peak pulsed fractions of 10--15\%, 
and marginal evidence
for an increase of pulsed fraction with energy.
The pulse profiles vary with both energy and time (\cite{CWilson97}).

%\newpage
%\begin{figure}
\psfig{file=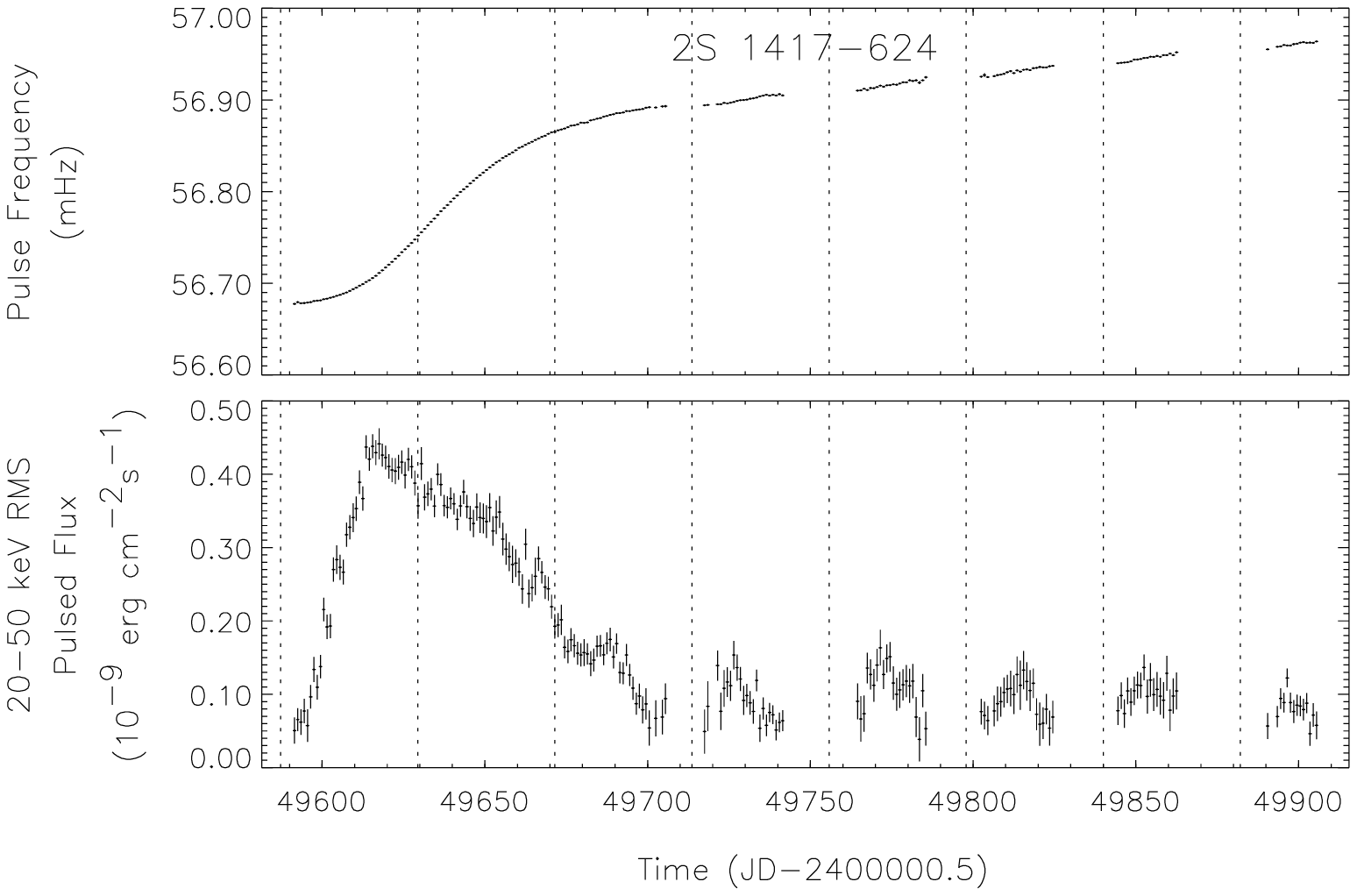}
\figcaption{{\final}
2S~1417--624 frequency and pulsed flux measurements from BATSE.
The intrinsic spin frequencies, which have been orbitally corrected
using parameters from Finger, Wilson \& Chakrabarty (1996) 
\protect\nocite{Finger1417}, were
determined at 1--day intervals by epoch-folding the 20-50 keV DISCLA
data \efold.  The pulsed fluxes were determined at 1--day
intervals by assuming a Comptonized spectrum model of the form
$F(E)=AE^{\lambda}\exp (-E/kT)$, with $\lambda = 1.6$ and $kT = 11.9$
keV (\protect\cite{Finger1417})  \flux.  The orbit was determined
assuming a correlation between pulsed flux and accretion torque,
which could potentially introduce modulations in the apparent rate of spin-up
during the sequence of outbursts following the main outburst.
\label{2s1417freqflux}}
%\end{figure}

{\em 2S\,1417--624.} ---  {\it SAS-3} discovered 17.6~s
pulsations from 2S\,1417--624 in 1978 (\cite{Kelley81}) and the
companion was later identified to be a 17th magnitude OB star
(\cite{Grindlay84}).  Detailed discussion of BATSE observations have
appeared elsewhere (Finger, Wilson \& Chakrabarty 1996).  BATSE
observed a large outburst of 2S 1417--624 from 1994 August 29 --
December 11 (MJD 49593--49697), followed by a sequence of five
smaller outbursts of diminishing amplitudes occuring every $\sim$40
days (\cite{Finger1417}), and two later outbursts (not shown).
At the peak of the initial outburst,
pulsations were detected up to 100 keV.  The pulse profile is
double-peaked and the ratio of the flux in the two peaks
changed systematically during the initial outburst. The binary orbit
was measured by a pulse timing analysis,
assuming that the accretion torque was correlated with the
measured pulsed flux. During the large outburst the spin-up
rate reached $\dot\nu \simeq 4 \times 10^{-11}$Hz s$^{-1}$.  
%{\nopfrac}

%\newpage
\psfig{file=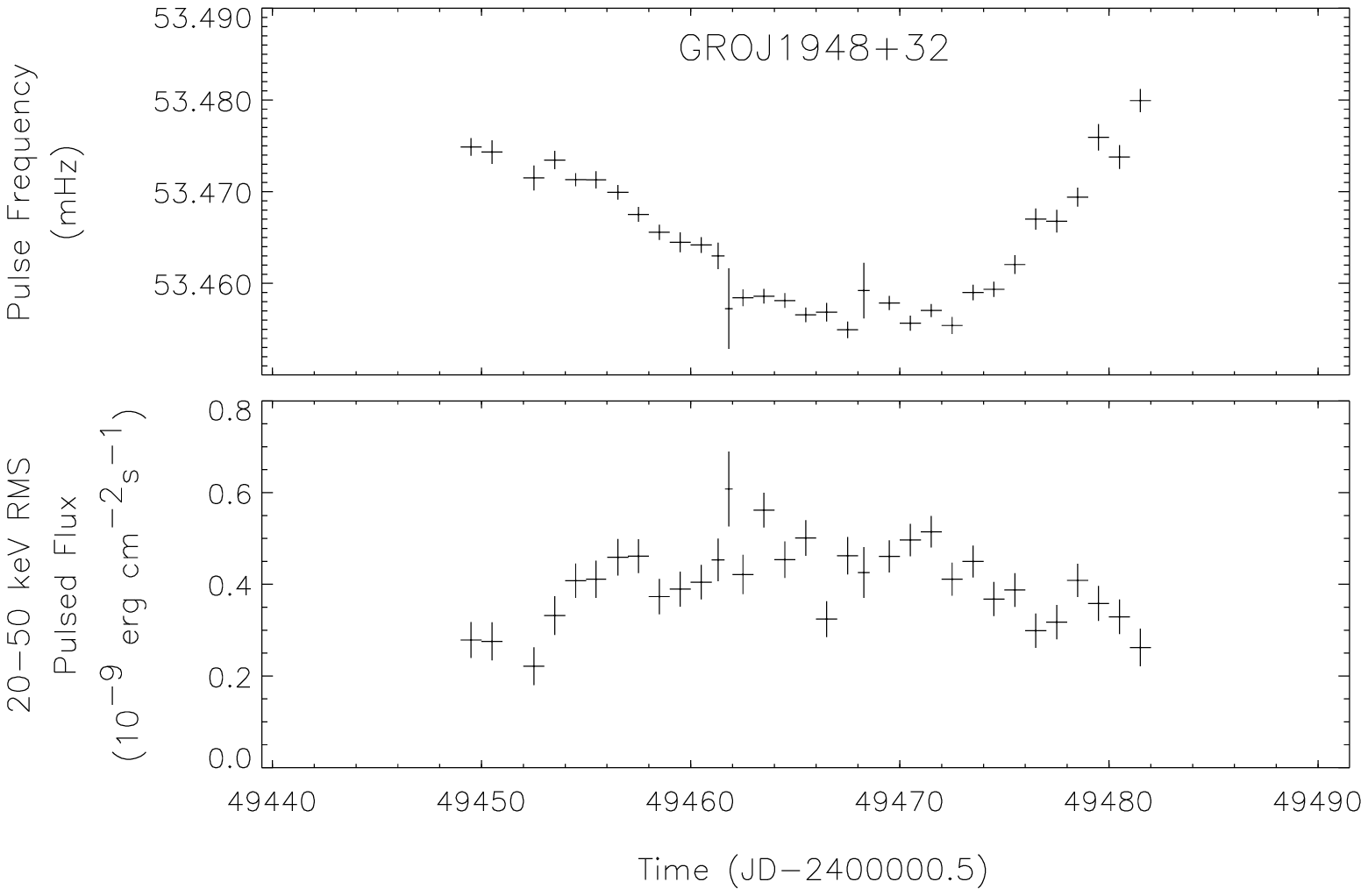}
\figcaption{{\final}
GRO~J1948+32 frequency and pulsed flux measurements from BATSE.
The pulse frequencies, which has not been orbitally corrected as the
orbital parameters are unknown, were determined at 1-day intervals
from the power spectra of the 20--70 keV CONT data \powspec.
The pulsed fluxes were obtained by assuming an exponential spectrum
with an e--folding energy of 15~keV \flux .}

{\em GRO J1948+32.} ---
  BATSE discovered and observed a single, 35 day outburst from the
18.7~s X-ray pulsar GRO~J1948+32 from 6 April to 12 May 1994 (MJD
49448--49482) and localized the source to within 10 deg$^2$
(\cite{Chakrabarty95}).  The pulse frequency showed a modulation
suggestive of orbital variation over less than a full cycle.
The 20--75 keV pulsed flux reached a maximum of 50 mCrab
on the 5th day of the outburst.
There is evidence for spectral variability uncorrelated with time or
intensity (\cite{Chakrabarty95}). The system is probably a Be
transient, although an orbit could not be uniquely measured and the
companion has not been identified.  
%{\nopfrac}

%\begin{figure}
\psfig{file=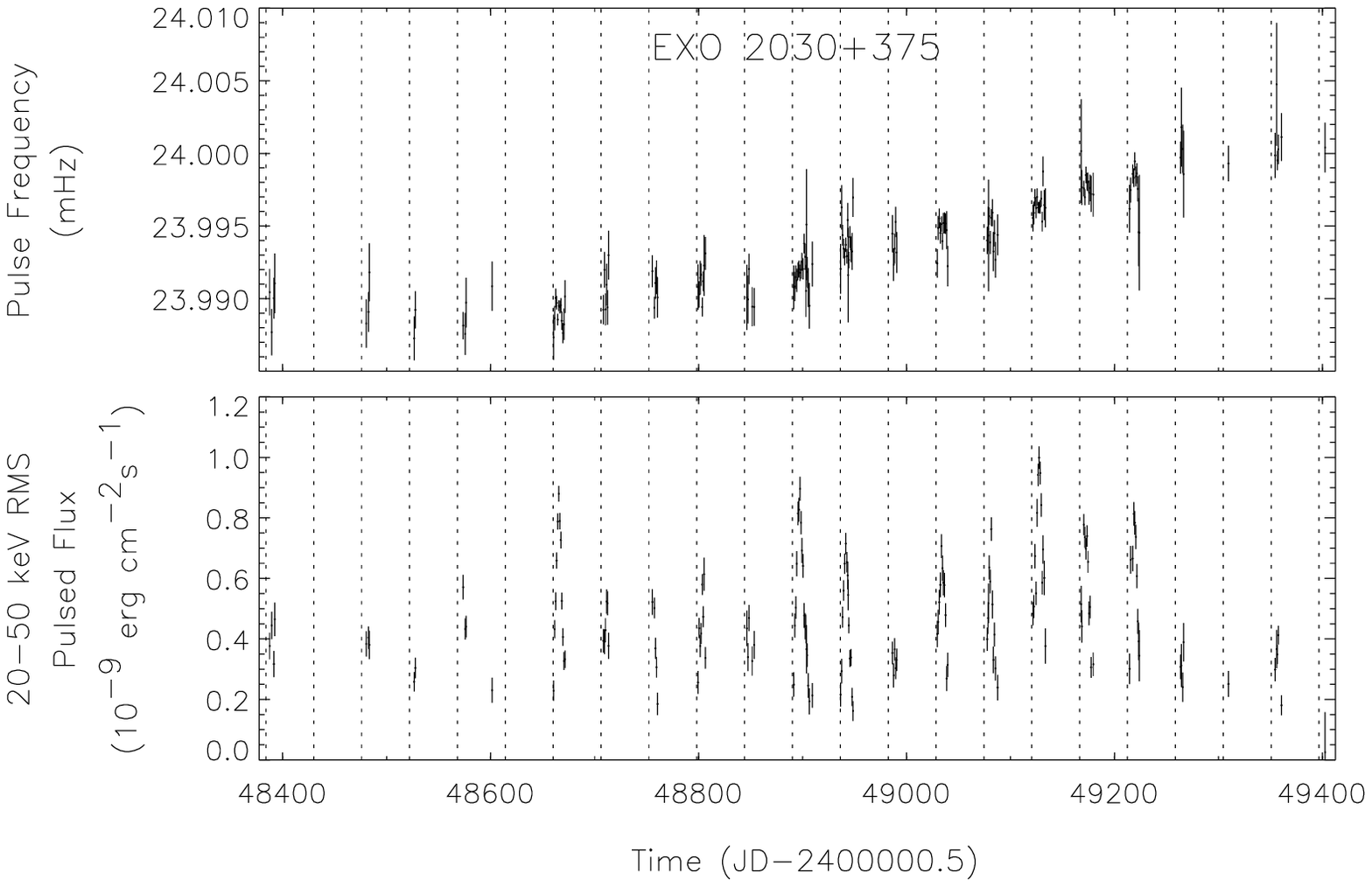}
\figcaption{{\final}
EXO 2030+375 frequency and flux measurements from BATSE.  The
intrinsic spin frequencies, which have been orbitally corrected using
parameters from Stollberg \etal (1994), were determined at 1-day
intervals from the power spectra of the 20--70 keV CONT data \powspec. 
The pulsed fluxes were determined at 1--day intervals by
assuming an exponential spectrum with an e--folding energy of 20~keV \flux. 
\label{exo2030freqflux} }
%\end{figure}

{\em EXO 2030+375.} ---
{\it EXOSAT} discovered 41.7~s pulsations from EXO~2030+375 during a
strong outburst of $\sim$80~d duration starting in May 1985, and observed a
smaller outburst in October 1985 (\cite{Parmar89}).  The companion was
later identified as a B0 Ve star (\cite{Coe88}).  The {\it
EXOSAT} observations found an orbital period of $\approx$46\,d and a
strong correlation of both the accretion torque and pulse shape with
luminosity, although the orbit and accretion torque could not be
separately measured. 

Detailed discussions of the BATSE observations of EXO 2030+375 have
appeared elsewhere (\cite{Stollberg93b,Stollberg94}).  
During the interval 1992 February 8 -- 1993 August 26 (MJD 48661--49226),
13 consecutive outbursts of EXO 2030+375 were seen with durations
of 7--19\,d, spaced at approximately 46\,d intervals 
(\cite{Wilson92,Stollberg94}).  A few detections of marginal statistical 
significance preceeded and followed the sequence of outbursts.
Over these 13 outbursts, EXO 2030+375 spun up at a mean rate of
$\dot{\nu}_{\rm s}\simeq1.3\times10^{-13}$\,Hz\,s$^{-1}$.  
The pulse profile is
double peaked with no evidence for spectral
differences between the two peaks (\cite{Stollberg93b}) and no 
pulse profile variations as were seen by EXOSAT (\cite{Parmar89b}).
This sequence of outbursts has allowed the first unambiguous
determination of the orbital parameters (\cite{Stollberg94}), shown
in Table {\taborbits}, 
indicating that the outbursts all began at or shortly after
periastron passage.  The orbit measured with BATSE has been used to
determine the correlation between luminosity, $L$, and accretion
torque, $N$, in the {\it EXOSAT} May--August 1985 outburst, yielding a
functional dependence $N\propto L^{1.2}$ (\cite{Reynolds96}).

  The source was quiescent for 2.5 years before being detected by
BATSE again in April and May 1996 (\cite{Stollberg96}).  These two
outbursts occurred $\sim$5\,d prior to periastron passage.
The latest outbursts were detected in July and November 1996.
The spin frequency of the latest
outbursts indicate that during quiescence EXO 2030+375 had spun down
at a rate $\dot{\nu}\simeq-3.4\times10^{-14}$Hz\,s$^{-1}$.  Simultaneous
30--70 keV pulsed and Earth occultation DC flux measurements on MJD
49120--49131 yielded a peak-to-peak
pulsed fraction of $0.36(5)$ (\cite{Stollberg94}).

%\newpage
%\begin{figure}
\psfig{file=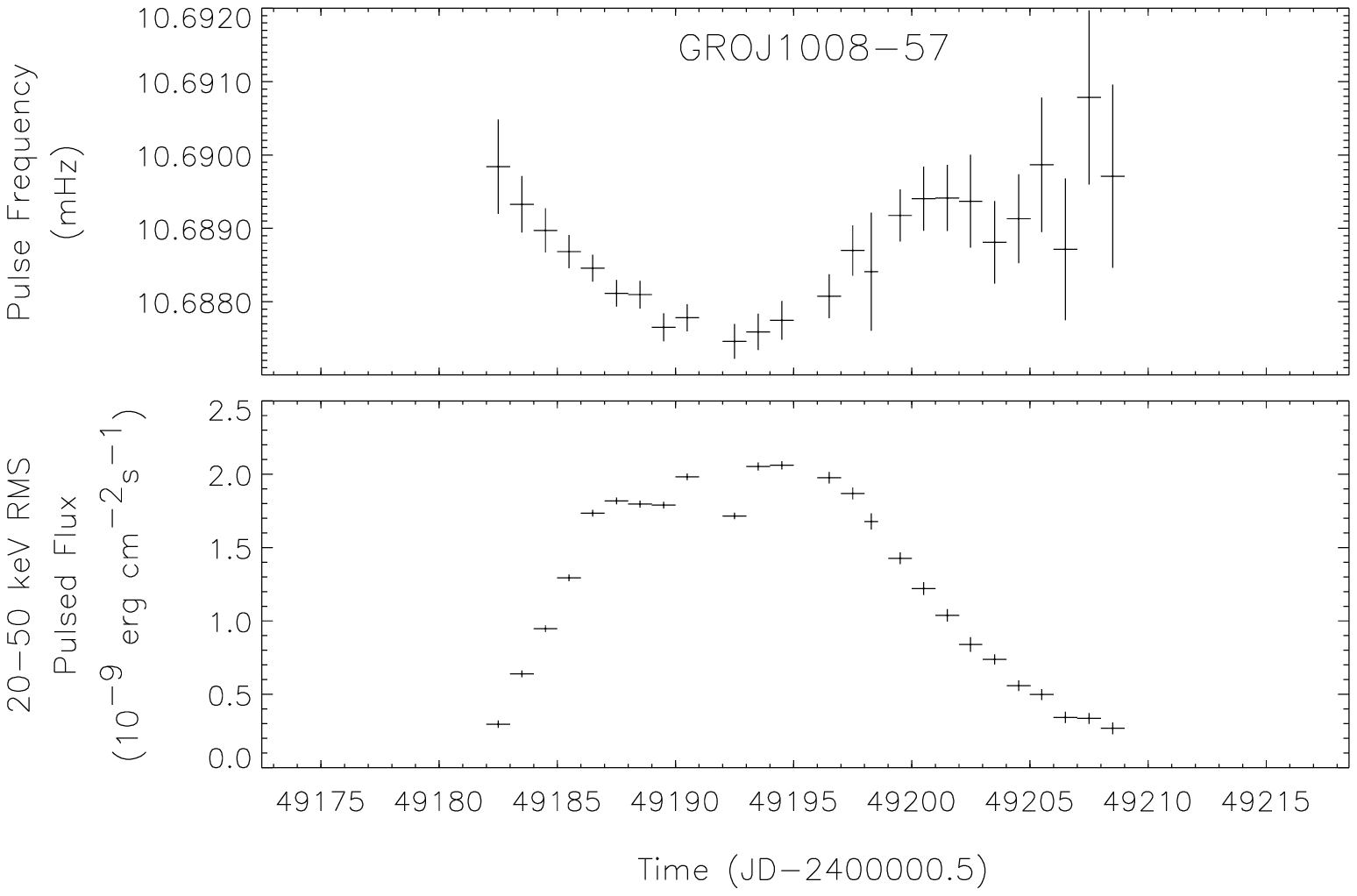}
\figcaption{{\final}
GRO~J1008-57 frequency and flux measurements from BATSE.  The
pulse frequencies, which have not been orbitally corrected as the
orbital parameters are unknown, were determined at 1-day intervals
from the power spectra of the 20--70 keV CONT data \powspec.
The pulsed fluxes were determined at 1--day intervals by assuming an
exponential spectrum with an e--folding energy of 20~keV  \flux. 
\label{j1008freqflux}}
%\end{figure}

\nocite{Coe94_J1008,Wilson94c}
{\em GRO J1008--57.} ---
  BATSE discovered 93.5~s pulsations and observed a 33 day outburst
from J1008--57 (\cite{Stollberg93a}) from 14 July to 16 August 1993
(MJD 49182--49215). A preliminary discussion of the BATSE observations
of GRO J1008--57 appeared in Wilson \etal (1994b). The source
localization to 2.5$^\circ$ by the Earth-occultation technique
(\cite{Stollberg93a}) and later by OSSE
(\cite{Grove93}), ASCA (\cite{Tanaka93}), and ROSAT (\cite{Petre93}) 
to $15^{\prime}$.  Coe \etal (1994a) later identified the companion
to be a Be star.
GRO~J1008--57 has a hard spectrum,
with pulsations observed from 20--160 keV. The peak-to-peak
pulsed fractions,
averaged over the interval MJD 49186--49195, are
0.66(9) (20--30 keV), 0.65(7) (30--40 keV), 0.69(7) (40--50
keV), and 0.76(15) (50--70 keV). Four additional outbursts,
not shown in Figure~\ref{j1008freqflux}, were
observed during March 1994, November 1994, and March 1996. The very
weak and short duration later outbursts occurred at multiples of
$\approx$248 days, indicating that this may be the orbital period of the
system.  

%\newpage
%\begin{figure}
\psfig{file=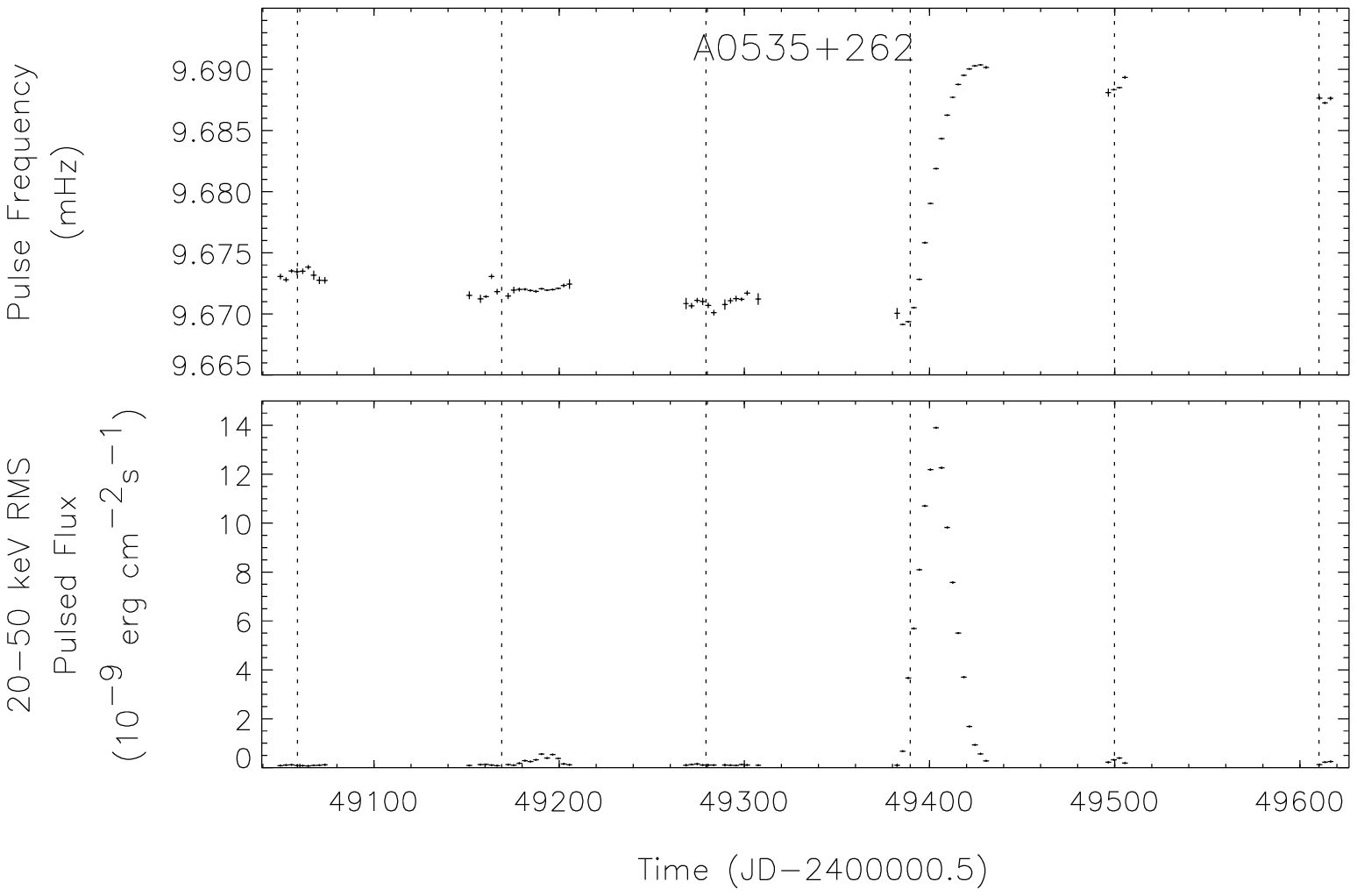}
\figcaption{A0535+262 frequency and pulsed flux measurements 
from BATSE.  The frequencies, which have been orbitally corrected
using orbital parameters from \protect\cite{Finger94e}, 
were determined at 3--day intervals from fits to phase measurements
made using the 20--50 keV DISCLA data  \phase.  
The pulsed fluxes were measured at 1--day intervals by 
assuming an exponential spectrum with an e--folding energy of 20 keV 
\flux.
\label{a0535freqflux}}

%\end{figure}

{\em A 0535+26.} ---
{\it Ariel 5} discovered 103~s pulsations from A 0535+26 in 1975
(\cite{Rosenberg75,Coe75}) and its companion is the Be star HDE 245770
(\cite{Stier76}, \cite{Hutchings78}). The 111~d orbital period of A
0535+26 was first inferred from the spacing of X-ray outbursts
(\cite{Nagase82}). The binary undergoes frequent outbursts with a wide
range of intensities, the brightest reaching 3 Crab in the 2--10 keV
band (\cite{Giovannelli92}).

Detailed discussions of BATSE observations of A 0535+26 have appeared
elsewhere (\cite{Finger94e,Finger96c}).  BATSE has observed 6
outbursts spaced roughly at the orbital period, the 4th of which is a
``giant'' outburst that occurred from 28 January 1994--20 March 1994
(MJD 49380--49430) and reached a peak flux of 8 Crab in the BATSE
energy band.  There was little or no spin-up during the normal
outbursts, spin-down between outbursts, but rapid spin-up during the giant
outburst, suggesting accretion from a disk. 
The giant outburst showed
enough dynamic range that the relation between accretion torque and 
pulsed flux could be tested directly (see \S \ref{sect:torque-lum}).

BATSE has provided the first measurement of the binary orbit, and
detection of Quasi-Periodic Oscillations (QPO) during the giant
outburst (\cite{Finger94e,Finger96c}).  A cyclotron
absorbsion line at 110 keV was reported by OSSE (\cite{Grove95a}), 
which is also evident in the BATSE pulsed flux spectrum.
The pulse shape is complex and highly variable with both
energy and intensity as shown in Figure \ref{fig:A0535pulses}.  Fluxes
could be measured with the occultation method only during the giant
outburst, and these yielded a 20--50 keV peak-to-peak pulsed fraction of
$> 0.8$ at low flux and decreased to about 0.3 at the highest flux.

%\newpage
%\begin{figure}

\centerline{
\psfig{file=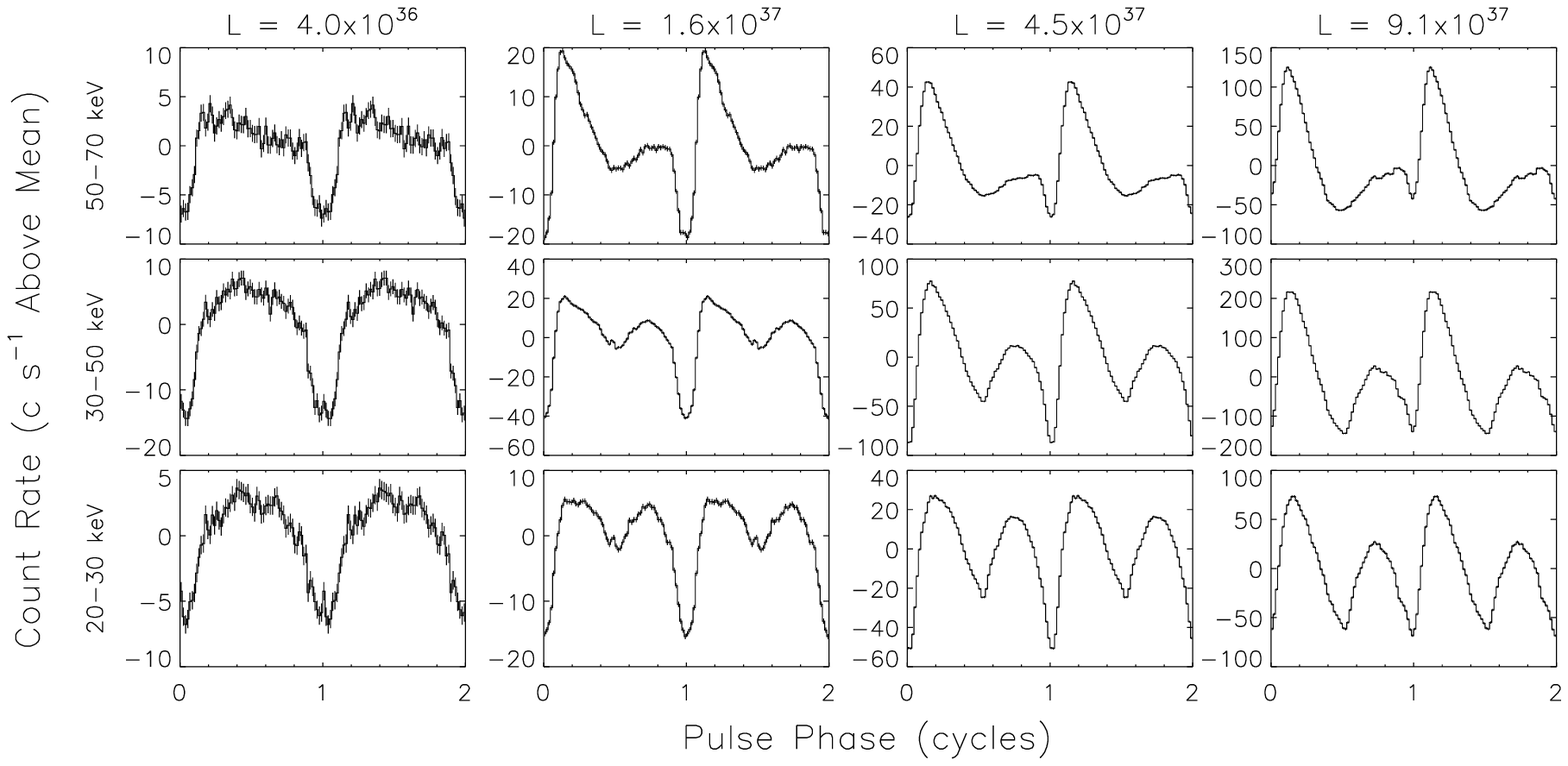}
}
\figcaption{Pulse Profile as a function of pulsed flux and energy 
for A0535+26.
Pulse profiles of A0535+262 during the giant outburst in
February--March 1994, obtained by epoch-folding CONT data. Profiles in
three energy bands are given for four time intervals. The mean
luminosity $L$ in ergs s$^{-1}$ is given for each time interval. The
time intervals are February 15.1-17.6 ($L=9.1 \times10^{37}$), February
25.0-March 1.5 ($L=4.5 \times10^{37}$), March 5.0-8.6
($L=1.6 \times10^{37}$), and March 13.1-15.6
($L=4.0 \times10^{36}$). Luminosities were calculated from 20-100 keV
fluxes based on occultation measurements by assuming a distance of 2
kpc and assuming the 20--100\,keV band contains 45\% of the bolometric flux.
\label{fig:A0535pulses}}
%\end{figure}

%\newpage
\psfig{file=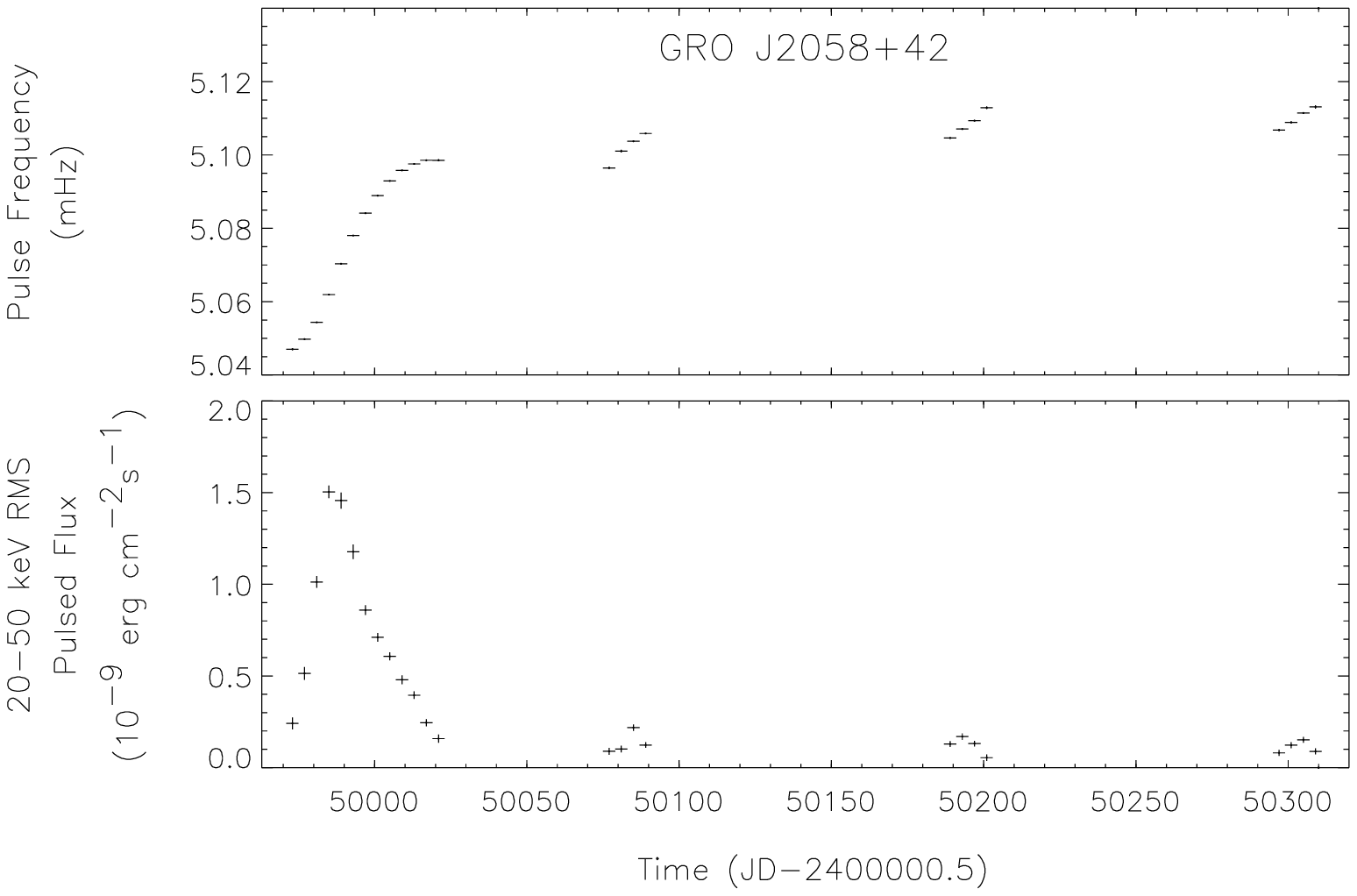}
\figcaption{{\final}
GRO J2058+42 frequency and flux measurements from BATSE.
The spin frequencies, which have not been orbitally corrected as
the orbital parameters are unknown, were determined at 4--day intervals
by epoch folding the 20--50 keV DISCLA data at a range
of trial frequencies  \efold.
The pulsed fluxes were determined at 4--day intervals 
by assuming an exponential spectrum   
with an e--folding energy of 20 keV  \flux.
\label{J2058freqflux}}

{\em GRO J2058+42.} ---
BATSE discovered 198~s pulsations and observed an intitial 46 day outburst 
from GRO~J2058+42 (\cite{Colleen95}) from 1995 September 14 to
October 30 (MJD 49974-50020). The source was localized to a
$1\arcdeg\times4\arcdeg$ error box with BATSE using both pulsed 
and Earth occultation data.
OSSE scans further reduced the size of the error box 
to $30^{\prime} \times 60^{\prime}$
(\cite{Grove95b}), and target-of-opportunity scan with the RXTE PCA in 
November 1996 reduced the error region to a 4\arcmin~circle 
(\cite{CWilson96}).  The optical counterpart has not been determined.
The total flux, as measured by Earth occultation, peaked 
at about 300 mCrab (20--50 keV). 
The large initial outburst was followed by a sequence of 
4 much smaller outbursts with pulsed 20--50 keV
fluxes peaking at 15-20 mCrab, the first 3 of which are shown in 
Figure~\ref{J2058freqflux}.  The 
outbursts were spaced by $\approx$ 110 days, which is 
likely to be the orbital period.

%\newpage

\psfig{file=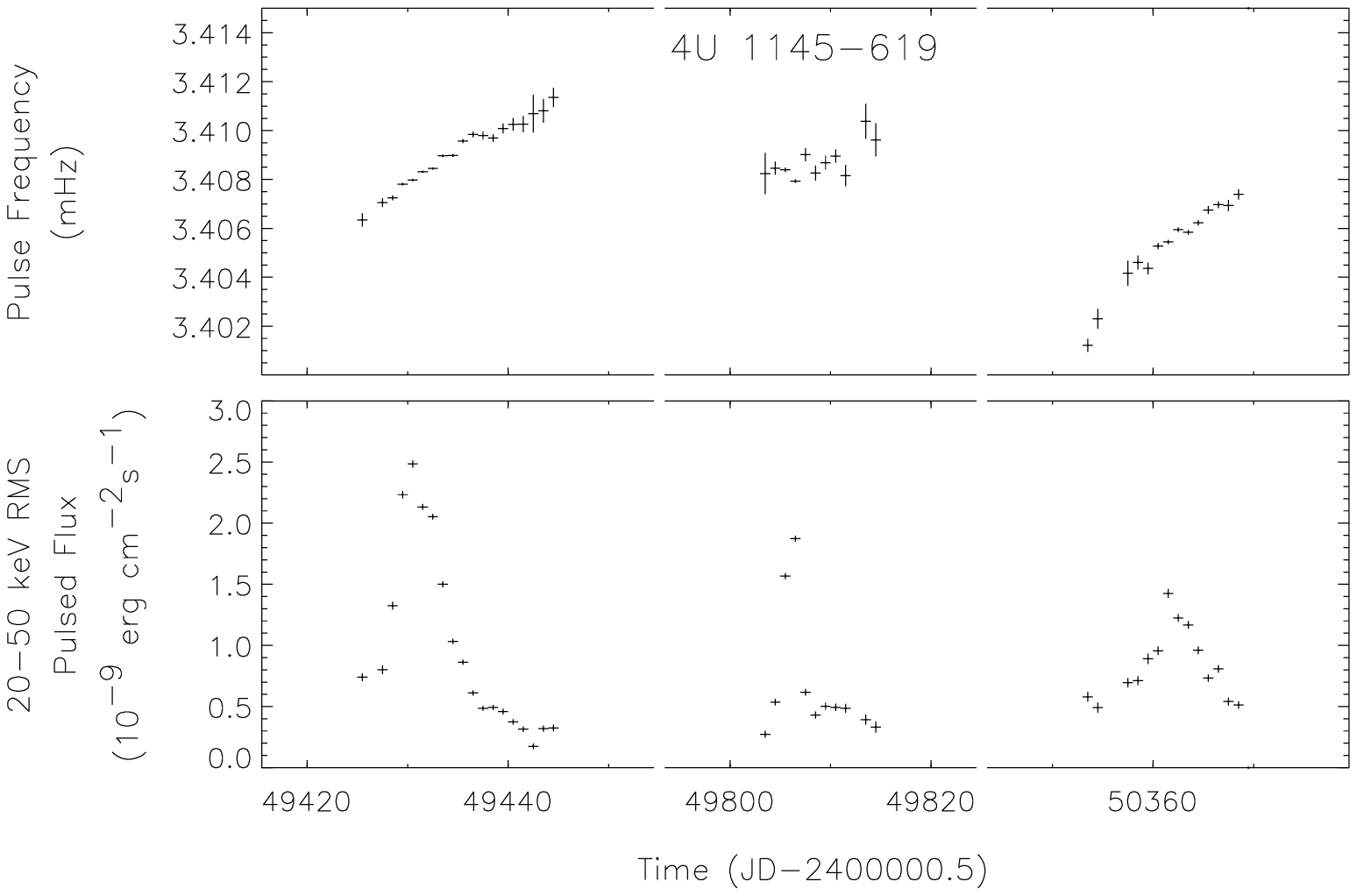}
\figcaption{{\final} 
4U~1145-619 frequency and pulsed flux measurements from BATSE.
The pulse frequencies, which have not been orbitally corrected
as not all the orbital parameters are known, were
determined at 1-day intervals by epoch folding the 20--50 keV
DISCLA data.  The pulsed fluxes were determined at
1--day intervals by assuming an exponential spectrum with a
e--folding energy of 15~keV  \flux.  }

{\em 4U 1145--619.} ---
{\it Ariel 5} discovered 292.5~s pulsations from 4U1145--619 in 1977
(\cite{White78}). The companion is the 9th magnitude Be star Hen 715
(\cite{Dower78,HH80,Bianchi80}), which exhibits emission lines and has
an equatorial rotational velocity of $v \sin i = 290 {\rm km}~s^{-1}$
(Hammerschlag-Hensberge \etal 1980, Bianchi and Bernacca 1980). 
%(G. Hammerschlag-Hensberge \etal Astr Ap 1980 85, 119, 
%L. Bianchi and P.L. Bernacca Astr Ap 1980 89, 214.)
An orbital period of 186.5~d was inferred from the recurrence times of
outbursts, which typically last $\approx$ 10~d
(\cite{Watson81,Priedhorsky83}).
%W.C. Priedhorsky and J. Terrell, 1983, ApJ 272, 709;
%M.G. Watson, R.S. Warwick and M.J. Ricketts 1981, MNRAS 195, 197.)
Pulse frequency variations over multiple EXOSAT observations
imply an eccentricity of $e \gsim 0.6$ (\cite{Cook87a}).
To date, BATSE has observed 7 outbursts, of which
three are shown.  %{\nopfrac}
The separation between the BATSE outbursts is in good agreement
with the 186.5~d period.

%\newpage

\psfig{file=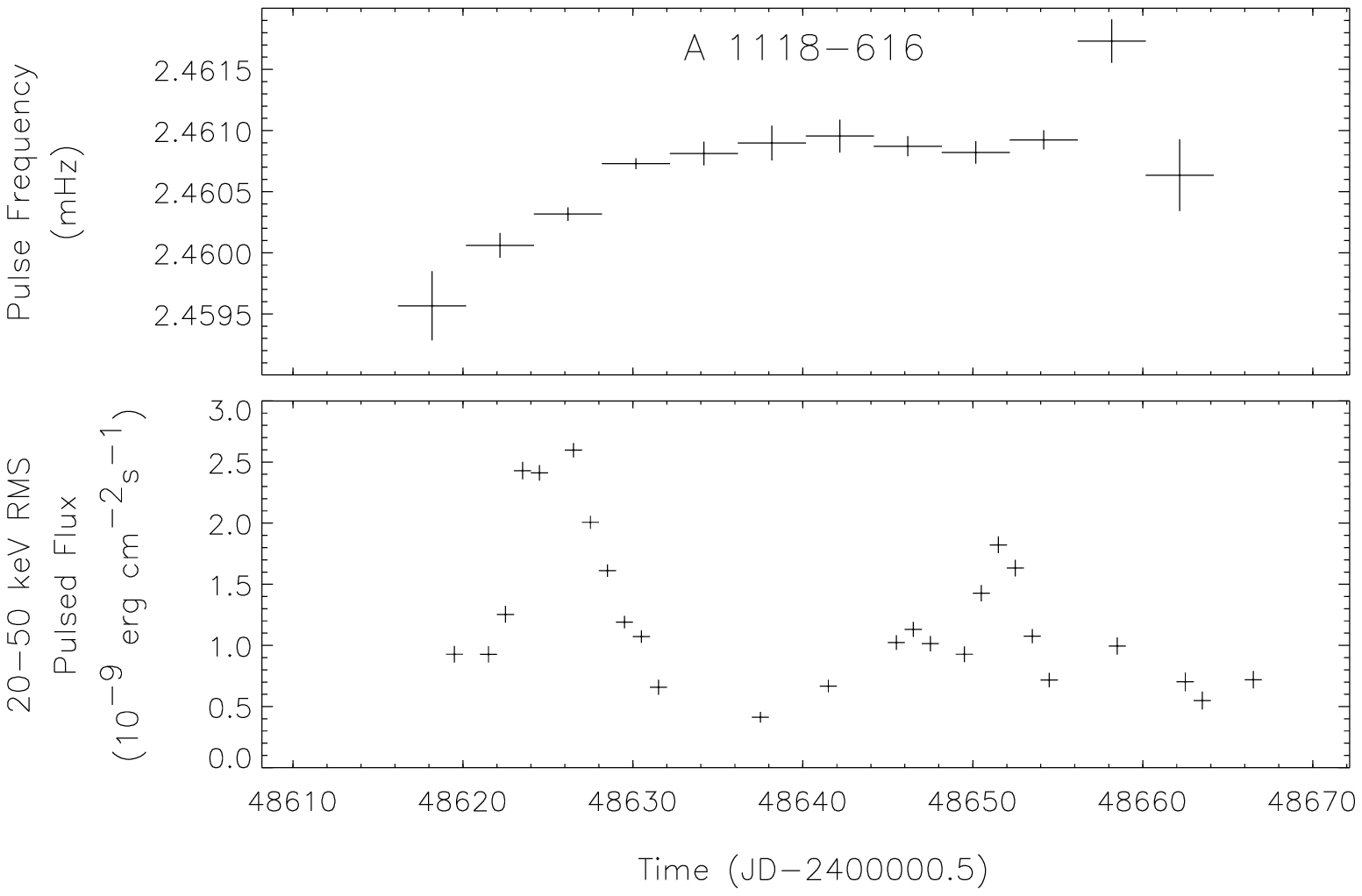}
\figcaption{{\final} A~1118--616 
frequency and pulsed flux measurements from BATSE.
The spin frequencies, which have not been orbitally corrected 
as the orbital parameters are unknown,
were determined by Coe \etal (1994b)
at 4--day intervals by epoch folding the 20--50 keV
DISCLA data at a range of trial frequencies  \efold.
The pulsed fluxes were determined at 1--day intervals  
by assuming an expotential spectrum
with an e--folding energy of 15~keV  \flux.}

\nocite{Coe94_A1118}

{\em A 1118--616.} ---
{\it Ariel 5} discovered 406.5~s pulsations from A~1118-616 in 1974 
(\cite{Ives75}) and the optical companion was later identified 
to be the Be star He 3-640/Wray 793 (\cite{Chevalier75,Heinze76,Wray76}).
%(Henize 1976, ApJS 30, 491, J.D. Wray 1976, PhD thesis, Northwestern U).
Since the initial discovery, no outbursts were
observed until BATSE detected one from 1991 December 30 -- 1992 January 10 
(MJD 48621--48633).  This outburst reached a 20--70 keV
pulsed intensity of $\approx$14 $\cps$ on 1992 January 3, followed by
approximately 50 days of erratic flaring behavior with a maximum on 
1992 February 1(MJD 48654) (\cite{Coe94_A1118}).  The WATCH experiment on {\it
GRANAT} independently detected and monitored the outburst
(\cite{Lund94}), which was also observed by the IUE and ground-based
telescopes (\cite{Coe94_A1118}).  The X-ray outburst was accompanied by an
 increase in $H_{\alpha}$ emission and an IR excess,
indicative of an extended disk around the companion star (\cite{Coe94_A1118}).
Pulsed emission is detected from 20--100 keV at the peak of the
outburst.  %{\nopfrac}

%%%%%%%%%%%%%%%%%%%%%%%%%%%%%%%%%%%%%%%%%%%%%%%%%%%%%%%%%%%%%%%%%%%%%%%%%%%%%%%

%\input discussion.tex

\section{DISCUSSION}

The long-term, continuous all-sky monitoring of accreting pulsars
by BATSE is providing new insight into these systems.
In \S 5.1, we show how BATSE observations have yielded a
qualitatively different picture of the spin behavior of disk-fed
pulsars on long timescales ($\sim$years) than understood from earlier
measurements. BATSE has also been able to test theories of accretion
torque on short timescales ($\sim$days) in transient pulsars (\S 5.2).  BATSE
observations of accretion torques in transient and wind-fed systems
show evidence of spin down during quiescence and of disk formation in
a predominantly wind-fed binary (\S 5.3).  Continuous monitoring
of persistent systems makes it possible to
quantify the variability of
accretion torques on timescales of months to years using power spectra (\S
5.4). BATSE's continuous monitoring capability 
has also provided new
insights into the properties of binaries containing pulsars 
which undergo transient
outbursts (\S~\ref{sect:trans}), 
the population of Be transient pulsars (\S~\ref{sect:pop}),
and the evolution of B-star binaries into
Be-transient accreting binary pulsars (\S~\ref{sect:be_orbits}).

\subsection{The Long Term Spin Evolution of Disk-Fed Pulsars}

The picture of long-term pulsar spin evolution developed in the
mid-1970s was based on sparse measurements provided by pointed
observations of $\sim$10 objects (\cite{RappaportJoss77a,GhoshLamb79}).
In particular, the spin behavior of Cen X-3 and Her X-1 at that time
suggested that the simple spin-up torque estimate in equation
(\ref{eq:torque}) was sometimes inadequate: these pulsars were
apparently spinning up on a timescale much longer than predicted by
equation (\ref{eq:tspin}). Moreover, both sources also underwent short
episodes of spin-down, indicating that angular momentum was actually
being lost by the pulsar while it continued to accrete. The continuous
pulse monitoring by BATSE, however, reveals that these early
observations sometimes gave a false impression of the strength and
continuity of the accretion torque.

The frequency history of the 4.8\,s pulsar Cen X-3, shown in Figure
\ref{fig:cenx3}, is an example where BATSE observations reveal a
strikingly different picture of pulsar spin behavior than previously
hypothesized.  Prior to 1991, the long-term frequency evolution (Figure
\ref{fig:cenx3}a) had been described as secular spin-up at $\dot \nu \simeq
8\times10^{-13}$\,Hz\,s$^{-1}$ (a factor of $\sim$5 slower than
predicted by equation (\ref{eq:torque})), superposed with fluctuations
and short episodes of spin-down. In contrast, the more frequently
sampled BATSE data (Figure \ref{fig:cenx3}b) show that Cen X-3
exhibits 10$-$100\,d intervals of steady spin-up and spin-down at a
much larger rate, consistent with equation~\ref{eq:torque}.
Figure~\ref{fig:torque-histogram} is a histogram of torques observed
in Cen X-3 showing a roughly bimodal distribution of torque states,
with the average spin-up torque ($\sim+7\times10^{-12}$\,Hz\,s$^{-1}$)
larger in magnitude than the average spin-down torque
($\sim-3\times10^{-12}$\,Hz\,s$^{-1}$).  Transitions between spin up and
spin down occur on a timescale more rapid than BATSE can resolve
($\la$10\,d).  The long-term spin-up rate inferred
from the pre-BATSE data is not representative of the instantaneous
torque; its small value is a consequence of the frequent transitions
between spin up and spin down.

 Interestingly, this switching behavior is very common.  At least 4
out of the 8 persistent pulsars observed by BATSE show torque
reversals between steady spin-up and steady spin-down.  The 7.6\,s
pulsar 4U~1626--67 underwent a reversal to smooth spin down at a rate
$\dot \nu\simeq-7\times10^{-13}$\,Hz\,s$^{-1}$ in 1991 after two decades of
smooth spin up with $\dot \nu \simeq +8.5\times10^{-13}$\,Hz\,s$^{-1}$
(\cite{Chakrabarty97a}). Most surprisingly, the final torque is
nearly equal in magnitude but opposite in sign.  A similar transition
to spin down was observed in the 120\,s pulsar GX 1+4 in 1988
(Makishima et al. 1988) after more than a decade of steady spin up
(Figure \ref{fig:freqmosaic}). Again, the spin down rate
($\sim3.7\times10^{-12}$\,Hz\,s$^{-1}$) is close in magnitude to the spin
up rate. In the 38\,s pulsar OAO~1657--415, both the duration
and strength of torque episodes are very close to those seen in Cen
X-3 (Chakrabarty et al. 1993). Of the remaining four systems, Her X-1 is
sampled infrequently at 35 day intervals so that we cannot measure its
torque on short timescales, while the other three (4U~1538--52, GX
301--2, Vela X-1) are wind-fed pulsars.

%\begin{figure}
\vbox{
\hbox{\psfig{figure=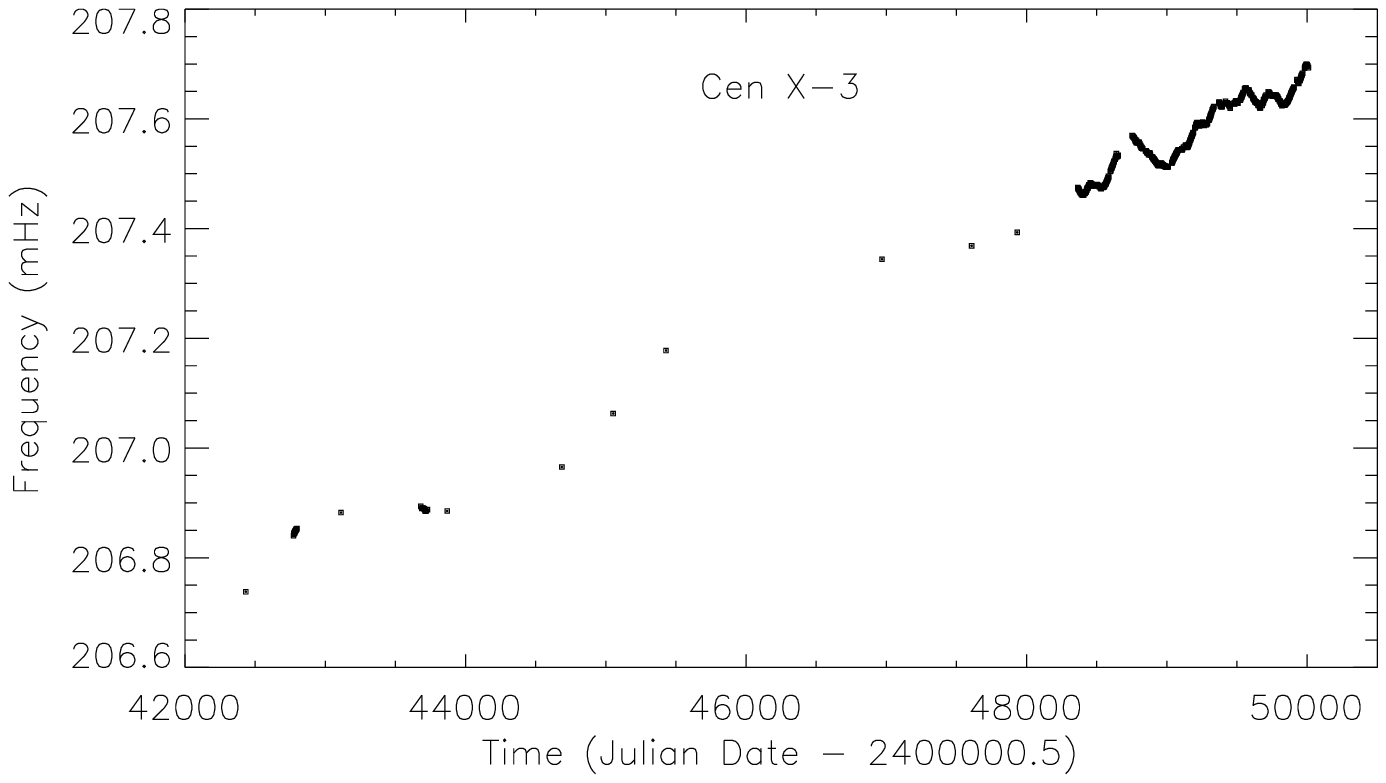}}\vspace*{0.2in}
\hbox{\psfig{figure=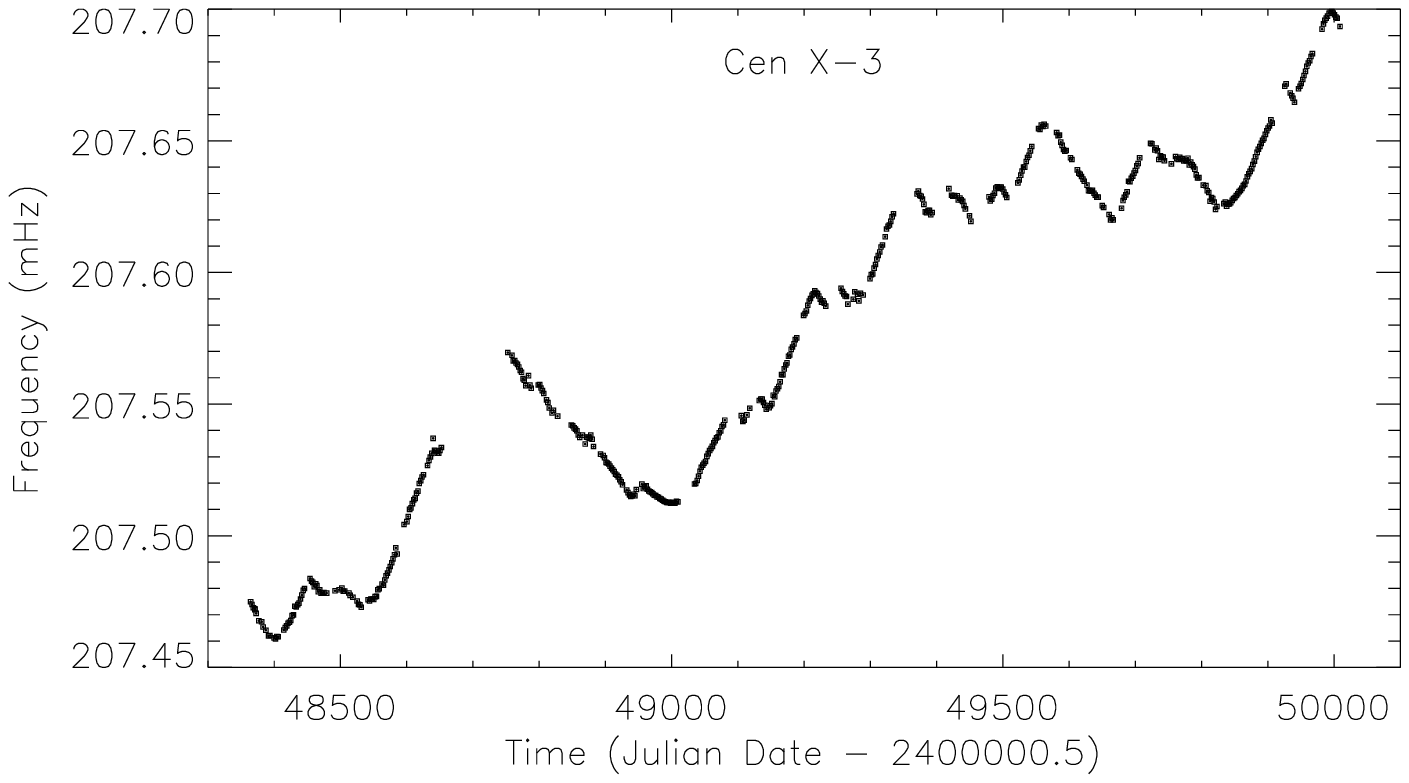}}
}
\figcaption{Frequency histories of Cen X-3
. Upper panel: The long-term
frequency history of Cen X-3.  Lower panel: High resolution BATSE
measurement of the intrinsic spin frequencies of Cen X-3.
\label{fig:cenx3}}
%\end{figure} 

%\begin{figure}
\centerline{
\psfig{file=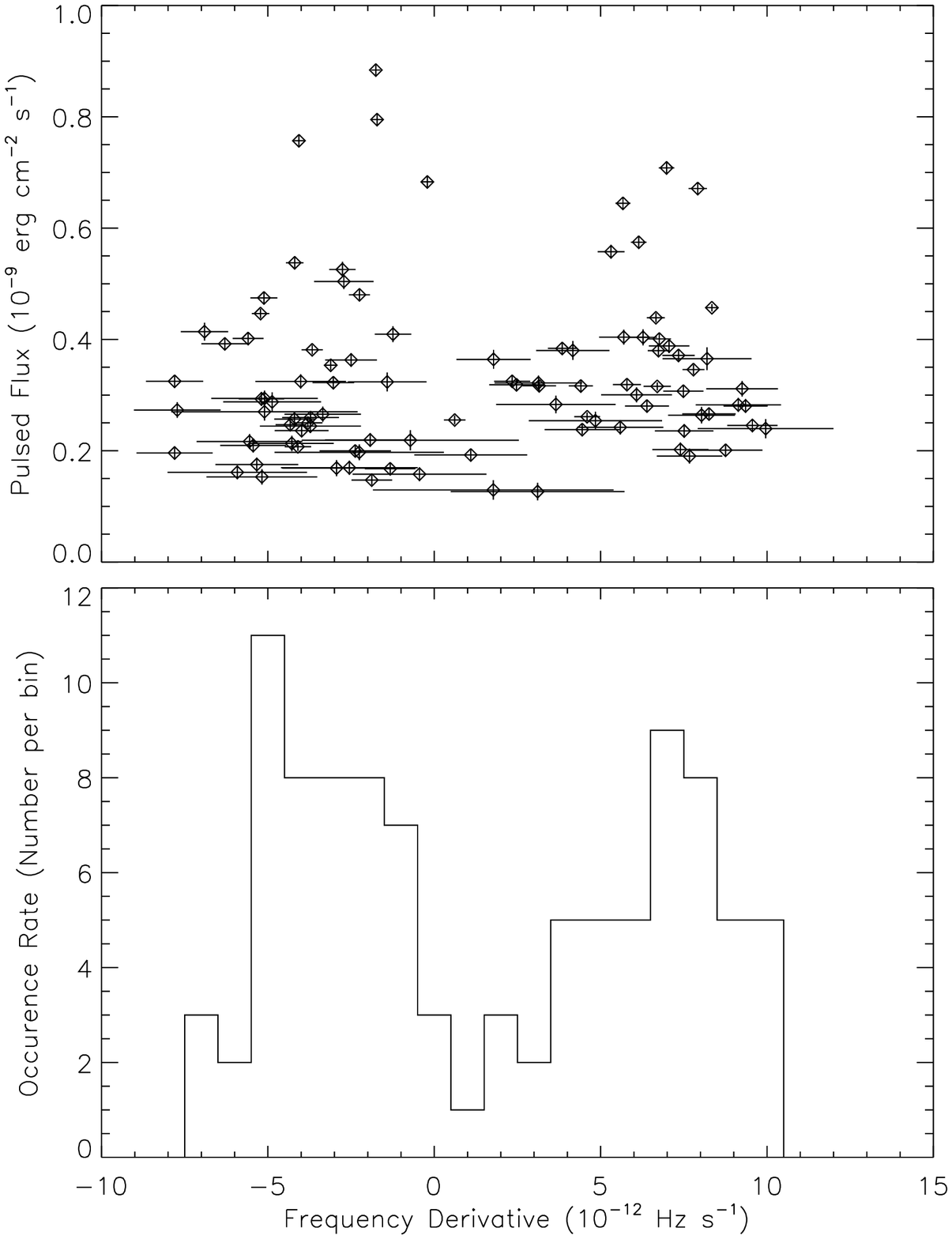,height=6in}
}
\figcaption{The Cen X-3 spin-up is plotted versus the 20--50 keV pulsed flux 
in the top panel. The bottom panel shows the Cen X-3 spin-up rate 
distribution. The fluxes used are 10 day averages. The spin-up rates are from
linear fits to the frequency measurements within the same 10 day intervals.
No clear correlation is seen between spin-up rate and flux. The spin-up 
rate distribution is clearly bi-model.
\label{fig:torque-histogram}}
%\end{figure}

   There are at least two classes of models that might explain
instantaneous spin-down in disk-fed pulsars, and both involve the
magnetic interaction between the accretion disk and the stellar
magnetosphere.  Ghosh and Lamb (1979) argued that Her X-1 and Cen X-3
must be near an equilibrium where the star rotates at a spin frequency
nearly equal to the Keplerian frequency of the magnetosphere,
$\Omega_{\rm spin} \simeq \Omega_{\rm K}(r_{\rm m})=(GM/r_{\rm
m}^3)^{1/2}$.  They found that additional {\it negative} torques would
then act on the star: magnetic field lines that thread the disk beyond
the corotation radius (where the disk rotates more slowly than the
star) are swept back in a trailing spiral and transport angular
momentum outward. Stars sufficiently close to equilibrium can spin
down while continuing to accrete. Other models attempt to explain
spin-down via the loss of angular momentum in a magnetohydrodynamic
outflow (\cite{Anzer80,Arons84,Lovelace94}).  Outflowing material
moves along rigid magnetic field lines like beads on a wire, gaining
angular momentum from the star as it is forced to corotate.  This
results in a stellar spin-down torque $N \sim \dot M_{\rm w}
\sqrt{GMr_\alpha}$, where $\dot M_{\rm w}$ is the loss rate in the
wind and $r_\alpha$ is the Alfven radius in the flow beyond which the
magnetic field is dynamically unimportant.

  The BATSE observations pose a number of difficulties for these
models. To produce the bimodal torque behavior we observe, most if not all
near-equilibrium models apparently require step-function-like changes
in the mass accretion rate --- finely tuned just so that the two
torques states have comparable magnitude, but opposite sign.  The many
transitions in Cen X-3, for example, {\it always} alternate between
torques of opposite sign: How does the companion star know just how to
change its mass transfer rate so that transitions between two torques
of the same sign never occur?  It seems especially implausible in a
system like 4U 1626--67, where the average mass accretion rate is likely
determined by the loss of orbital angular momentum via gravitational
radiation, that the companion would switch to such a finely-tuned mass
transfer rate that the spin-down torque would have nearly the same
magnitude as the previous spin-up torque (\cite{Chakrabarty97a}).
Since the timescale for angular-momentum loss via gravitational radiation
is much longer than the timescale of BATSE observations, changes in mass
accretion rate must be due to physical changes in the star or accretion disk.

Our observations suggest that disk-accreting pulsars are subject to
instantaneous torques of magnitude $\approx N_0\equiv \dot M
(GMr_{\rm co})^{1/2}$ (see equation \ref{eq:maxtorque}) and
only differentiate themselves by the timescale for reversals of
sign. We see some (e.g. Cen X-3) that switch within $\sim 10-100$ days,
whereas others (e.g. 4U 1626--67 and GX 1+4) switch once in 10--20
years. The primary theoretical issues are then identifying the physics
that sets this timescale and understanding why the magnitudes of the
spin-up and spin-down torques are so similar.

   It is intriguing to apply our picture of the long-term evolution of
disk-fed pulsars to those we cannot observe with BATSE. First, it
makes it more plausible that one of the class of pulsars which are
spinning down (1E 2259+586, 1E 1048.1--5937, 4U 0142+61, see Mereghetti
\& Stella 1995) might eventually switch to spin-up. The pulsar 1E
1048.1--5937 has the shortest spin-down time amongst these
($t_{\rm sd}\simeq10^4$~yr) and might be the most likely one to undergo
a torque reversal. There is already some evidence for a brief torque
reversal in 1E 2259+586 (\cite{Baykal96}).  The long-term torque
inferred for LMC X-4 is nearly a factor of 100 lower than $N_0$,
suggesting that this pulsar may be undergoing rapid switching like Cen
X-3.

\nocite{Mereghetti95}

\subsection{Torque and Luminosity of Transient Pulsars}
\label{sect:torque-lum}

 Short-term instantaneous torque measurements --- not long-term
averages --- are necessary to test accretion torque theory.  All such
theories predict that the magnetospheric radius should decrease as
$\dot M$ increases, and the simplest version (equation
{\ref{eq:alfven}) predicts $r_{\rm m}
\propto \dot M^{-2/7}$ for $r_{\rm m} < r_{\rm co}$.  This implies
that a pulsar 
should spin-up at a rate $\dot \nu \propto \dot M^{6/7}$. In
principle, we can test this prediction by measuring the correlation
between torque and bolometric luminosity.  Luminous outbursts in
GRO~J1744--28 and A~0535+262 seen with BATSE showed enough dynamic
range that the relation between torque and {\it observed} flux could
be tested directly (\cite{Finger96a,Finger1417}).  In addition, the
orbital parameters of EXO 2030+375 measured with BATSE made it
possible to compute the accretion torque from a luminous 1985 outburst
seen by {\em EXOSAT} (\cite{Reynolds96}).  
Figure ~\ref{fig:torque-lum} shows the spin-up rate of 
A~0535+26 versus 20--100\,keV flux (upper panel),
and the spin-up rate of 
GRO~J1744--28 versus 20--50\,keV RMS pulsed flux (lower panel).  Also plotted
are the power laws $\dot \nu \propto
F_{\rm obs}^\gamma$ with $\gamma=6/7$ (dotted line) and the best-fit
power laws (dashed line).  The best fit
index for A0535+262 is $0.951(26)$ and that for GRO J1744-28 is
$0.957(26)$.  A similar fit to the 1985 outburst of EXO 2030+375 gave 
$\gamma\simeq1.2$ (\cite{Reynolds96}).

 All three systems suggest a larger $\gamma$ than the naive prediction
of $\gamma =6/7$. In particular, the system with $\gamma>1$,
indicates an {\it increase} in $r_m$ with $\dot M$, if it is
proportional to the measured flux.  However, one must
remember that BATSE does not measure bolometric flux, but only the
20--50\,keV pulsed flux. In addition, the large changes in beaming
fraction implied by the changing pulse profiles need to be modeled and
accounted for. The observed flux may thus be related nonlinearly to
the mass accretion rate and thus contaminate the measurement of
$\gamma$. {\em EXOSAT} measured the 1--10 keV flux from EXO~2030+375;
since this should be a good tracer of the bolometric flux, only 
changing beaming could have affected their measured $\gamma$.

%\end{figure}

  The difficulties in determining the mass accretion rate from the
observed flux point out the necessity for testing the scaling $r_{\rm m}
\propto \dot M^{-2/7}$ in a way that does not depend on the uncertain
bolometric corrections.  An indirect test was possible during a giant
outburst in A~0535+26, when a simultaneous quasi-periodic oscillation
(QPO) was detected (\cite{Finger96c}).  The centroid frequency of the
QPO was strongly correlated with the observed spin-up torque and
luminosity, varying in the range $\nu_{\rm QPO}=$ 30--70
mHz. Interpreting the QPO frequency as the Kepler frequency at the
inner disk boundary (\cite{Alpar85,Lamb85}), 
we expect $\nu_{\rm K}=(GM/4\pi^2 r^3_{\rm
m})^{1/2} \propto \dot M^{3/7}$.  Consequently, with this
interpretation for the QPO one expects $\dot \nu = \dot
M\sqrt{GMr_{\rm m}} \propto \nu^2_{\rm QPO}$.  This predicted
relationship agrees with the observed trend in the data
(\cite{Finger96c}). An alternative interpretation of the QPO as the
beat frequency between the inner disk and the rotating magnetosphere,
$\nu_{\rm QPO} = \nu_{\rm K}-\nu$, gives an equally good fit.  This
correlation between the torque and QPO frequency is the strongest
evidence to date supporting the simple spin-up accretion torque model
described in \S 3.

To summarize, the observational evidence supporting the simplest
picture of accretion torques described in \S 3 is mixed. In all
cases of persistent disk-fed pulsars, the {\it magnitude} of the
accretion torque is consistent with the large lever-arm of an
extended magnetosphere. The observed correlation between torque and
flux, however, does not confirm the expected scaling $\dot \nu \propto
\dot M^{6/7}$. It is presently unclear if this disagreement can be due 
to bolometric and/or beaming corrections. On the other hand, if one
presumes that the observed QPO in the outburst of A~0535+26 scales
with the Keplerian frequency at the magnetosphere, then the data are
consistent with the expected magnetosphere relationship, $r_{\rm m} \propto
\dot M^{-2/7}$.  Further progress on these important issues requires
simultaneous torque, bolometric flux, and pulse profile
measurements. Since BATSE continuously monitors the torque, a
series of well-timed observations with a broad-band X-ray telescope
for many of these objects is needed.

%\begin{figure}
\centerline{
\psfig{figure=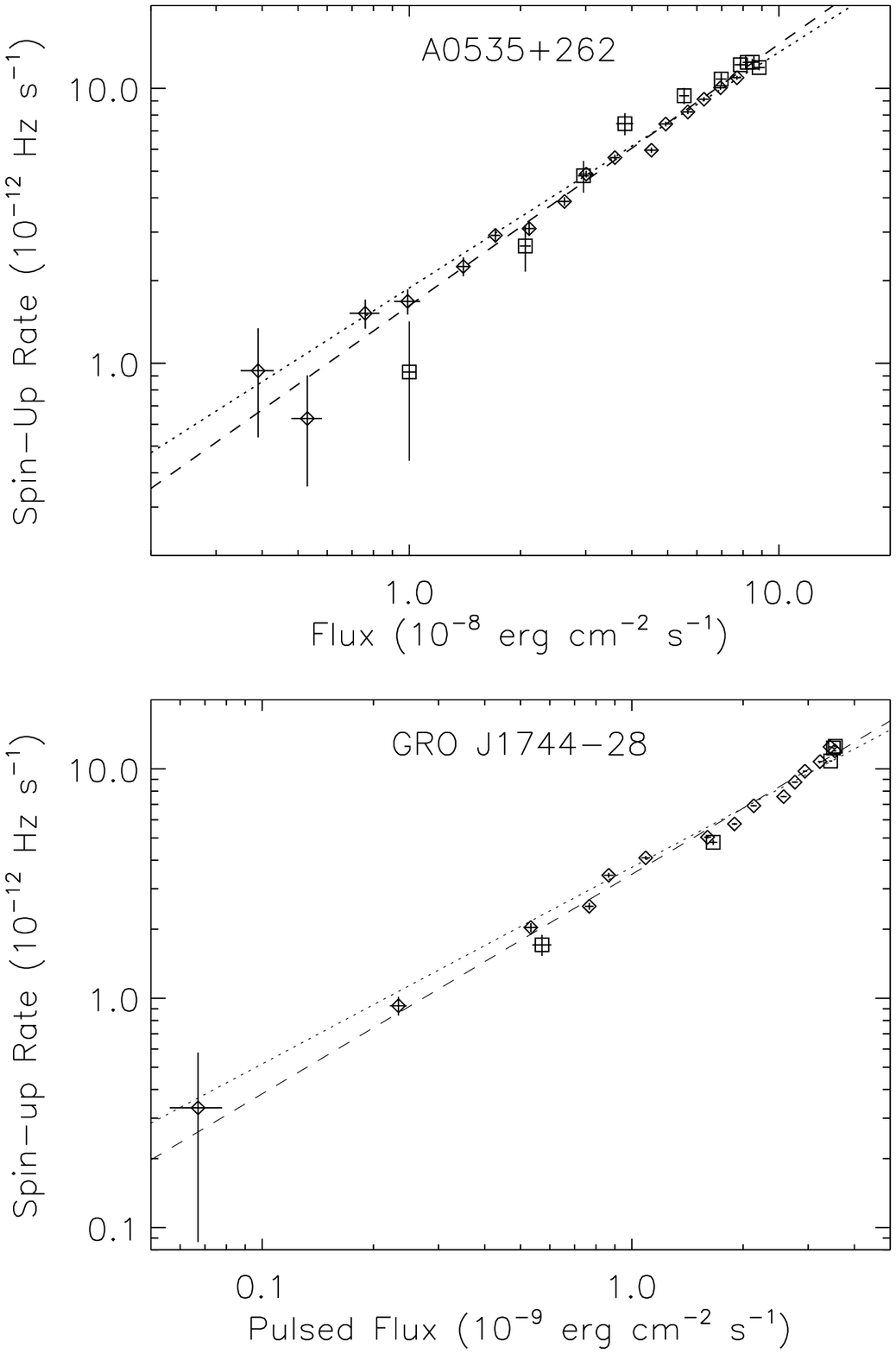,height=6in}
}
\figcaption{
Observed relationships between flux and pulsar spin-up rate $\dot \nu$.
The upper panel shows the spin-up rate of A0535+262 during the 1994 giant
outburst versus the measured total 
20-100 keV flux, determined from Earth occultation measurements. 
The bottem panel shows
the spin-up rate of GRO J1744-28 during the December 1995-March 1996
outburst versus the 20-50 keV R.M.S. pulsed flux. For both sources the
square symbols are from the outburst rise, and the diamond symbols are from
the outburst decline. The dotted curves are power-laws with the expected
index of 6/7, while the dashed curves are best fit powers-laws. The best fit
index for A0535+262 is $0.951(26)$ and that for GRO J1744-28 is
$0.957(26)$.
\label{fig:torque-lum}
}

\subsection{Accretion Torques in Transient and Wind-Fed Pulsars}

 Before BATSE, it was already known that the spin frequency in some
transients decreases between outbursts. We have unambiguously seen
spin-down between outbursts in A~0535+26 at a rate $\dot \nu=-2.2(6)
\times 10^{-13} \ {\rm Hz \ s^{-1}}$ (\cite{Finger94e}).
This may be due to the propeller effect (\cite{Illarionov75}), when
$\dot M$ becomes small enough so that the magnetospheric radius
exceeds the corotation radius.  Accretion is then centrifugally
inhibited and material may become attached to magnetic field lines and
flung away, removing angular momentum and causing the star to spin
down. Unfortunately, we can only make these measurements in those
binaries where the orbit is known, which among the Be transients are
still few.

  In 1984, after a decade of erratic spin behavior, the persistently
accreting pulsar GX~301--2 appeared to be spinning up steadily at
$\dot\nu\simeq 2\times 10^{-13}$\,Hz\,s$^{-1}$.  This trend was based
on only three measurements, however. Continuous BATSE observations found that
GX~301--2 was generally undergoing the stochastic torque fluctuations
expected from a wind accretor.  However, two dramatic episodes of spin
up of $\sim$20\,d duration at $\approx 5\times10^{-12}$\,Hz\,s$^{-1}$, 
comparable to the spin-up rates
in the disk-fed systems Cen X-3 and GX~1+4,
% OAO~1657--415 used to be in above list, but was removed since
% it is not unequivocally disk-fed.  DK
occurred accompanied by
enhanced luminosity (\cite{Koh97}). Moreover, these two spin up
events produced a net change in spin frequency consistent with the
long-term trend previously reported.  These observations strongly
suggest that GX~301--2 is primarily a wind fed pulsar and that the
secular trend is due to a few short but large spin-up episodes,
possibly caused by the creation of transient accretion disks. This
result blurs the common distinction between disk-fed and wind-fed
pulsars.

\subsection{Power Spectra of Torque Fluctuations}
 
\nocite{Lamb78}
\nocite{dekool93}

The power density spectrum of torque fluctuations in accreting pulsars can
potentially provide a probe of both the accretion flow and the internal
structure of neutron stars (\cite{Lamb78}). The most crucial requirement for
this type of study is a lengthy time baseline of precise timing observations,
with sampling over a wide range of timescales.  
Pre-BATSE estimates have been
made of the power spectrum of the spin frequency derivative, 
$\dot \nu$, of Vela~X-1
(\cite{Deeter89}), a wind  accretor, and of Her~X-1 (\cite{Deeter81}), a 
low-mass disk-fed system.
Both showed the power spectrum of $\dot \nu$,
$P_{\dot \nu}(f)$, to be flat,
indicating white torque noise with the neutron star responding as a solid body
in the range of accessable analysis frequencies, $f$. To avoid confusion
with the spin frequency, $\nu$, we use the term `analysis frequency' for 
the argument of the power  spectrum.  The power spectral
density $P_{\dot \nu}(f)$ gives the contribution to the variance of
$\dot\nu$ per unit analysis frequency  at analysis frequency $f$.
White torque noise is expected in wind-fed systems, where simulations show
that transient accretion disks with alternating rotational sense 
form and dissipate on $\sim$hours timescales (\cite{Taam88,Fryxell89}).

Studies of torque noise have also been made using time domain analyses, 
which are equivalent to
estimating simple power spectral models. A study was made by Baykal and
\"{O}gelman (1993\nocite{Baykal93}) applying a time domain model of the
frequency noise, a first order Markov process, to the published frequency
histories of a wide range of accretion-powered pulsars. This model has two
parameters, the noise strength and the correlation time. For a correlation time
of zero the model represents white noise in spin frequency  or equivalently
blue noise ($P_{\dot \nu}\propto~f^2$ ) in $\dot \nu$.
For an infinite correlation time the model represents a random walk in
spin-frequency, or equivalently white noise in $\dot \nu$.
For the systems Her~X-1, Cen~X-3, and~Vela X-1 they concluded that
the $\dot \nu$ noise was white. Applying this assumption to the
other sources, they found the noise strength correlated with source luminosity
and long-term spin-up rate. De Kool and Anzer (1993) studied how the size of
frequency changes in accretion pulsars depended on the time between
measurements. They concluded the frequency behavior of Vela~X-1 was consistent
with a random walk in spin frequency, or equivalently that the power spectrum
of frequency fluctuations was $P_\nu(f)\propto f^{-2}$.  They found
the frequency behavior Her~X-1, and Cen~X-3
consistent with random walks plus long-term linear trends.

For each of the 8 persistent sources monitored by 
BATSE we have estimated the power spectrum of the spin frequency
derivative by applying the Deeter polynomial estimator method 
(\cite{Deeter84}) to our frequency measurements. These power spectra
are shown in Figure~\ref{fig:powerspectra}. 
We plot $P_{\dot\nu}(f)$, 
the contribution per Hertz to the variance in $\dot\nu$ as a function
of analysis frequency, $f$.  $P_{\dot\nu}(f)$ is normalized such that
$\int_0^\infty P_{\dot\nu}(f)df = \left<(\dot\nu - \bar{\dot\nu})^2\right>$.
The power due to measurement noise has been subtracted
from the estimates and is shown independently by the square symbols.

The square root of the integrated power over a range in analysis 
frequency gives the root-mean-square (RMS) amplitude of variations
in $\dot \nu$ in that frequency range. This is shown in
Table \ref{tab:rmsfdot}, where the integration range 
$[f_1, f_2]$ in analysis
frequency is chosen as the range where
measurement errors do not dominate.  

Each estimate is made by
dividing the spin frequency measurements into intervals of duration
$T$ and fitting the frequencies with a quadratic polynomial in time.  
The square of the second order term is divided by the value it would have
for unit strength white noise in $\dot\nu$, defined as $P_{\dot\nu}(f)=1$.
The average over intervals is the power estimate.  The procedure is repeated
for different durations $T$ to obtain a power spectrum.  
This polynomial estimation techique is 
essentially equivalent to using a polynomial instead
of a sinusoid to estimate the power at each timescale $T$.
While correctly addressing
the difficulties cause by non-uniformly sampled data and red-noise power
spectral components, this technique produces power spectra of low resolution.

The frequency response of this estimate
of $P_{\dot\nu}(f)$ peaks near an analysis frequency $f \sim 1/T$. 
We plot $P_{\dot\nu}(f)$ 
at the logarithmic mean analysis frequency of the estimator response.

These quadratic estimators are by design independent of linear trends in
frequency. Chakrabarty et al. (1997) found a quadratic trend in the frequency
of 4U 1626-67 which was too large to be due to the measured torque noise. For
4U 1626-67 we have therefore instead used cubic estimators,  making the power
spectral estimates independent of  quadratic trends in the frequency.

For Vela X-1 and Her X-1 we find $P_{\dot\nu}$  consistent
with white torque noise in agreement with previous results
(\cite{Deeter89,Deeter81}).   The power spectra of 4U 1538--52 
(\cite{Rubin97})
and GX 301--2 are also consistent with white torque noise.  In contrast the
power spectra of Cen~X-3  (\cite{Finger94d}),  OAO~1657--415, and GX~1+4
(\cite{Chakrabarty97b}) show red torque noise with $P_{\dot\nu}$
varying approximately as $f^{-1}$. These red power
spectra imply long-term correlations in the torque,  which are evident
in the BATSE frequency histories. Due to the low noise level in 4U~1626--67 
only limited conclusions can be reached about the shape of its power spectrum
\cite{Chakrabarty97a}.  Because Her~X-1 is sampled only once per 35\,d cycle,
we can only measure $P_{\dot\nu}$ for $f\lesssim 2 \times 10^{-7}$\,Hz.
These power spectra have poor frequency resolution, and
unresolved narrow features may be present, affecting the continuum
shape.

The measured red torque noise in Cen X-3 contradicts the
conclusions based on time-domain analyses of published frequencies. In
retrospect it is clear that the model used by Baykal and \"{O}gelman (1993) 
cannot represent a red $\dot \nu$ spectrum, 
and therefore cannot discriminate
between white and red torque noise. The de Kool and Anzer (1993) result may
just be due to the poor sampling in the published frequency history.  Since
many of the power spectra we have measured have red torque noise, the meaning
of the noise strengths determined by Baykal and \"{O}gelman (1993), which
assumed a random walk in $\dot \nu$, is now unclear.  
The sampling in the
frequency histories of the pulsars they examined differs from source to source,
and hence $P_{\dot\nu}$ is being address in a different range of 
analysis frequencies for each source. The results for sources can thus only be
intercompared if the power spectra are all white. For red power spectra,
correlations between luminosity and sampling density could lead to correlations
between luminosity and estimated noise strength.

As a probe of the nature of the accretion flow, the low-resolution power
spectra presented here are a mixed success. The sources known to be wind
accretors (Vela~X-1, 4U~1538--52, GX~301--2) 
have power spectra consistent with 
white torque noise with strengths in the range 
$10^{-20}-10^{-18}$\,Hz$^2$\,s$^{-2}$\,Hz$^{-1}$. 
For disk-fed pulsars with low-mass companions (Her~X-1, 4U~1626--67) 
the power spectra are consistent with
white $\dot \nu$ noise with strengths in the range 
$10^{-21}-10^{-18}$\,Hz$^2$\,s$^{-2}$\,Hz$^{-1}$.  However, we cannot rule
out red noise in either system.
The low power in 4U~1626--67 precludes our determining the slope
of $P_{\dot\nu}$.
In the case of Her~X-1, the power spectrum does not span as large a range
in analysis frequency as in the other sources.
The one known disk-fed pulsar with a supergiant companion, Cen X-3,
has a red $\dot \nu$ power spectrum, reaching powers of
$10^{-16}-10^{-18}$\,Hz$^2$\,s$^{-2}$\,Hz$^{-1}$ at low frequencies.
For GX 1+4 and OAO 1657-415 there is no evidence independent of their 
frequency histories that reveal the presence or absence of accretion disks.

\begin{figure}
\centerline{
\psfig{file=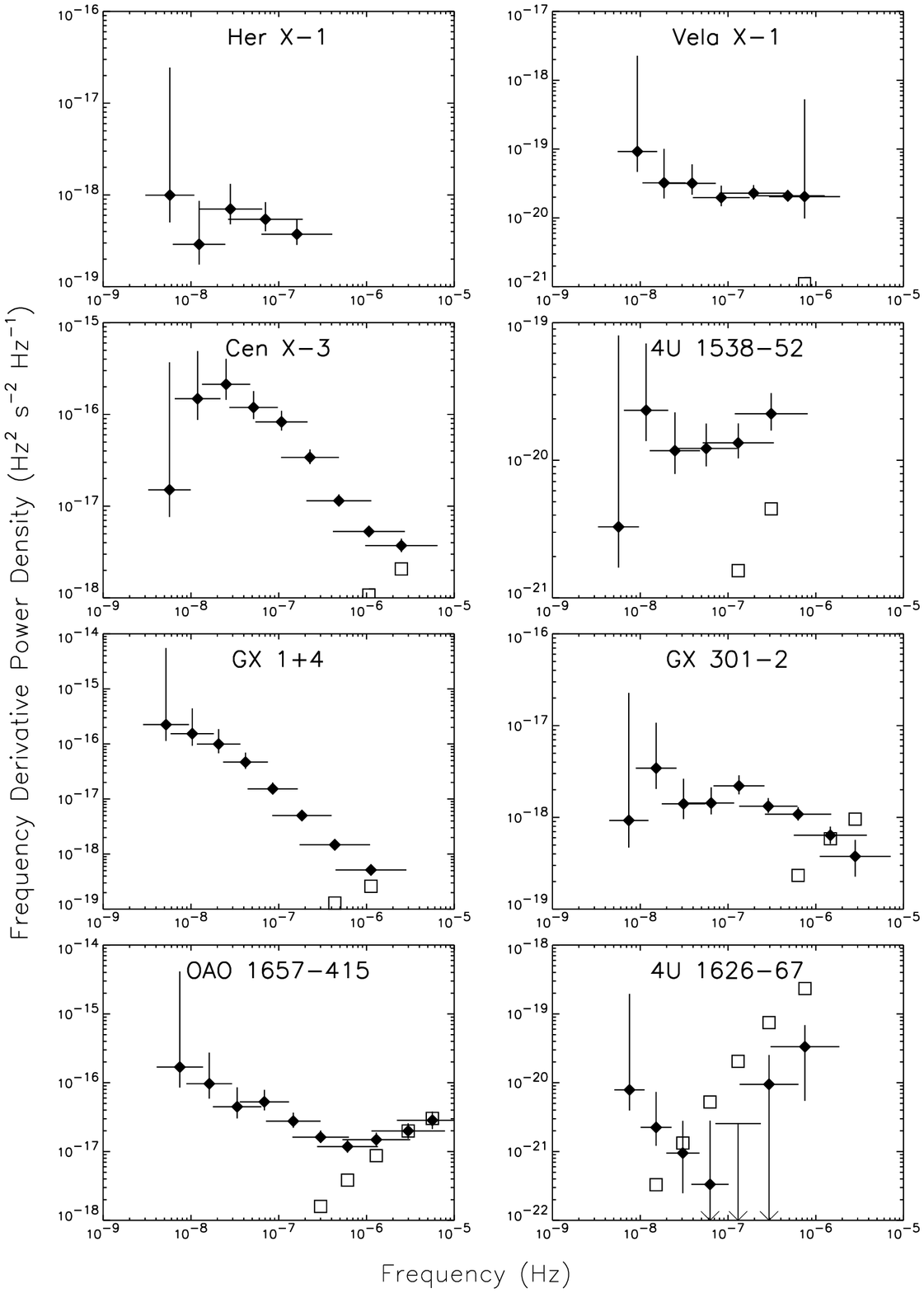,height=6.2in}
}
\figcaption{Power Density Spectra of torque fluctuations in persistent
pulsars, computed using methods described in Deeter \etal (1987).  The errors
bars on power indicate the 68\% confidence region. The error bars on analysis
frequency give the RMS log frequency width of the estimator response. Open
squares indicating power from apparent torque variations introduced by
counting noise, which have been subtracted from all measured
values.  The integrated RMS torque variation for these measurements
are provided in Table \ref{tab:rmsfdot}.
\label{fig:powerspectra}
}
\end{figure}

\setcounter{table}{4}
\begin{deluxetable}{lccc}
\tablecaption{RMS Spin Derivatives for Persistently Accreting Neutron Stars
\label{tab:rmsfdot}
}

\tablehead{\colhead{Source} &
	   \colhead{$f_1^a$} &
	   \colhead{$f_2^b$} &
	   \colhead{$\dot \nu_{\rm rms}^c$}
%	   \colhead{$\dot \nu_{\rm char}^{d}$} &
%	   \colhead{ {$\dot \nu_{\rm rms} / \dot \nu_{\rm char}$}}
}

\startdata
Cen X-3      &  3.4  &  60.0  & 65.7(35)\nl%  & 50  & 1.3 \nl
OAO 1657--415 &  12.0 &  19.0  & 55.8(36)\nl%  & 50  & 1.1 \nl
GX 1+4       & 7.3   & 20.0  & 26.2(31)\nl%  & 60  & 0.4  \nl
GX 301--2     & 5.0   & 21.0   & 13.9(8) \nl
%				 & 1(30)$^e$     & 14(0.5)$^e$\nl
Her X--1      & 3.4   & 3.8  & 3.7(9)\nl%   & 2  & 1.8   \nl
Vela X--1     & 6.9   & 6.6  & 1.2(3)\nl%   & 0.5 & 2.4   \nl
4U 1538-52   & 4.0   & 6.0  &  1.0(1)\nl%   & 0.5 & 2.0   \nl
4U 1626-67   & 11.0  & 0.38 &  0.06(3)\nl% & 8 & 0.008   \nl
\enddata

\tablenotetext{a}{Lower analysis frequency for calculation of
$\dot \nu_{\rm rms}$, in units of $10^{-9}$ Hz}

\tablenotetext{b}{Upper analysis frequency for calculation of
$\dot \nu_{\rm rms}$, in units of $10^{-7}$ Hz}

\tablenotetext{c}{RMS frequency derivative in units of $10^{-13}$ Hz~s$^{-1}$,
obtained by integrating the power spectral density, $P_{\dot\nu}(f)$,
from $f_1$ to $f_2$.}

\end{deluxetable}

\clearpage

\nocite{Deeter87}
%\bibliography{ref}

\nocite{Deeter87}

%%%%%%%%%%%%%%%%%%%%%%%%%%%%%%%%%%%%%%%%%%%%%%%%%%%%%%%%%%%%%%%%%%%%%%%%%%%%%%%

%\input trans.tex

\subsection{Transients Outbursts in Be Systems}
\label{sect:trans}

  More than 50 outbursts from 12 transient pulsars were detected with 
BATSE in the first five years of observations, the times of which 
are shown in Figure \ref{fig:outburst_times}. 
One of these transient sources, GRO~J1744--28, 
has a low mass companion. Seven of the remaining pulsars have known Be
star companions. No optical counterparts have been identified for four
of the other sources; however their temporal behavior suggests that
the companions are Be stars. Accreting neutron stars in Be systems typically
have long periods and eccentric orbits. The source of accreting material
is the slow, dense, stellar wind which is thought to be confined to 
the equatorial plane of the rapidly rotating Be star. 
Evidence for this equatorial disk in Be stars comes from observations 
of hydrogen and helium emission lines, and an IR continuum excess. 
For a review of these Be/X-ray binary systems, see van den Heuvel \&
Rappaport (1987) or
Apparao (1994). 
\nocite{Apparao94,vdHeuvel87}

A striking feature of the long-term light-curves of these pulsars is
the frequent occurrence of a series of outbursts. An example is the
long series of outbursts from EXO~2030+375 shown in Figure
\ref{exo2030freqflux}, each outburst beginning soon after periastron
(shown by dotted lines). Other examples are A~0535+262 (shown in
Figure \ref{a0535freqflux}), GRO~J2058+42 (Figure
\ref{J2058freqflux}), 2S~1417--624 (Figure \ref{2s1417freqflux}),
and the series of outbursts of GS~0834--430 (Figure
\ref{gs0834freqflux}) which begin with orbital spacing, but don't 
end with it. Another feature of the lightcurves are the occasional
``giant'' outbursts. An example is the giant outburst of A~0535+262
(Figure \ref{a0535freqflux}) which occurred in 1994 February/March.
Other examples are the first outburst observed from 2S~1417--624
(Figure \ref{2s1417freqflux}) and the first outburst observed from
GRO~J2058+42. As well as being bright, these giant outbursts have high
spin-up rates, longer durations, and while often beginning at the same
orbital phase as the smaller outburst, tend to peak at a later phase.

\nocite{Stella86}
\nocite{Motch91}

These two types of outbursts have been noticed previously. Stella, White
\& Rosner (1986) contrasted the 1973 outburst of V~0332+53 (which lasted
over three binary orbits) with a series of three smaller outbursts
detected in 1983--1984. They defined two classes of outburst activity:
class I was periodically occurring outbursts associated with periastron
passage; and class II was irregular transient activity, with higher
luminosity and outbursts peaking at arbitrary orbital phases.  Motch
et al. (1991) classified outbursts of A~0535+262 as giant,
normal, or missing (i.e. no detection at the expected X-ray
maxima). The more luminous giant outbursts peak at a phase delayed
relative to the mean normal outburst X-ray maximum by up to 0.3
orbital cycles, and were associated with large pulse period changes.
Prior to BATSE no association had been observed between giant and normal
outbursts.
BATSE has found that many of the giant outbursts are in the middle of,
or followed by, a series of normal outbursts.
A sequence of normal outbursts from 4U~0115+634 has now been seen by both
BATSE and {\it RXTE}. Prior to these observations only isolated giant outbursts
had been seen.

Several authors have suggested that transient accretion disks are
formed during the giant or class II outbursts
(\cite{Kriss83,Stella86,Motch91}). This helps explain the large and
steady spin-up rates seen during the giant outbursts, which are
difficult to explain with direct wind accretion. BATSE has observed
peak spin-up rates of $4.3\times10^{-11}$\,Hz\,s$^{-1}$ (2S~1417--624),
$3.8\times10^{-11}$\,Hz\,s$^{-1}$ (GRO~J1750--27),
$1.2\times10^{-11}$\,Hz\,s$^{-1}$ (A~0535+26), and
$8\times10^{-12}$\,Hz\,s$^{-1}$ (4U~0115+634). The discovery outburst
of EXO~2030+375 found it spinning up at a rate of
$2.2\times10^{-11}$\,Hz\,s$^{-1}$ (\cite{Reynolds96}). Additional
evidence for disk accretion occurring during giant outbursts is
provided by the BATSE observations of beat or Keplerian frequency QPO
during a giant outburst of A~0535+262 (\cite{Finger96c}).
Optical observations have so far been unable to provide evidence of
accretion disks during the giant outbursts.

  Given that an accretion disk seems to be present, it is natural to
ask about its fate. Is it completely consumed at the end of a giant
outburst, or is some portion of it left? Accretion may be
centrifugally inhibited at the end of the outburst, when the
magnetosphere lies outside of the corotation radius. It is unclear how
efficient this mechanism is when the magnetosphere is still close to
the corotation radius (\cite{Spruit93}). If the ejected material does
not acquire escape velocity it might not leave the system, but may continue
to circulate around the neutron star.

If a disk can be sustained between giant outbursts, then it is
plausible that one is present during normal outbursts.  In this case
the repeating normal outbursts might be explained by the large tidal
torques experienced by the disk during periastron passage. The angular
momentum of material flowing into a disk must eventually be removed by
tidal torques from the companion, and these torques increase rapidly
with decreasing pulsar-companion separation (\cite{Papaloizou77}).
The enhanced torque in the outer disk shrinks the disk and increases
the mass accretion rate there.  This results in a wave of new material
that will reach the inner disk on a global viscous timescale
($\sim$weeks for typical binary parameters).  This could explain the
series of normal outburst that were observed following the giant
outburst in 2S~1417--624 (all of which peaked near apastron) as well as
the sequence of outbursts in GRO~J2058+42. It may also explain series
of normal outbursts that are not preceded by a giant outburst, such as
those seen from EXO~2030+375.

What causes the giant outbursts? Possibilities that have been
investigated are episodes of enhanced Be disk density, or reduced Be
disk expansion velocity. However, these should result in consistent
correlations between optical and hard X-ray activity, which is
typically not seen. For example, {\em UBVRIJHK} band photometry of the
HD2457700/A~0535+262 system over the past decade (Clark et al., in
preparation) reveals no correlation between the photometric
lightcurves and hard X-ray outbursts. 
Recently it has been proposed that the thermal disk instablity thought 
to cause dwarf nova outbursts also is at work in soft x-ray transients 
(van Paradijs 1996; King, Kolb \& Burderi 1996). This instablity should 
also affect accretion disks around Be/X-ray pulsars, and could be the 
cause of the giant outbursts.  For an accretion disk
to be vulnerable to this instability, the outer portion of the disk
must be below the hydrogen ionization temperature ($T_{\rm H}\approx
6500$ K) while the disk accumulates. For A~0535+262 we find that for a
steady accretion rate of $3\times 10^{-10} \ M_\odot \ {\rm yr^{-1}}$,
corresponding to the average luminosity during the 600 day interval
during which outbursts were observed by BATSE, the portion of the
(X-ray heated) disk beyond $10^{11}$ cm would still be neutral. The
disk would extend to approximate 90\% of the Roche lobe at periastron,
or $2 \times 10^{12}$ cm, and would therefore be subject to this
instability once a critical amount of matter has accumulated.

\nocite{Smak83,vanParadijs96,King96}

%%%%%%%%%%%%%%%%%%%%%%%%%%%%%%%%%%%%%%%%%%%%%%%%%%%%%%%%%%%%%%%%%%%%%%%%%%%%%%%

%\input pop.tex

\subsection{The Population of Be Transients}
\label{sect:pop}

The Galactic population of Be--transients has been estimated
before by Rappaport \& van~den~Heuvel (1982) and
Meurs \& van~den~Heuvel (1989), who both arrived at a number
of several thousand.  However, the sparse and non--uniform coverage
of pre--BATSE intruments resulted in several non--quantifiable 
ambiguities in their analysis.
The continuous, uniform and all--sky coverage provide by BATSE
alleviated some of these problems and enables us to 
check previous estimates of the Galactic Be--transient population.

BATSE has detected 11 transients with high mass companions (mostly Be
stars) between 1991 April and 1997 January (MJD 48370--50464). 
This is a complete sample at 20--50 keV of transient sources with
pulsed fluxes in excess of
$F_{\rm min}\approx 2\times 10^{-10}$\,erg\,cm$^{-2}$\,s$^{-1}$.  
They have a mean Galactic latitude of $1.3^\circ$,
a mean absolute Galactic longitude, $|\ell|$, of $81.3^\circ$, and
are concentrated at galactic longitudes $ 60 \lesssim |\ell| \lesssim 90$.
This may be due to clustering in nearby spiral arms.  Of these
11, 7 have exhibited giant outbursts as described in the previous
section and their galactic locations are plotted
in Figure~\ref{fig:transients}.  Of these 7, giant outbursts 
in A~0355+26, 2S~1417--624, 4U~0115+63,
4U~1145--619 and GRO~J2058+42 were identified by their high pulsed flux
and spin-up rate relative to other outbursts from these sources.
The single outburst from GRO~J1750--27 was identified as a giant from its 
peak spin up rate ($\dot \nu \sim 4\times10^{-11}$\,Hz\,s$^{-1}$), and
the single outburst from A~1118--616 from
its duration ($\sim 50$\,d) and its large $\dot \nu
\sim 2\times10^{-12}$\,Hz\,s$^{-1}$ compared
with the largest expected orbital signature (the orbit is unknown), although
comparable rates are seen in normal outbursts of 2S~1417--624 and 
GS~0834--430.
Using the distances inferred from the optical counterparts
where they are available (see Nagase 1989), we find that the peak
20--50 keV pulsed luminosities of these outbursts are in the range of
$(3-10)\times 10^{36} \ {\rm erg \ s^{-1}}$. This implies that we can
detect giant outbursts at distances of at least 11.5 kpc, roughly
consistent with the giant outburst detection of GRO J1750--27, near the
Galactic center.

%\begin{figure}
\centerline{
\psfig{file=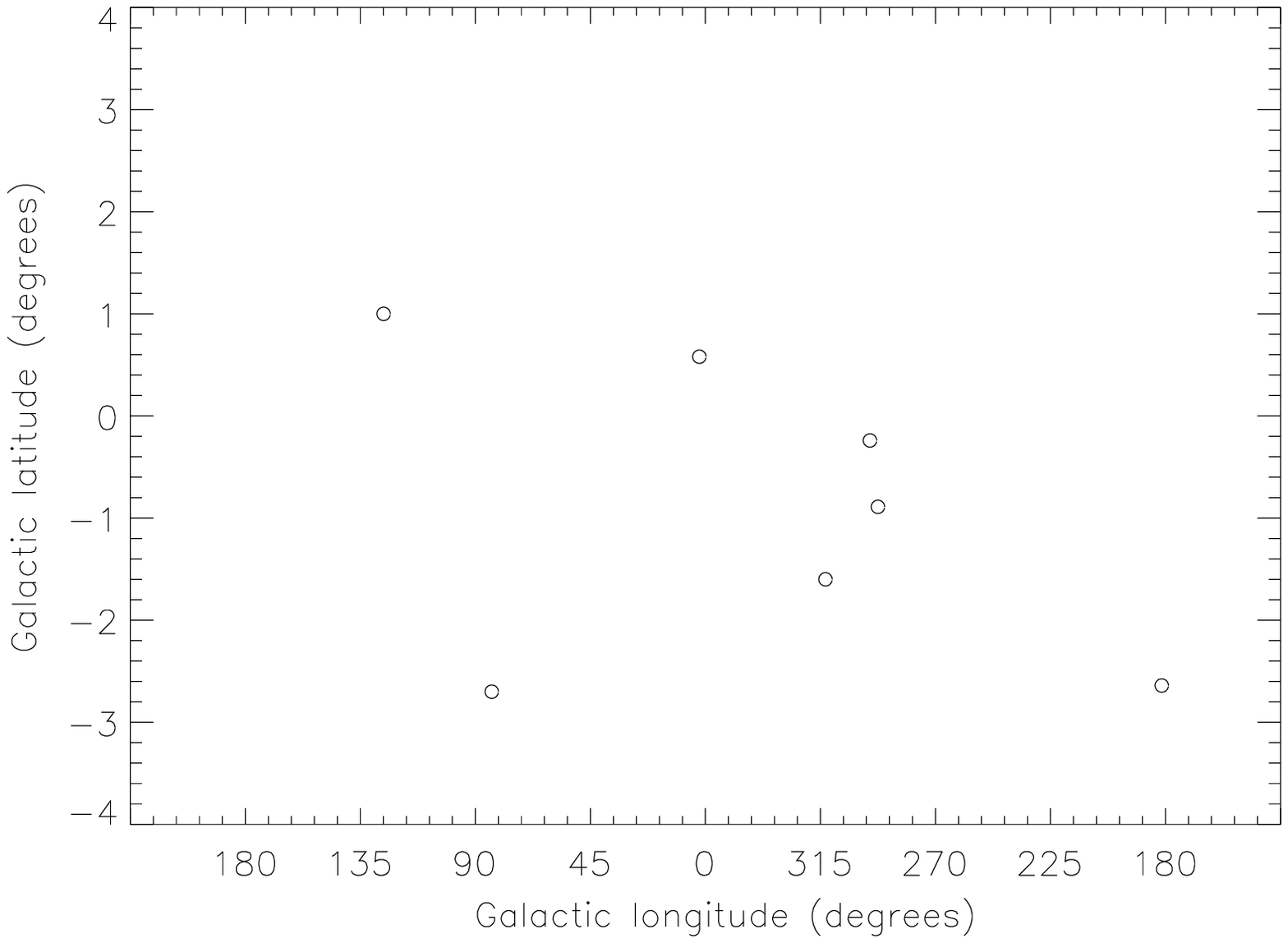}
}
\figcaption{Galactic location of BATSE-detected giant
outbursts from Be transients. The circles denote the location within
the galaxy of those Be transients which were detected by BATSE
during giant outbursts.\label{fig:transients}}
%\end{figure}

  Although these outbursts may not be standard candles, it is 
interesting to ask what sampling distance one would infer from their
distribution in $\ell$ and $b$, shown in Figure~\ref{fig:cumdistri}
if they are standard candles.  Given
the limited data set, we take a simple model for the Galactic
distribution. We assume that the Be transients are distributed as
$\exp(-|z|/z_0)$ in the direction perpendicular to the plane and, like
the matter in the Galaxy, fall off radially $\exp(-r/r_0)$ away from
the Galactic center, where $r_0=3.5$ kpc (\cite{Pence78}). 

For a Galactic
center distance of 8.5~kpc, we find that the acceptable fits 
($\gsim$90\% confidence) to the
cumulative $b$ distribution require a sampling distance in the range 
of $(35-50)z_0$ (see Figure \ref{fig:cumdistri}).  If the Be binaries
have the scale height $z_0=100$~pc of massive stars ({\cite{Miller79}),
then the sampling distance inferred from the latitude
distribution is 3--5 kpc. This is consistent with the observed excess
of objects in the direction of the Galactic center versus the
anti-center, as the sampling distance is of order the
exponential scale length in the disk population, $r_0$. 
However, kick velocities of $v
\sim 450 \pm 90$ km~s$^{-1}$ are typically imparted to neutron stars
during the supernova (\cite{Lyne94}), potentially increasing the scale
height of those which remain in binaries up to $\approx$ 140~pc
(\cite{Brandt95}). This helps to make the sampling distance more
consistent with our first estimate, but still a bit short. The
resolution of this discrepancy may be that the giant
outbursts are not standard candles.
We note also that our assumption of $\pm b$ symmetry 
is not strictly correct because the sun
is known to lie a vertical distance of ${\zsun} \approx$ 15~pc above 
the Galactic plane (\cite{Cohen95}).
Nevertheless, since ${\zsun}$ is small compared to both the sampling
distance and the expected scale height of Be transients, including
the effects of ${\zsun}$ will negatively shift the peak 
of the latitude distribution to $|b| \lesssim 0.2 \arcdeg$, 
which does not affect
our results, especially in light of other larger uncertainties.

%\begin{figure}
\centerline{
\psfig{file=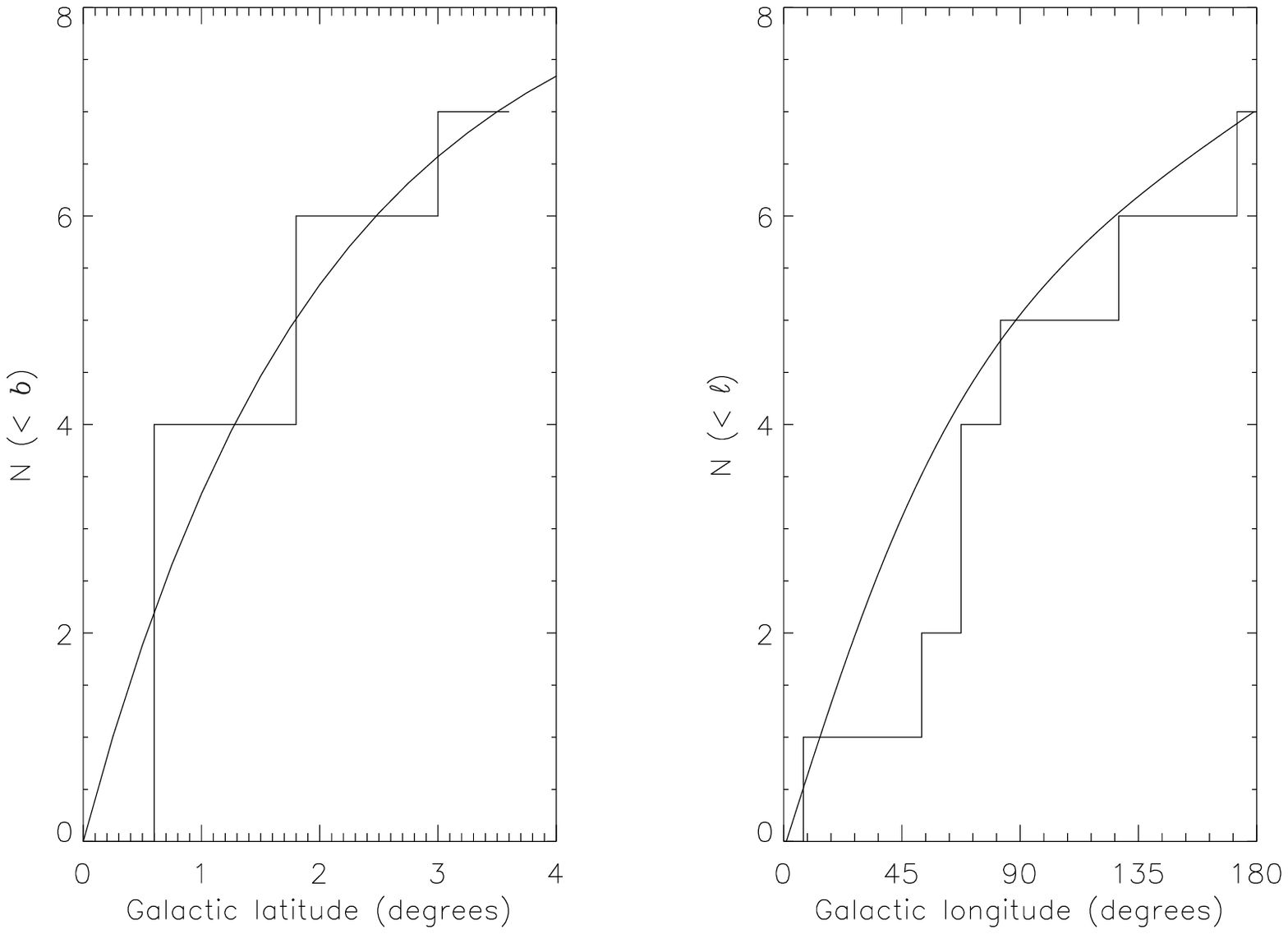,height=4in}
}
\figcaption{Cumulative $b$ distribution (left panel)
and cumulative $\ell$ distribution (right panel) of 
the 7 Be-transient systems detected by BATSE during giant outbursts:
A~0535+262, A~1118-61, GRO~J1750-27, GRO~J2058+42, 2S~1417-624, 4U~0115+634
and 4U~1145-619.
The histogram and solid line respectively represent the actual data
and the model prediction. Since our modeling has $\pm \ell$ and $\pm b$
symmetry, we use the absolute values of $\ell$ and $b$ in this
analysis.  A sampling distance of $40 z_0$, with $z_0 = 100~pc$ 
gives the best fit to the $b$ distribution and yields 
an acceptable Kolmogorov-Smirnov
statistic.  The relatively poor correspondence between the $\ell$
distribution and the model is due to the clustering of transients at
$60^{\circ} \lesssim \ell \lesssim 90^{\circ}$.
\label{fig:cumdistri}}
%\end{figure}

BATSE detected 8 giant outbursts from 7 Be--transients in 4 years.
The repetition by one of these sources enables
us to estimate the recurrence time scale of giant outbursts.  
Define $N_{\rm loc}$ to be the total number of transients which 
exist within a distance $R_{\rm loc}$ of Earth, and assume
that all the 7 systems from which BATSE had detected giant outbursts
are within distance $R_{\rm loc}$ of Earth.
By modelling the frequency of giant outbursts as a Poisson process,
we inferred that the most probable recurrence time scale
of giant outbursts in each transient $\approx$ 12.5~years, which 
imply that $N_{\rm loc} \approx 25$.
This allows us to compute the
proportionality constant in the density distribution of Be--transients
which, when integrated over the Galactic disk, yields an estimate of 
the total number of Be transients in the Galaxy, $N_{\rm tot}$.
As shown in Figure \ref{fig:cumdistri}, $R_{\rm loc} = 40z_0$ provides
the best fit to the observed cumulative $b$ distribution.
For $z_0=100$~pc, $R_{\rm m}=4$~kpc, and we obtain
$N_{\rm tot} \approx 1300$
while for $z_0=200$~pc, $R_{\rm m}=8$~kpc, we obtain
$ N_{\rm tot} \approx 250 $.
If BATSE could indeed sample out to $R_{\rm loc} = 11.5$~kpc, the 
estimated distance to GRO~J1750--27, $ N_{\rm tot} \approx 130 $ for
for $z_0=100$~pc.
Since our estimate of the recurrence time scale hinges upon a single
transient which exhibited more than one outburst, these estimates should
be considered crude.  However, they are consistent with estimates
from evolutionary models of the total number of Be/neutron-star
binaries, $\sim 10^4$, most of which are quiescent 
(\cite{Meurs89}).

%%%%%%%%%%%%%%%%%%%%%%%%%%%%%%%%%%%%%%%%%%%%%%%%%%%%%%%%%%%%%%%%%%%%%%%%%%%%%%%

%\input be_orbits.tex

\subsection{Be/X-ray Pulsar Orbits}
\label{sect:be_orbits}

BATSE has more then doubled the number of 
orbits that have been determined for Be/X-ray binaries, increasing the number
from 4 to 8.  See table~\taborbits.
We include in this classification GRO J1750--27, 2S~1553-54 and GS 0834--430, 
which we suspect
have Be star companions. With this size sample we can begin to make
comparisons between the observed distribution of orbital elements and our
expectations. 

A main sequence B star has a mass in the range of 4--16 $M_\odot$. The
measured X-ray mass functions $f_{\rm x}(M)$ for the Be/X-ray pulsars
should be consistent with masses in this range. If we assume a common
neutron-star mass of $1.4M_\odot$, we can use the mass function,
$f_{\rm x}(M)$, to determine lower limits to the masses of the
companions.  If we further assume that the systems we see have
randomly distributed orientations relative to our line of sight, we
can use the distribution of mass functions to
determine the distribution of companion
masses.  In Figure~\ref{fig:P_f_M} we
compare the cumulative mass-function distribution, $N[<f_{\rm x}(M)]$
with the distribution we would get assuming a constant companion mass,
$M_{\rm c}$, a constant neutron star mass, $M_{\rm x}=1.4M_\odot$,
and random orientation.  With the observed distribution is plotted the
theoretical distributions for 4$M_\odot$, 6$M_\odot$ and 10$M_\odot$
companions. None of these curves can be said to fit the data, however
masses in the 6--12 $M_\odot$ range are clearly called for. To explore
the width of the companion mass distribution a maximum likelihood fit
was made to the observed mass functions, using a companion mass
distribution that was uniform in log between two limiting masses. The
best model had masses in the range of 6.7--13.1 $M_\odot$, however due
to the limited statistics the distribution width was poorly
constrained, with the 50\%-confidence region containing lower mass
limits from 2--8 $M_\odot$ and upper mass limits from 12.2--21
$M_\odot$.

Be/X-ray binaries are thought to be the result from the evolution of a
binary system of two B stars (van den Heuvel \& Rappaport 1987). The
progenitor of the neutron star is initially more massive. First it
transfers mass to its companion by Roche-lobe overflow due to
hydrogen-shell burning, resulting in a helium star orbiting a rapidly
rotating Be star. Then the helium star transfers mass due to
helium-shell burning. Finally the helium star undergoes a supernova
explosion.  The velocity kick and the mass loss experienced in the
supernova explosion can result in a wide eccentric orbit (or a
disrupted system). Since the orbits are wide, orbital changes are slow
compared to the evolutionary timescale of the Be star. The observed
Be/X-ray system orbits are therefore fossils of supernovae in Be/helium
binaries.
\nocite{vandenHeuvel87}

%\begin{figure}
%\vspace*{4in}
\centerline{
\psfig{file=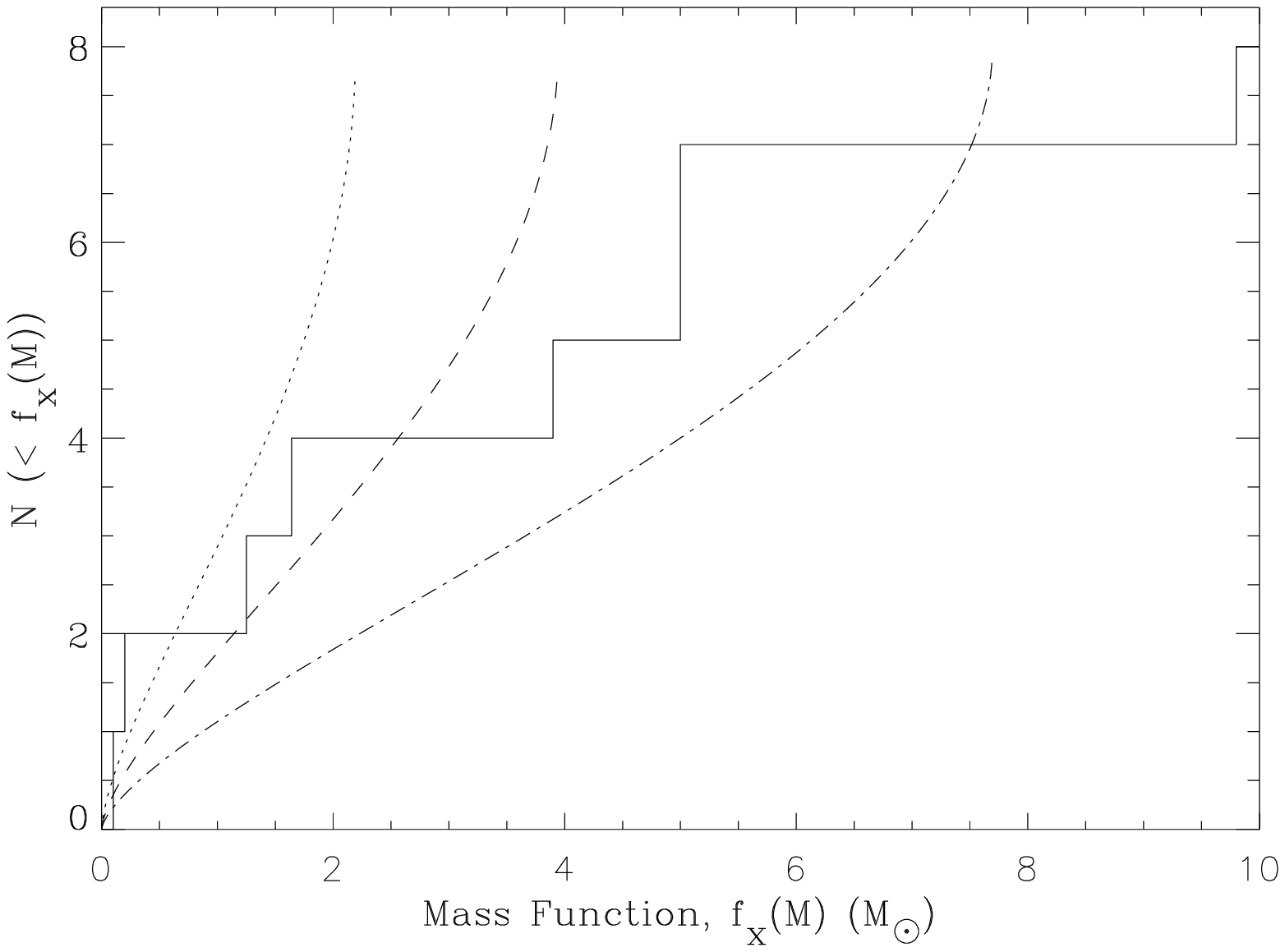,height=4in}
}
\figcaption{Cumulative distribution of the mass functions of the Be transient
pulsars. The measurements are given by the stair-step curve. The remaining
curves give the expected distribution for random orientation and a constant
companion mass of $4M_\odot$ (dotted), $6M_\odot$ (dashed), and $10M_\odot$
(dot-dashed).
\label{fig:P_f_M}}
%\end{figure}

The widest resulting orbits (and therefore those of longest period)
should be the most eccentric, and we therefore expect eccentricity to
be correlated with orbital period. In Figure \ref{fig:Be_e_porb} we
plot the orbital periods and eccentricities of the Be systems for
which these have been determined. Only a weak correlation is
present. The possible range of the orbital period $P_{\rm init}$ of
the pre-supernova system, assuming a initial circular orbit and an
asymmetric supernova explosion, is
$P_{\rm orb} (1-e)^{3/2} \beta^{1/2} < P_{\rm init} < P_{\rm orb}
(1+e)^{3/2} \beta^{1/2}$ where $\beta$ is the ratio of the current
system mass to the pre-supernova system mass. This is shown in figure
\ref{fig:Be_e_porb} for each source, assuming $\beta = 0.9$. The
initial period distribution could have been much narrower than the 
distribution of $P_{\rm orb}$, but must
still have significant width. This may explain the weakness of the
correlation.

%\begin{figure}
\centerline{
\psfig{file=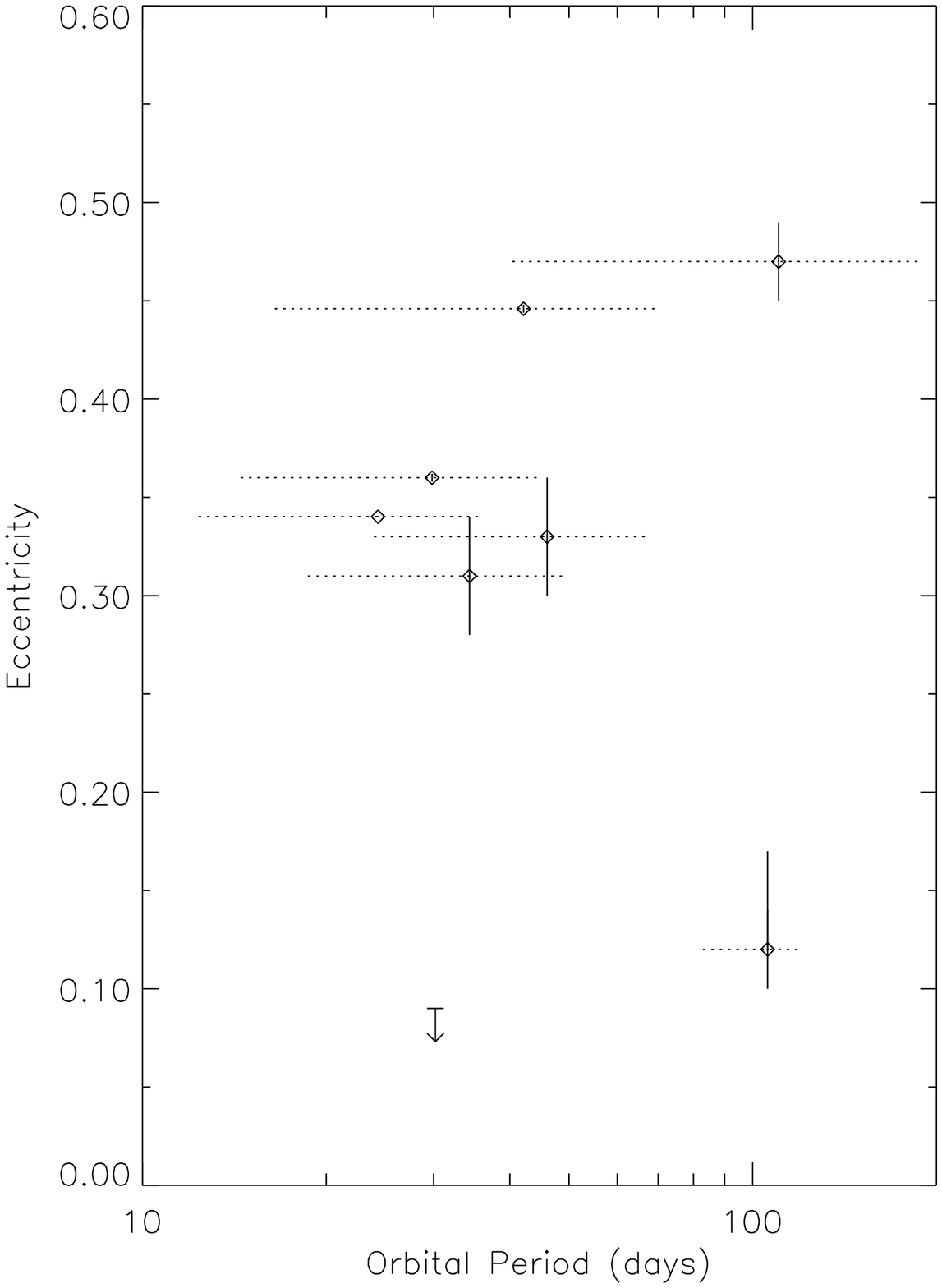,height=4in}
}
\figcaption{The eccentricities of Be/X-ray pulsars systems plotted against
the orbital period. The dotted lines give the possible range of the 
orbital period for each system prior to the supernova that formed the
neutron star, assuming an initially 
circular orbit and an asymmetric supernova explosion.
\label{fig:Be_e_porb} }
%\end{figure}

The kick velocities are likely to be larger than the orbital
velocities in the pre-supernova systems ($\approx$100 km
s$^{-1}$). Lyne and Lorimer (1994) found from a study of radio pulsar
proper motions a mean kick velocity of $450~\pm~{\rm 90 km}~{\rm
s}^{-1}$. The majority of systems are therefore disrupted. Kalogera
(1996) gives analytic expressions for the distribution of orbital
parameters of the undisrupted systems, assuming an initially circular
orbit and a Gaussian kick velocity distribution. The form of the
eccentricity distribution in the limit of large kick velocity relative
to orbital velocity is found to be independent of all other
parameters. In Figure~\ref{fig:Be_e_dist} the cumulative distribution
of the observed eccentricities is compared to the predicted cumulative
distribution, which has been normalized to a mean of eight sources
with $e<0.5$.

The observations and predicted distribution agree reasonably for
$e<0.5$, but poorly above.  No sources with $e > 0.5$ are observed, but
30 are expected. We think it unlikely that this is due to an error in our
assumptions about the kick distribution. More likely this is evidence for
strong selection effects against high eccentricities. High-eccentricity
orbits will typically undergo outbursts only near periastron, and may
only be observed in isolated outbursts, making orbit determination difficult
or impossible.  If this is the explanation, then nearly all of the Be/X-ray
pulsar for which no orbit has been determined (20 sources) must have 
high eccentricity ($e > 0.5$).  It is also intriguing to note that 
the eccentricities are high 
($e=0.8698$ in PSR B1259-63 Johnston et al. 1992  and 
$e=0.8080$ in PSR J0045-7319 Kaspi et al. 1994) in the two
known Be/radio pulsar binaries.

%\begin{figure}
\centerline{
\psfig{file=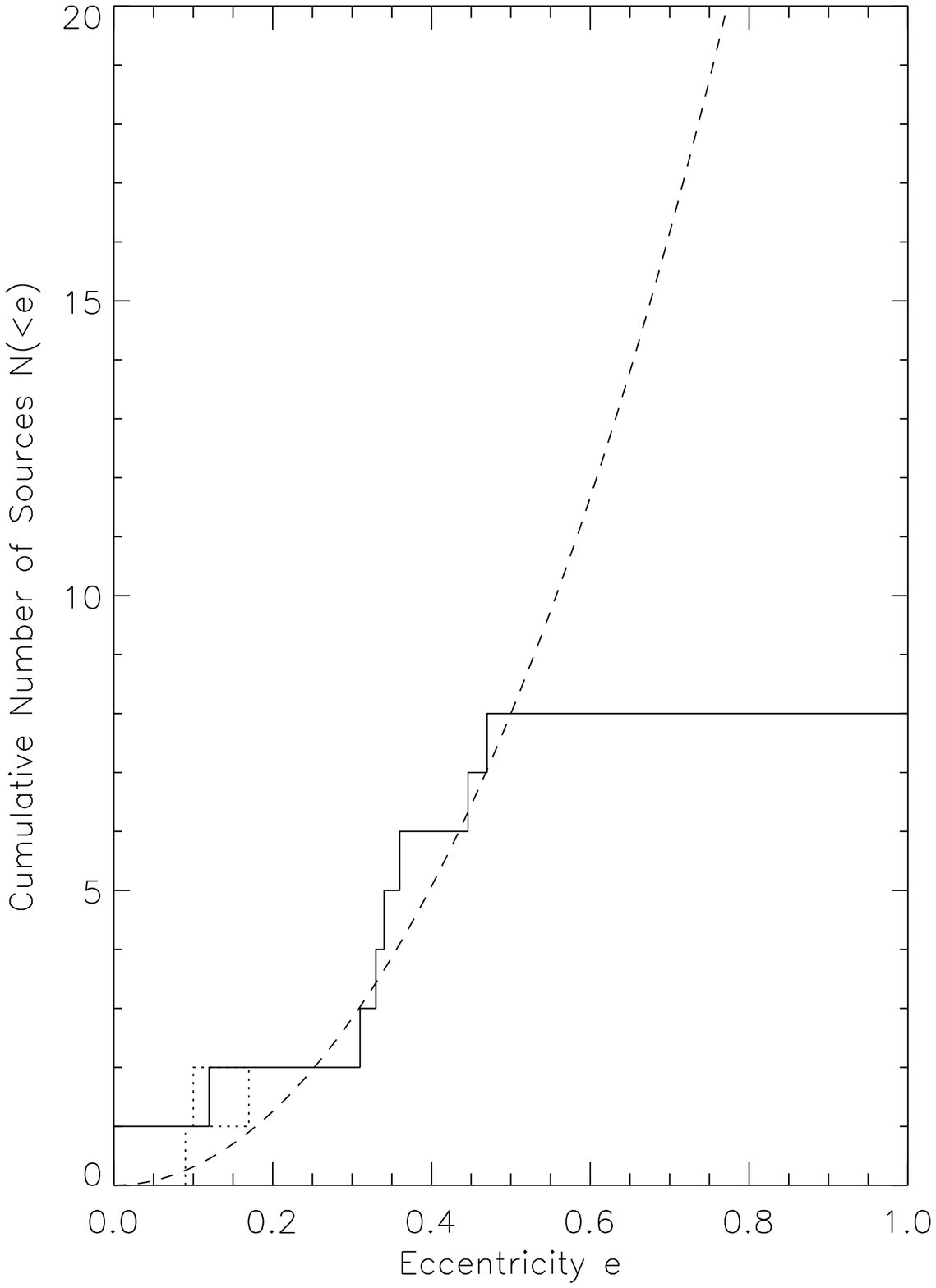,height=4in}
}
\figcaption{The cumulative distribution of Be/X-ray pulsar binary
eccentricities. The observed distribution is given by the solid line. The
predicted distribution, normalized for a total of eight sources with $e<0.5$,
is given by the dashed line (see text). The dotted lines given the eccentricity
range allowed by the upper-limit for the eccentricity of 2S 1553-54 and the
errors on the eccentricity of GS 0834-430.
\label{fig:Be_e_dist}
}
%\end{figure}
\nocite{Lyne94}
\nocite{Kalogera96}
\nocite{Johnston92}
\nocite{Kaspi94}

%%%%%%%%%%%%%%%%%%%%%%%%%%%%%%%%%%%%%%%%%%%%%%%%%%%%%%%%%%%%%%%%%%%%%%%%%%%%%%%

%\input summary.tex

\section{SUMMARY}

We have presented five years of continuous
pulse timing and flux observations of
accreting binary pulsars with the BATSE instrument on CGRO.
This is the most detailed and complete history of spin frequency behavior and 
outburst activity in accreting pulsars to date, and presents a qualitatively
different picture of accreting binary pulsars
than understood from the sparse histories previously available.
Frequencies and fluxes presented in this paper,
along with daily folded pulsed profiles, are being made available through
the {\it Compton Observatory} Science Support Center
(http://cossc.gsfc.nasa.gov/cossc/COSSC\_HOME.html).

The standard picture of accreting pulsars was developed over twenty years
ago, and has been largely accepted and applied to other 
systems containing accreting magnetic stars.
The BATSE data allow us to
test these theories critically, and in many cases the
observed behavior is unexpected and difficult to explain.
The accretion torque behavior seen in persistent disk-fed systems was
particularly surprising. The slow, long-term spin-up trend in 
Cen X-3, long considered an example of a pulsar near equilibrium
(\S~5.1 Figure~\ref{fig:cenx3} upper panel),
is, in fact, the result of alternating
10--100\,d intervals of steady spin-up and spin-down 
(\S~5.1 Figure~\ref{fig:cenx3} lower panel, \S~4.2 Figure~\ref{cenx3freqflux}).
The torque displays rapid transitions 
between  spin-up and spin-down with a magnitude $\sim 5$ times larger than 
the long-term spin-up torque (\S~5.1 Figure~\ref{fig:torque-histogram}).  
This switching behavior is also seen in OAO~1657--415
(Figure~\ref{OAO1657freqflux}), and in the long-term behavior of 
4U~1626--67 and GX~1+4 (Figure~\ref{fig:freqmosaic}).
We propose that this behavior is common in disk-fed pulsars
(\S~5.1). 

Observing the predicted correlation between spin-up rate and bolometric
luminosity, $\dot \nu \propto L^{6/7}$, has long been considered
a critical test of accretion torque theory (\S~3).
We have observed $\dot \nu$ and 20--50\,keV pulsed flux, $F_{\rm pulsed}$, in
outbursts of two transients, A~0535+26 
(Figure~\ref{a0535freqflux}) and GRO~J1744--28 
(Figure~\ref{fig:j1744freqflux}).
We find $\dot\nu\propto F_{\rm pulsed}^\gamma$ with 
$\gamma = 0.951(26)$ for A~0535+26 and $\gamma = 0.957(26)$ for
GRO~J1744--28 (Figure~\ref{fig:torque-lum}, \S~\ref{sect:torque-lum}).  
EXOSAT observations of EXO~2030+375, using the orbit measured with BATSE,
yielded $\gamma\simeq1.2$ (\cite{Reynolds96}).
This disagrees with 
the predicted scaling of magnetospheric radius with mass accretion rate,
$r_{\rm m}\propto\dot M^{-2/7}$.  However,
the apparent contradiction may be that $F_{\rm pulsed}$
is not a good indicator of bolometric luminosity or $\dot M$.
In fact, an indirect test of the 
$r_{\rm m}$--$\dot M$ relation is provided by the observed correlation between
the frequency of quasi-periodic oscillations seen with BATSE in A~0535+26 and
$\dot\nu$, and is consistent with the predicted scaling (\S 5.2, 
\cite{Finger96c}).

Power density spectra of torque fluctuations in the 8 persistently
bright pulsars monitored with BATSE (Figure~\ref{fig:powerspectra})
show that pulsars known to wind-fed
(Vela~X-1, GX~301--2, 4U~1538--52) are flat.
Their spin-frequency behavior can be described as a random sequence
of independent torque fluctuations.  In contrast, 
the disk-fed pulsar Cen~X-3 
has an approximately $1/f$ power spectrum of torque fluctuations, 
as do GX~1+4 and OAO~1657--415,
indicating the presence of correlations between accretion torques on long time
scales (\S~5.4).  It is unclear if such correlations are a signature
of disk accretion however, 
as both 4U~1626--67 and Her~X-1 show flat power spectra,
and the accretion mechanism in GX~1+4 and OAO~1657--415 is unknown.
Furthermore, two dramatic spin-up episodes in
GX~301--2 of $\sim$20\,d duration (Figure~\ref{fig:gx301freqflux})
suggest that transient accretion disks sometimes form
in primarily wind-fed pulsars (\S~5.3).
The $1/f$ power spectrum of Cen~X-3
is in conflict with previous measurements,
and calls for a reexamination of the relation of noise strength with luminosity
and spin-up rate observed prior to BATSE.  

Most of the Be transient pulsars observed with BATSE
show a sequence of ``normal'' outbursts spaced
at the orbital period (Figure~\ref{fig:outburst_times}, \S~4.3).
Also observed are occasional ``giant'' outbursts, characterized by
high spin up rates, strongly correlated with flux and of 
of longer duration
(e.g. Figures~\ref{a0535freqflux}, \ref{2s1417freqflux},
\ref{J2058freqflux}), which suggests accretion from a 
transient disk (\S~5.5).
The repeating normal outbursts might be explained by the large tidal
torques experienced by the disk during periastron passage (\S~5.5).

We have estimated the total number of Be/X-ray transients in the Galaxy, 
and find there are $\sim$100 --1000 such systems, depending 
upon their assumed Galactic scale height and the BATSE sampling 
radius (\S~5.6).  
BATSE observations have doubled the number of measured 
Be/X-ray pulsar binary orbits, from 4 to 8 (Table~{\taborbits}).
The distribution of orbital eccentricities (Figure~\ref{fig:Be_e_dist})
agrees well with predictions
from radio-pulsar proper motions for $e<0.5$, and suggests that nearly all 
of the Be/X-ray pulsars for which no orbit has been determined must have 
high eccentricity (\S~5.7).

There are several important topics, not touched upon here, which will
eventually benefit from BATSE observations. These include the
time evolution of orbital parameters (e.g., orbital decay, apsidal
motion), a more complete analysis of the population of Be transients,
and searches for rapidly rotating pulsars from analysis of higher
time resolution data.

%%%%%%%%%%%%%%%%%%%%%%%%%%%%%%%%%%%%%%%%%%%%%%%%%%%%%%%%%%%%%%%%%%%%%%%%%%%%%%%

%\input acknowledgements.tex

\acknowledgements{We acknowledge John Grunsfeld for important
contributions to the early stage of this project, and Ed Brown Saul Rappaport
for helpful comments and for carefully reading the manuscript.
This work was funded in part by NASA grants NAG 5-1458, NAG 5-3293, and
NAGW-4517. D.C. was supported at Caltech by a NASA GSRP Graduate
Fellowship under grant NGT-51184. The NASA Compton Postdoctoral
Fellowship program supported D.C. (NAG 5-3109), L.B. (NAG 5-2666), and
R.W.N. (NAG 5-3119). L.B. was also supported by Caltech's Lee
A. DuBridge Fellowship, funded by the Weingart Foundation; and by the
Alfred P. Sloan Foundation.}

%%%%%%%%%%%%%%%%%%%%%%%%%%%%%%%%%%%%%%%%%%%%%%%%%%%%%%%%%%%%%%%%%%%%%%%%%%%%%%%

%\input appa.tex

\addcontentsline{toc}{section}{APPENDICES}
\appendix
\addcontentsline{toc}{section}{A\hspace{0.2cm} PULSED OBSERVATIONS WITH BATSE}
\renewcommand{\theequation}{A\arabic{equation}}
\renewcommand{\thesubsection}{A.\arabic{subsection}}
\section*{APPENDIX A \\PULSED OBSERVATIONS WITH BATSE}

\nocite{Fishman89,Horack91}

  The {\em Compton Gamma-Ray Observatory} was launched on 1991 April 5
into a 400 km orbit inclined 28.5$^\circ$ with respect to the Earth's
equator that precesses about the Earth's polar axis with a period of
$\approx$53 days. BATSE consists of eight identical uncollimated
detector modules arranged on the corners of the {\em Compton}
spacecraft (Fishman et al. 1989; Horack 1991).  Each detector module
contains a large-area detector (LAD) and a smaller spectroscopy
detector.  Our pulsar studies deal entirely with data from the LADs,
each of which contains a NaI(Tl) scintillation crystal 1.27 cm thick
and 50.8 cm in diameter, viewed in a light collection housing by three
12.7 cm diameter photomultiplier tubes. The LADs are shielded in front
by a 1 mm aluminum window and a plastic scintillator for charged
particle detection, and have an effective energy range of 20 keV--1.8
MeV and an energy resolution of about 35\% at 100 keV. Below 30 keV,
the sensitivity is severely attenuated by the aluminum and plastic
shielding.

Scintillation pulses from the LADs are processed in two parallel
paths: a fast, four-channel discriminator circuit and a slower
multi-channel pulse height analyzer. Calibration is maintained during
flight by an automatic correction scheme which adjusts the detector
gain so as to keep the 511 keV line from the gamma-ray background in
the same channel. There are many different BATSE data products
available. The three which we use for pulsar investigations are:
\begin{itemize}

\item {\bf DISCLA} data: Count rate samples for each detector
from the discriminator circuit in 4 energy channels at a time
resolution of 1.024\,s.

\item {\bf CONT} data: Count rate samples for each detector from the 
pulse-height analyzer in 16 energy channels at a time resolution of
2.048\,s.

\item {\bf PSR} data: Count rate samples for a programmable combination of 
detectors folded into 64 phase bins with a programmable folding period
in 16 energy channels and a programmable collection times, typically
8--16\,s.

\end{itemize}
The CONT and DISCLA data types are available continuously and are used
for slowly rotating pulsars.  We use the PSR data for pulsars with
spin periods less than or comparable to the DISCLA and CONT sampling
rates (e.g., Her~X-1 and GRO~J1744--28).  Approximate energy channel
boundaries for the DISCLA and CONT data and the typical count rates
(for background-dominated observations) are given in
Table~\ref{tab:channels}.  Figure \ref{fig:cont} shows the variation
of the energy channel boundaries over the various LADs for CONT channels
0--7.

\nocite{Pendleton95}
Most of our accreting pulsar studies are carried out below 100 keV, so
we principally use DISCLA channel 1 (20--60\,keV) and CONT
channels 1--6 (25--125\,keV). Pendleton et al. (1995) have
determined the response of the LADs to monochromatic photons of
various energies. The effective area at normal incidence for full
energy deposition peaks near 1500\,cm$^2$ around 100\,keV. The
effective area drops below 100 keV due to the absorption of low energy
photons in the aluminum window and the charged-particle detector and
above 100 keV due to the increased transparency of the LAD detector.
Each of the LADs has slightly different energy channels (see Figure
\ref{fig:cont}) which we take into account when constructing the
response matrices.

Prior to scientific analysis, the BATSE data undergo four stages of
processing. Each step except the first must be performed separately
for each source. {\it 1. Background subtraction:} The detector
background is removed using a phenomenological model. In addition,
quality flags are applied at this stage, and spikes removed.  {\it
2. Detector weighting:} Detectors with significant projected area to
the source are weighted and summed to optimize the signal-to-noise
ratio for timing analysis.  {\it 3. Barycenter correction:} Photon
arrival times are corrected for spacecraft orbital motion to Barycentric
Dynamical Time (TDB) using the Jet Propulsion Laboratory
DE-200 solar-system ephemeris (\cite{Standish92})
The orbit of the pulsar is also removed for
systems where it is known. {\it 4. Earth occultation windowing:}
Intervals where the source is occulted by the Earth are removed.
Steps 1 and 2 are particular to the BATSE data, so we describe them in
detail here.

Background subtraction is not performed on PSR data, and the selected 
detectors are uniformly weighted.  Barycentric corrections are applied
once per collection (8--16\,s) using the midpoint time, with the
times of phase bin edges computed relative to the midpoint time.

\subsection{Background Subtraction}

Because the LADs are uncollimated and non-imaging, the background
count rate includes contributions from the diffuse Galactic and cosmic
background, atmospheric gamma rays, a prompt local background caused
by interactions of primary cosmic rays with detector materials,
activation of radionuclides in the detectors by particles during
passages through the South Atlantic magnetic anomaly (SAA) that
results in a delayed internal background, and discrete source
contributions.  A detailed review of the gamma-ray background for
low-Earth orbit instruments in general is given by Dean, Lei, \&
Knight (1991), while a discussion of the background for BATSE in
particular is given by Rubin et al. (1996). 

\nocite{Dean91,Rubin96}

The diffuse cosmic gamma-ray background is the strongest component
below 300\,keV. It is isotropic, and varies for each detector
depending on the fraction of the field of view that is occulted by the
Earth. The atmospheric and prompt backgrounds depend upon the position
of the spacecraft in the Earth's magnetic field and vary with the
local geomagnetic cutoff rigidity.  Primary cosmic ray protons below
the local cutoff energy do not penetrate the Earth's field at a given
geomagnetic latitude.  Bombardment by cosmic rays and trapped protons
in the SAA produce radionuclides that decay with various lifetimes,
notably $^{128}$I ($\tau_{1/2}=25$\,min), creating a delayed internal
background.  Sharp changes in the count rate also occur when bright,
discrete sources cross the limb of the Earth.  In addition, there are
data gaps due to brief telemetry outages or errors, and from turning
off the detector high voltage during SAA passages.

 These background variations introduce power at a wide range of
Fourier frequencies and reduce the sensitivity to pulsed signals
relative to the Poisson counting limit. Strong quasi-sinusoidal
variations with $\approx 93$ min period are mainly due to orbital
modulation of sky area visible to the detectors.  In addition, the
spikes (mainly due to charged-particle-induced phosphorescence in the
NaI) and gaps in the data introduce variability at all analysis
frequencies, and are one of the major sources of noise power.  Data
quality flags set by the BATSE operation team are initially used to
reject data containing gamma-ray bursts, phosphorescence events and
other rapid background variations.  Additional quality control that
identifies spikes and gaps is a crucial step in our analysis of BATSE
data. Three separate background-subtraction techniques have been
developed, each of which involves modeling the background variations.

\nocite{Rossi70}

The most sophisticated technique is the use of a phenomenological
model that includes each of the known contributions (Rubin et
al. 1996). The atmospheric and prompt components of the background
depend upon the geomagnetic rigidity.  The cosmic, atmospheric, and
prompt components all vary with the direction of the detector relative
to the Earth. The delayed component can be modeled as an exponential
decay following each SAA passage.  Finally, discrete sources
contribute a component that we approximate as the product of the
effective area to the source and a step function that is zero when the
source is occulted by the Earth and unity when it is not.  The
resulting explicit form of the function used to fit the background
independently for each LAD is
\begin{equation}
M(t_i) = \sum_{j=0}^3(a_j+b_jL)P_j(\cos\theta_i)+
f_k\exp\left[{-(t_i-t_{\rm SAA})\over \tau_k}\right] +
\sum_{l=1}^{n_{occ}}g_lT_{li} 
\end{equation}
where $i$ labels the time bin corresponding to $t_i$, $L$ is the
McIlwain $L$-shell parameter (a measure of the location in the Earth's
magnetosphere; see, e.g., Rossi \& Olbert 1970), $P_j$ is the Legendre
polynomial of order $j$, $t_{\rm SAA}$ is the end time of the most
recent SAA passage, $\tau_k$ is the decay time of $^{128}$I, and
$T_{li}$ is the atmospheric transmission function of source $l$ at the
mean energy of bin $i$.  The coefficients $a_j$, $b_j$, $f_k$ and
$g_l$ are determined by a linear least-squares fit to the raw data. We
obtain acceptable fits for segments of length 0.125\,day and then
subtract the best-fit model from the raw data.  This technique assumes
the presence of periodic behavior at harmonics of the orbital period,
so some attenuation of low frequency signals is inevitable in the
fitting process.

  Figure \ref{fig:powerback} shows the power spectrum of the CONT
channel 1 data before and after this background subtraction. Following
background subtraction, the noise power is consistent with the Poisson
level on time scales $\lesssim 80$\,s. This background model performs
somewhat better than the other techniques at longer time scales and
allows us to probe deeper at lower pulse frequencies than would
otherwise be possible.  Histories of OAO~1657--415, Vela~X-1, 4U~1538--52,
GX~301--2, GS~0834--430, GRO~J1948+32, EXO~2030+375,
4U~1145--619 and GRO~J1008--57 in
\S~4 were constucted using this background-subtraction technique, as were
data sets at the COSSC.

A more ad hoc model for the background can be produced by applying
a simple frequency-domain digital filter to the data
(\cite{Chakrabarty93,Deeptothesis}).   To construct the background
model, we first remove impulsive spikes and interpolate over gaps in
the raw data and then perform smoothing in the frequency domain by 
multiplying the Fourier transform of the interpolated time series by a
low-pass filter defined as $R(\nu<\nu_0)=[1+\cos(\pi\nu/\nu_0)]/2$ for
$\nu<\nu_0$ and $R(\nu>\nu_0)=0$ for $\nu>\nu_0$, with
$\nu_0=1.6\times 10^{-3}$\,Hz. The inverse transform of this product
is a good approximation to the orbital background variation, which is
then subtracted from the raw data after reintroducing the original gap
structure.  Histories of 4U~1626--67 and GX~1+4 in \S~4 were
constucted using this background subtraction technique. 

Finally, an empirical background model is often used in analysis of the DISCLA
data. This is a spline function with quadratics in time, in segments
normally of 300 s length, connected with continuity in value and slope
at segment boundaries within contiguous sets of data. This model is
fit simultaneously with a Fourier representation of the pulsed signal, 
described further in Appendix B.1.2.  Histories of GRO~J1744--28,
Cen~X-3, 4U~0115+63, 2S~1417--624, A~0535+26, 4U~1145--619 and A~1118--616
in \S~4 were constructed using this technique.

\subsection{Optimal Combination of Detectors}

  The octahedral arrangement of the BATSE detector planes makes a
pulsar visible to four BATSE detectors during any given spacecraft
pointing (which typically changes every two weeks).  The sensitivity
to a given pulsar then depends on the angular response of the
detectors. A discussion of techniques for extracting
signals from multiple BATSE detectors appears in Chakrabarty (1996). We
summarize that discussion here, then present an optimal scheme for
weighting detectors when the energy spectrum of the source is known.

 The dominant effect is the projected detector area along the line of
sight to the source, which varies as $\cos\theta$ (where $\theta$ is
the viewing angle between the detector normal and the source). The
response to 100 keV photons is approximately $\cos\theta$, but
diverges from this for energies both above and below 100 keV for
different reasons.  At higher energies, the photon attenuation length
in NaI becomes comparable to the thickness of the crystal, so that not
all incident photons are captured. Hence, the decrease in projected
geometric area with increasing $\theta$ is partially offset by the
increase in path length through the detector, both of which have a
$\cos\theta$ dependence. This results in a relatively flat angular
response for $\theta\gtrsim 50^\circ$ at high energies.  At low
energies, the attenuation length in NaI is very short, making the
detector thickness irrelevant.  However, the path length through the
shielding in this case increases with $\theta$, and the resulting
attenuation, $\exp(-\tau(E)/\cos\theta)$ where $\tau(E)$ is the
optical thickness of the shield at energy $E$, of flux incident on the
LAD causes the response to fall more steeply than $\cos\theta$.

  A consequence of the variation of angular response with energy is
that the response to a pulsar depends upon its energy spectrum.
Typical accreting pulsar spectra can be represented as $dN/dE \propto
E^{-\alpha}$, where $2<\alpha<5$ in the 20--100 keV range. Figure
\ref{fig:angresp} shows the angular response to a 20--75 keV photon
power law spectrum for $\alpha=2$ and $\alpha=5$. For comparison,
$\cos\theta$ and $\cos^2 \theta$ are also plotted. The response falls
off more quickly than $\cos\theta$ since the large number of
low-energy photons dominate despite the attenuation by the shield. As
expected, this effect is more pronounced for the steeper power-law
index, where the proportion of incident lower energy photons is even
higher. In general, the integrated response varies $\approx
\cos\theta$ for small angles ($\theta\lesssim 25^\circ$), independent
of photon index. At larger angles, $\cos^2\theta$ is a more
conservative general assumption when the source spectrum is unknown.

Our timing analysis uses a weighted sum of the count rates from the
four detectors exposed to the source. If we consider a
background-limited observation and make the approximation
that the background rate is the same in all detectors,
isotropic, and governed by Poisson statistics, then the
resulting signal-to-noise ratio (SNR) is 
\begin{equation}
\label{eq-snr}
\mbox{\rm SNR} = \frac{C_S T\ \sum\,w_i(\theta_i)\,r_i(\theta_i)}{\sqrt{C_B T\
  \sum\,w^2_i(\theta_i)}}, 
\end{equation}
where $C_S$ is the source count rate in a single detector at normal
incidence, $C_B$ is the background count rate in each detector, $T$ is the
exposure time $\theta_i$ is the source viewing angle for the $i$th detector, 
$r_i(\theta_i)$ is
the angular response function of the $i$th detector, and $w_i(\theta_i)$ is the
detector weighting function that we are optimizing. 

BATSE is more sensitive to some areas of the sky than others.  The
highest sensitivity at low energies
is at the eight points in the sky that lie on the
BATSE detector normals (i.e., the direction vector for normal
incidence), while the lowest sensitivity is at six points in the sky
that lie equidistant from the eight detector normals.  The
weighting function for summing detectors that maximizes the
signal-to-noise ratio as parametrized in Equation (\ref{eq-snr})
depends upon where in the sky the source is relative to the detector
normals.  Because the angular response, $r(\theta)$, depends upon the
intrinsic energy spectrum of the source and is different for each
detector, determining the optimal weighting coefficients 
requires the full response matrix of each
exposed detector and an assumed incident energy spectrum. 
For DISCLA data, where we utilize a single energy channel, we use an optimal
detector weighting given by

\begin{equation} w_i = r_i(E_f) \left(
\sum_j r^2_j(E_f) \right)^{-1}~~.
\label{eq:optimal_weight}
\end{equation}  
Here $r_i(E_f)$ is the predicted count rate in DISCLA channel 1 for
detector $i$ assuming that the source has a photon spectrum $ F(E) =
(A/E) \exp(-E/E_f)$. This weighting scheme takes into account
differences between detectors (see figure~\ref{fig:cont}), unlike the
weighting scheme described below which assumes identical detectors and
$\cos^2\theta$ response.  The $e$-folding energies $E_f$ were selected
based on published spectra.

We also use a weighting scheme that is independent of the source
spectrum by approximating the angular response function as
$r(\theta)=\cos^2\theta$ and calculating sensitivity as a function of
sky position for weighting functions of the form $w(\theta)=\cos^n
\theta$, with $n=0, 1, 2, 3, 4$. The best overall sensitivity is
achieved by choosing the detector weighting adaptively (i.e.,
selecting whichever cosine power law optimizes sensitivity for a given
source location with respect to the detectors, rather than using a
single form of $w(\theta)$ for all source locations). However,
$w(\theta)=\cos^2 \theta$ weighting provides both reasonable average
sensitivity and spatial uniformity, is $\sim$15\% more sensitive on 
average than an unweighted sum of exposed detectors,
and is at worst $\sim$5\% poorer
than adaptive weighting.  We use $w(\theta)=\cos^2 \theta$ weighting 
for most frequency measurements with CONT data.

%\newpage
%\begin{figure}
\centerline{
\psfig{file=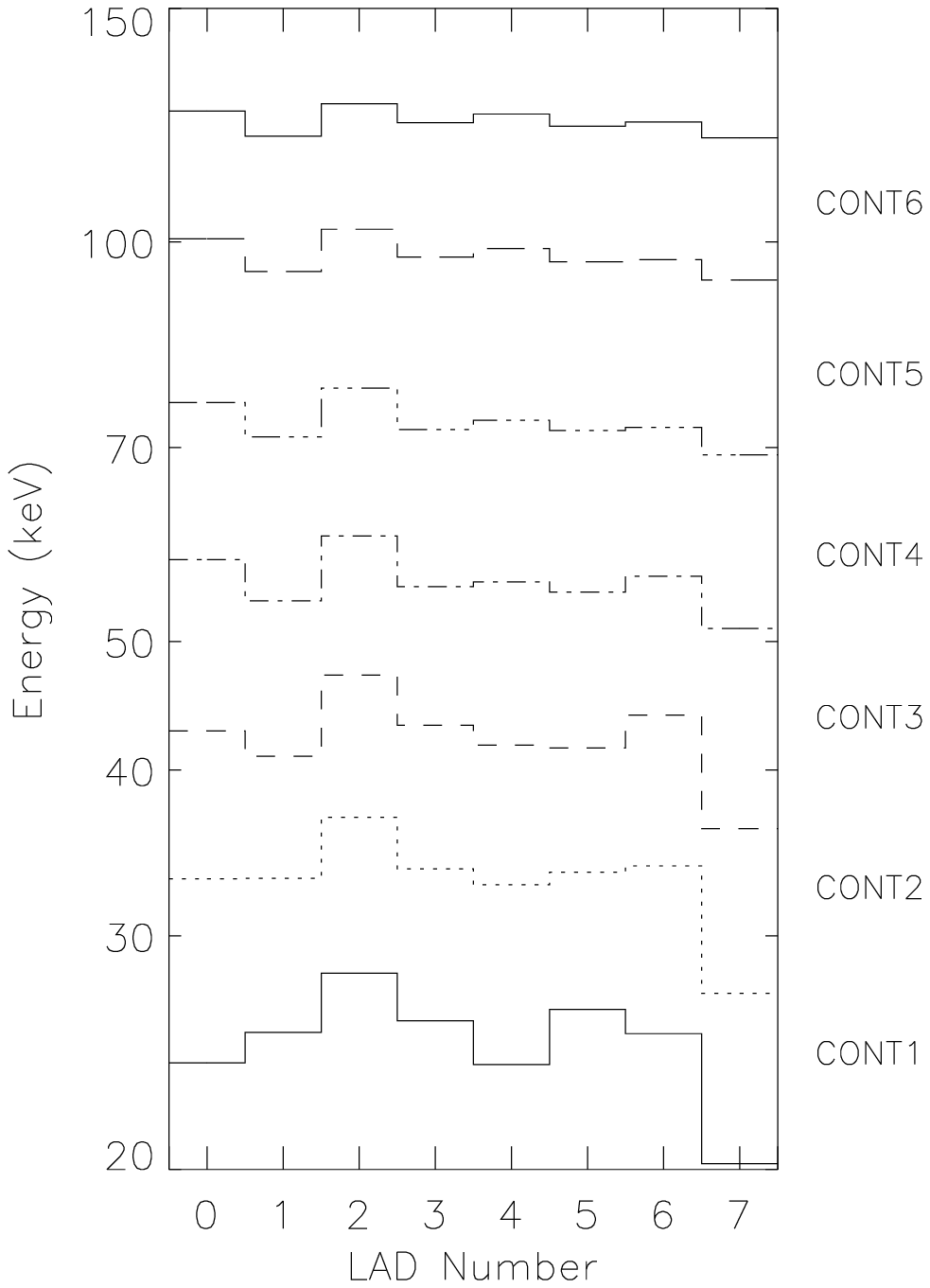,height=5.3in}
}
\figcaption{The energies of the Large Area Detector (LAD)
channel boundaries for the CONT data. 
We display the lower and upper boundaries 
of CONT channels 1--6 for all detectors.
The gain is stabilized onboard using the 511 keV background
feature. These bin energies apply to two different intervals, MJD 48406.11
-- 49400.69 and MJD 49419.69 -- 50062.82, which together comprise most of
the CGRO mission.
\label{fig:cont}
}
%\end{figure}

\newpage
%\begin{figure}
\centerline{
\psfig{file=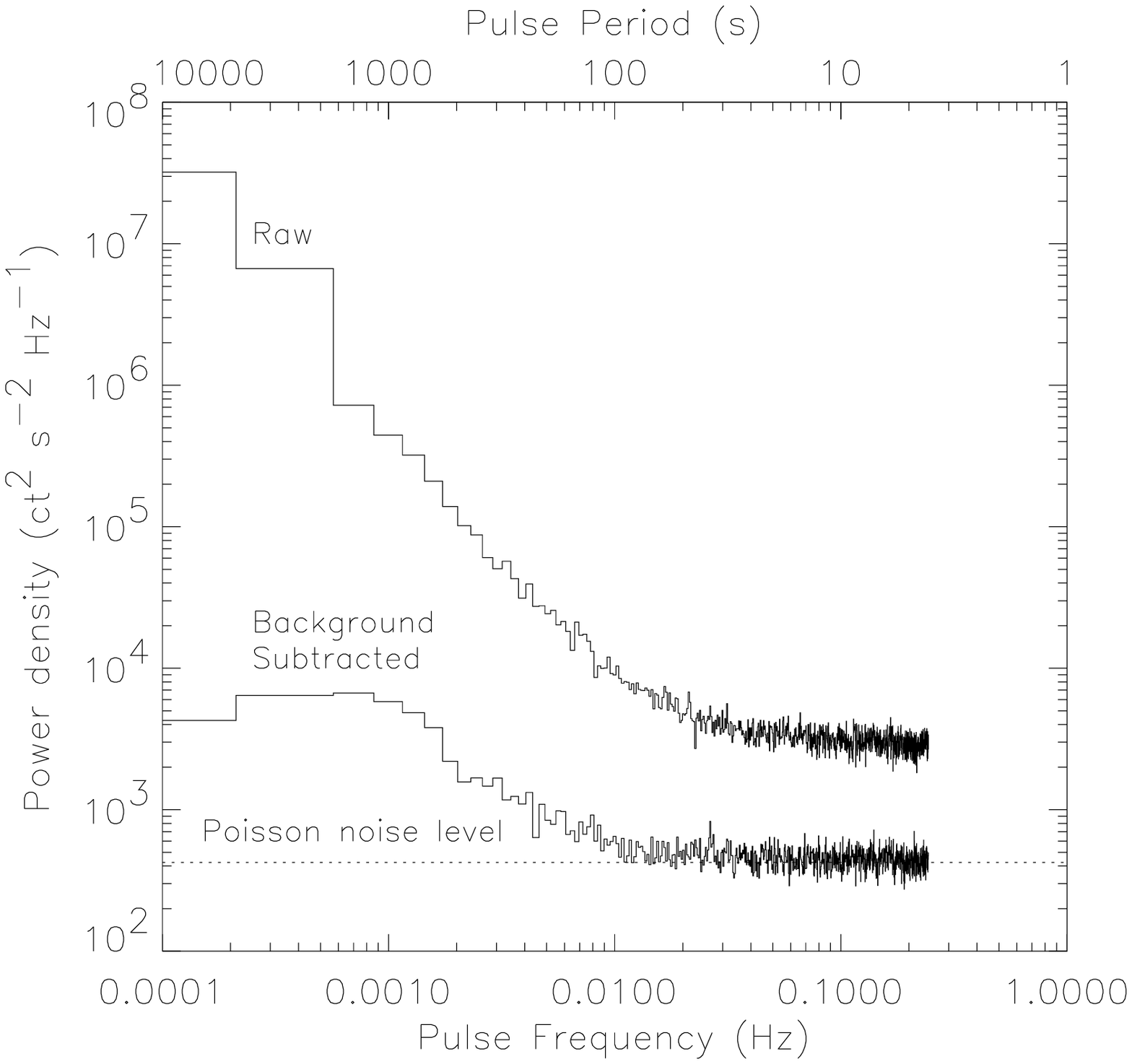,width=6in}
}
\figcaption{The typical power spectrum of the BATSE LAD background in
the 25--33 keV range.  The data shown are for CONT Channel 1 (25--33
keV) from LAD 2 on MJD 48851--48852 when the mean count rate was
$423.7\ \cps$. The top curve is the unprocessed raw data, while the
bottom curve is after background subtraction.
\label{fig:powerback}
}
%\end{figure}

\newpage
%\begin{figure}
\centerline{
\psfig{file=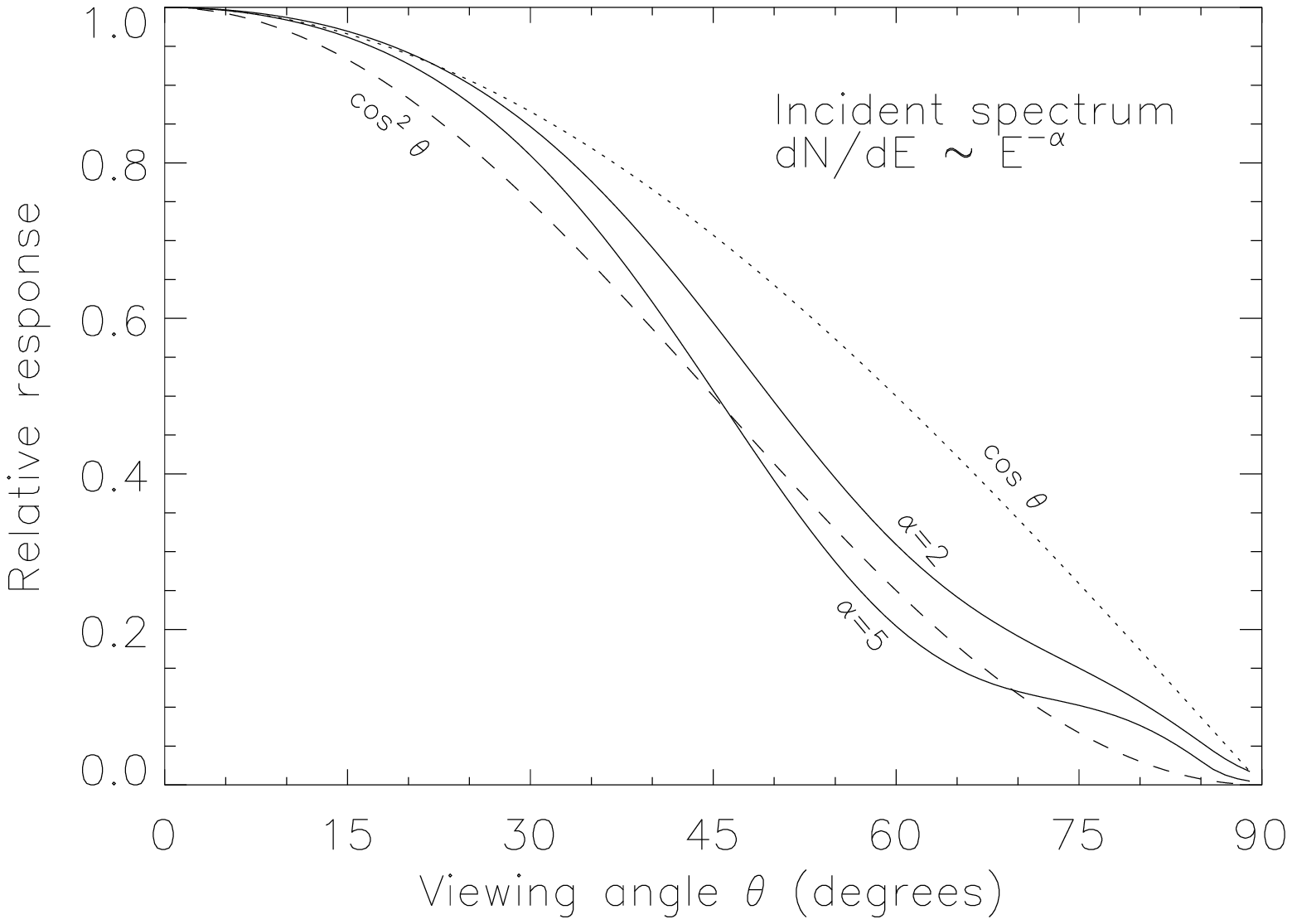,width=6in}
}
\figcaption{The 20--75~keV BATSE LAD angular response to an incident photon
power-law spectrum $dN/dE \propto E^{-\alpha}$, for $\alpha=2$ and
$\alpha=5$. For comparison, $\cos\theta$ response (dotted line) and
$\cos^2\theta$ response (dashed line) are also shown. 
\label{fig:angresp} 
}
%\end{figure}

%\input table_cont_edges.tex

\newpage
\setcounter{table}{5}
\begin{deluxetable}{ccc}
\tablecolumns{6}
\tablewidth{0pt}
\tablecaption{Energy Channels in BATSE DISCLA and CONT Data
\label{tab:channels}}
\tablehead{
\colhead{} & \colhead{Energy Range} & \colhead{Background Rate} \\
\colhead{Channel} & \colhead{(keV)} & \colhead{($\cps$)}}
\startdata
DISCLA 1 & 20--60 & 1500 \nl
DISCLA 2 & 60--110 & 1200 \nl
DISCLA 3 & 110--320 & 1000 \nl
DISCLA 4 & $>$320 & 700 \nl
 & &  \nl
CONT 0 &20--24  & 250 \nl
CONT 1 &24--33  & 450 \nl
CONT 2 &33--42  & 500 \nl
CONT 3 &42--55  & 500 \nl
CONT 4 &55--74  & 500 \nl
CONT 5 &74--99  & 450 \nl
CONT 6 &99--124 & 300 \nl
CONT 7 &124--165 & 300 \nl
CONT 8 &165--232 & 300 \nl
CONT 9 &232--318 & 200 \nl
CONT 10 &318--426&  130 \nl
CONT 11 &426--590&  130 \nl
CONT 12 &590--745&  50 \nl
CONT 13 &745--1103&  80 \nl
CONT 14 &1103--1828&  80 \nl
CONT 15 &$>$1828&  200 \nl
\enddata
\tablenotetext{}{NOTE: These channel boundaries are {\em approximate},
and are averaged over the eight detectors. Each detector has slightly
different edges, as shown in Figure \ref{fig:cont}. The CONT edges are
programmable; the displayed values are typical, and are computed
using the calibration discussed in \cite{Preece97}.  We have made use
of CONT data through channel 7 in our analyses.}
\end{deluxetable}

%%%%%%%%%%%%%%%%%%%%%%%%%%%%%%%%%%%%%%%%%%%%%%%%%%%%%%%%%%%%%%%%%%%%%%%%%%%%%%%

%\input appb.tex

\renewcommand{\theequation}{B\arabic{equation}}
\setcounter{subsection}{0}
\renewcommand{\thesubsection}{B.\arabic{subsection}}
\renewcommand{\thesubsubsection}{B.\arabic{subsection}.\arabic{subsubsection}}
\section*{APPENDIX B \\BATSE DATA ANALYSIS}
\addcontentsline{toc}{section}{B\hspace{0.2cm} BATSE DATA ANALYSIS}
In this appendix we discuss
how frequency and flux are computed following the preprocessing
described in appendix A.  Frequency estimation is covered first, followed
by pulsed flux.
%BATSE's forte is measuring spin frequency,
%which we discuss first, then pulsed flux.

\subsection{Frequency Estimation}

\subsubsection{Frequency Estimation using Power Spectra}
\label{appb:powspec}

Power spectra were used to compute the frequency histories of
4U~1626--67, GX~1+4, OAO~1657--415, GX~301--2, GS~0834--430,
GRO~J1948+32, EXO~2030+375 and  GRO~J1008--52 displayed in \S~4, 
as well as in data files at the COSSC.
Power spectra are computed using the Fast Fourier Transform (FFT).
Given a contiguous dataset of duration $T$ seconds in
$N$ uniform bins of duration $T/N$, the FFT returns $N/2+1$
statistically-independent estimates, $a_j$, of the amplitude and phase of the
variability at frequencies $\nu_j=j/T$, where $0\le j\le N/2$.  The 
normalized power at
$\nu_j$ is $P_j\equiv|a_j|^2/\bar P(\nu_j)$, where $\bar P(\nu)$ is the average
power, $\langle|a_j|^2\rangle$,
in the vicinity of $\nu_j$ (excluding frequencies 
immediately adjacent to $\nu_j$).  This normalization is important because
the background power is strongly frequency dependent 
(see Figure~\ref{fig:powerback}).  Thus defined, $\langle P_j \rangle=1$.
Data collected by BATSE are in uniform bins of duration 2.048\,s (CONT
data), 1.024\,s (DISCLA data) or shorter (PSR data).  
However, corrections of spacecraft times to the solar system barycenter 
causes the bins to become nonuniform.
We therefore define a new array with bins 
equispaced in arrival time
at the solar-system barycenter (or in the frame of the neutron
star if the orbit is known).  An individual time-series bin 
from the spacecraft will generally
overlap two such bins.  We can either treat each bin as a delta 
function at the bin center and add all
counts to the appropriate bin in the new array ({\it whole binning}) or split
counts between overlapping bins in proportion to the degree of
overlap ({\it bin splitting}). We choose bin splitting because it
performs significantly better than whole binning when $\nu\gsim N/2T$
while introducing negligible corrections to the power spectrum.  

The response, $R^2(\nu)$, of the power spectrum to a sinusoid of
frequency $\nu$ decreases with increasing frequency and also depends
upon the separation, $\Delta\nu$, between $\nu$ and the nearest
Fourier frequency, falling off as
\begin{equation}
R^2(\nu)={\rm sinc}^2(\pi\nu/2\nu_{\rm Nyq})
{\rm sinc}^2(\pi\Delta\nu T)
\label{eq:R}
\end{equation}
where ${\rm sinc}(x) = \sin(x)/x$ and
$\nu_{\rm Nyq}=N/2T$ is the Nyquist frequency.  Attenuation at
high frequency is intrinsic to binned data and constrains us to use
DISCLA and/or PSR data for fast pulsars, but attenuation with
$\Delta\nu$ can be circumvented by computing the power at a denser set
of frequencies.  By appending an array of zeroes of length $(m-1)N$ to
the array of data and computing the FFT, we obtain an overresolved Fourier
transform with $m$ times finer
frequency spacing, but with frequency bins that are no longer
statistically independent.  The signature of a sinusoid is no longer
in one or two bins, but distributed over $\sim m$ bins.  
To preserve the definition $\nu_j=j/T$, we allow $j$ to take on $m(N/2)+1$
values in increments of $\epsilon\equiv 1/m$.  If the
highest power occurs in frequency bin $j$, then Middleditch (1976) has
shown that the best estimate of the signal frequency 
is given by \nocite{Middleditch76}
\begin{equation}
\nu = \nu_j + {3 \over {4\pi^2\epsilon T}}
              \left( {{{P_{j+\epsilon}}-{P_{j-\epsilon}}}
              \over {\bar{P_j}}}\right).
\end{equation}
The uncertainty in the frequency 
(\cite{Middleditch76,MiddleditchNelson76})
%(Middleditch 1976;Middleditch \& Nelson 1976) 
is given by
\begin{equation}
\sigma_\nu\simeq {1 \over {2\pi T}} \sqrt{{6 \over {P_j}}},
\label{nu_0}
\end{equation}
In all cases we use the
dominant harmonic for frequency estimation.  For double-peaked pulsars
(see figure \ref{fig:pulsemosaic}), the second harmonic usually
dominates.

It is worth noting that $\sigma_\nu\propto T^{-3/2}$ (since $P\propto T$),
and hence $N_d$-day power spectra are $N_d^{3/2}$ times more precise than
1-day estimates.  However, transform
length cannot be increased indefinitely because
variations in $\nu$ will eventually cause a loss
of coherence when $T^2\gsim\langle\dot \nu \rangle$.
We call this the {\em decoherence time scale} (see Appendix A of
\cite{Chakrabarty97a}).

The frequency histories in this paper for 
OAO~1657--415, GS~0834--430, GRO~J1948+32, EXO~2030+375, GRO~J1008--52, 
and 4U~1145--619
were made
from daily power spectra of CONT data. These histories and others made
from daily power spectra are available in the datasets at the COSSC via:

http://cossc.gsfc.nasa.gov/cossc/COSSC\_HOME.html

\noindent
For the sources 4U~1626--67, GX~1+4 and GX~301--2, multi-day power 
spectra were used.

\subsubsection{Frequency Estimation using Epoch Folding}\
\label{appb:efold}

Spin frequencies were also estimated 
by epoch folding data 
at a range of trial frequencies.  We use the term ``epoch folding'' to 
describe any technique where a pulse phase is assigned to each time
bin based upon a model of the spin frequency.  In some cases we
impliment epoch folding as a fitting procedure.
This technique was used with DISCLA data for the
histories of Cen X-3, 4U 1538-52, 4U 0115+634, GRO J1750-27, 2S 1417-624,
GRO J2058+42, 4U~1145--619 and A1118-616  shown in \S~4.
Detection and determination of
the pulse frequency are based on the $Z^2_m$ test, which measures the
significance of the first $m$ Fourier amplitudes of the epoch-folded
pulse profile.  
	 
Rather then rebining the
data into uniform phase bins and then epoch folding them, we fit the data
using a background model and a low order Fourier expansion in the pulse
phase model.  Fitting does not require rebinning, and hence avoids the
loss of phase resolution inherent to rebinning techniques, which is
important when the low time resolution is $\gsim P_{\rm spin}$
(\cite{Deeter86}).  
The data
are divided into $1\sim10$\,d intervals, 
and each interval fit with a model
$C(t)$ of the form 
\begin{equation}
C(t) ~=~ M(t)~+~\sum_{k=1}^{n} \left[A_k \cos 2 \pi k \phi(t)
+B_k\sin 2\pi k\phi(t) \right].
\label{eq:C(t)}
\end{equation}  
where $M(t)$ is a model of the
background rate, the harmonic sum is a Fourier representation of
the pulse profile, and $\phi(t)$ is the pulse phase, where we generally
assume a constant frequency for each interval and use $n=3$.
The background is modeled
as a quadratic spline, with segments every 300\,s and with value
and slope continuous across segment boundaries (but not across
gaps). This background modeling procedure is approximately equivalent
to applying a high-pass filter with $\nu^6$ roll off and a roll-off
frequency of $\sim$0.002\,Hz (500\,s).
The $Z^2_3$ statistic is determined for each trial frequency for the
best set of fit parameters, $\{A_k,B_k\}$,
and values of $Z^2_3$ near the peak value
fit to determine the frequency.  
In the case of 4U~1538--52, because of its low frequency, we use the 
phenomenological background model described in Appendix A.1 rather than
a quadratic spline.

\subsubsection{Frequency Estimation from Fits to Pulse-Phase Measurements}
\label{appb:phase}

The frequencies histories of GRO J1744-28, Her X-1, Vela X-1, 
GRO~J1750-27 and A~0535+26
shown in \S~4 were determined from fits to pulse phase
measurements. Phases were determined (for orbital analyses) by epoch
folding data over short intervals ($\sim 0.5days$) and correlating the
resulting pulse profiles with pulse templates to determine phase offsets
from the epoch folding phase model. The total phases were then divided into
several day intervals and a linear fit in emission time made to the total
phases (model plus offset) in each interval to obtain frequencies.

\subsection{Pulsed Flux Estimation}

\subsubsection{What is Pulsed Flux and Pulsed Fraction?}

{\bf Pulsed flux} is the periodically varying part of the flux from a
source.  
Unlike average flux, pulsed flux is not uniquely defined, and we
employ two separate definitions for different purposes and for
different sources.  Let $F(\phi)$ be the flux of a pulse profile at
phase $\phi$, $0\leq\phi\leq1$, let $\bar{F}=\int_0^1F(\phi)d\phi$ be
the average flux, and let $F_{\rm min}={\rm min}[F(\phi)]$ be the
minimum flux.  We define peak-to-peak pulsed flux as $F_{\rm
pulsed}=\int_0^1(F(\phi)-F_{\rm min})d\phi$, and root-mean-square
(RMS) flux as $F_{\rm RMS}= \left[\int_0^1(F(\phi) - \bar{F})^2
d\phi\right]^{1/2}$.  Their relative values depend upon pulse shape.
For a square wave they are equal. For a sinusoid, which is a good
approximation to most accreting pulsars, $F_{\rm
pulsed}=\sqrt{2}F_{\rm RMS}$.  Peak-to-peak 
pulsed flux has intuitive appeal, but
it is more difficult to measure since $F_{\rm min}$ is harder to
determine than $\bar{F}$.  We generally use peak-to-peak 
pulsed flux for average
spectra of many days (see Table~4), 
where S/N is large, and RMS flux for daily estimates, as displayed in
\S~4.  {\bf Pulsed fraction} is the ratio of the pulsed flux to
the mean flux (pulsed + unpulsed).  It is in general energy dependent.

\subsubsection{Peak-to-Peak Pulsed Spectra and Flux}
\label{appb:spec}

\vskip 0.2cm
\centerline{\it Average energy spectrum and flux using CONT data}
\vskip 0.2cm

To estimate the energy spectrum of the pulsed
emission we first compute the pulse count rate in each energy channel
by epoch folding $\sim 1$\,d of data at a time 
into $N\lsim P_{\rm spin}/\tau$
phase bins, where $\tau=2.048$\,s (CONT data) or 1.024\,s (DISCLA data) 
and $P_{\rm spin}$ is determined
from overresolved power spectra or from folding, as
discussed in B.1.  A running average pulse profile is constructed by
aligning each day's profile with and adding it to the existing
running sum, weighted by its exposure (area times exposure time).
Alignment is performed by maximizing the cross
correlation between the daily profile and the running sum or using the
cross spectrum (see \cite{Koh97}).  The end result is an average
profile in each energy channel (for DISCLA we use only channel 1).  
Channels with a significant detection are used for spectral fitting.  Prior to
fitting the profiles are Fourier transformed and all harmonics higher
than $n_H\sim6$ set to zero to reduce the counting noise (see \cite{Deeter86}),
then inverse transformed.  The rate in the minimum phase bin is
subtracted from each bin and the resulting rates averaged to obtain
the pulsed count rate in each energy channel.

In the following discussion we denote the intrinsic strength of the 
pulsar signal as $s$, and the amplitude we measure as $a$.
If the signal-to-noise is low, then the measured amplitude, $a$, of the
pulsar signal, $s$, is significantly biased by the measurement noise, $n$,
such that $\langle a \rangle > s$.  To account for this bias and rigorously
estimate the measurement uncertainty we use the probability distribution
of the measured amplitude
(\cite{Thomas69,Goodman85})
\begin{equation}
p(a|s,n) = \frac{2a}{n^2}\exp\left[\frac{-(a^2+s^2)}{n^2}\right] 
I_0\left(\frac{2as}{n^2}\right) {\mbox{\rm\hspace*{0.5in}$(a>0)$}},
\label{eq:p(a)}
\end{equation}
where $I_0$ is the zeroth order Bessel function of the first kind.
This can be inverted to yield the probability distribution of $s$
(\cite{Deeptothesis}):
\begin{equation}
p(s|a,n) = \frac{2}{n}\sqrt{\frac{1}{\pi}} \exp\left[
\frac{-(a^2+2s^2)}{2n^2}\right] \frac{I_0 \left(\frac{2as}{n^2}
\right)}{I_0\left(\frac{a^2}{2n^2}\right)} 
{\mbox{\rm\hspace*{0.5in}$(s>0)$}} .
\label{eq:p(s)}
\end{equation}
If $a\gg n$, the probability distribution approaches a Gaussian with
variance $\sigma^2=n^2/2$.  Because the noise varies with time and
energy, we determine $n$ separately for each energy channel by
averaging the power in frequencies in the vicinity of the dominant
harmonic in each of the daily power spectra of all days used to
construct the average profile.

We determine the incident spectrum by folding standard models through the
detector response matrix and varying the model parameters to fit the
pulsed count rate in each energy channel.  
Pendleton \etal (1995) have computed BATSE detector
response matrices as a function of viewing angle, $\theta$.  The
energy edges of the CONT channels are independent of $\theta$ but vary
with detector (see Figure A.1).  Because a single average profile may
contain data from multiple spacecraft orientations (a typical
viewing period lasts 1--2 weeks), we compute an average
response matrix and average channel edges for the spectral fitting.  
A photon power-law model,
$dN/dE=C_{30}(E/30\,{\rm keV})^{-\alpha}$, and an exponential (EXP)
model $dN/dE=C_0/E\,\exp(-E/E_f)$, provide reasonable fits to the data.
In most cases the EXP model is superior.

\nocite{Pendleton95}

An alternative to aligning and averaging daily profiles is to perform
a {\it long coherent fold}.  This is only possible for consistently
bright, non-eclipsing systems where gaps in the data are short
compared with the decoherence time scale, or where the torque history
is smooth.  Spectra have been constructed for 4U~1626--67 and GX~1+4
in such a way (\cite{Chakrabarty97a,Chakrabarty97b}),

\vskip 0.2cm
\centerline{\it On-Peak Minus Off-Peak Pulsed Spectra and Flux}
\vskip 0.2cm

In another approach to pulsed flux estimation,
5--10\,d of CONT data are folded using a
phase model determined by fitting pulse phases measured using
epoch-folded DISCLA data.  An
``on-pulse'' and ``off-pulse'' phase interval are chosen by eye
and the average rate in the off-pulse interval
subtracted from each phase bin, yielding a rate differences in each
energy channel for each phase bin.  

\subsubsection{Daily RMS Flux}
\label{appb:flux}

Unlike profiles constructed from 10--200 days, daily profiles have low
$S/N$ for most sources most of the time.  Rather than attempting to
fit a spectral model, we generally assume a spectral shape and
determine the normalization.  Although we use an exponential model in
some cases and a power law in others, daily flux histories are not
particularly sensitive to the choice of model.

\vskip 0.2cm
\centerline{\it Daily RMS flux from CONT and DISCLA data}
\vskip 0.2cm

For most sources, an RMS flux, $F_{\rm RMS}$, 
is estimated daily by constructing
 pulse profiles in flux units.  First, the average rate is
subtracted from the folded profile from
each detector and energy channel.  Next, each phase
bin, $\phi$, of each daily pulse profile for each exposed detector and
 energy channel is modeled as a phase-dependent exponential
spectrum
\begin{equation}
F(E,\phi)=F_0(\phi)(E_0/E)\exp(-(E-E_0)/E_f),
\label{eq:F(E)}
\end{equation}
where $F_0(\phi)$ is the flux at $E_0$, which depends upon phase, and
where $E_f$ is assumed independent of phase.  The e--folding energy is not
adjusted in the daily fits but rather taken from longer term analysis.
The flux as a function of phase is parameterized in terms of a small 
number $m$ (3--6) of Fourier amplitudes;
\begin{equation}
F_0(\phi)=F_0^{\rm avg}+\sum_{k=1}^m
[A_k\cos(2\pi k\phi)+B_k\sin(2\pi k\phi)],
\label{eq:F_0}
\end{equation}
with the Fourier coefficients $A_k$ and $B_k$ determined in the fit.
Finally, the RMS flux $F_{\rm RMS}$ is determined as
\begin{equation}
F_{\rm RMS}=\left(0.5\sum_{k=1}^m[A_k^2+B_k^2]\right)^{1/2}.
\label{eq:F_RMS}
\end{equation}
This procedure assumes that the pulse shape is independent of energy,
which is not in general true.  Fluxes in the datasets at the COSSC
are determined using this procedure.

\vskip 0.2cm
\centerline{\it DISCLA flux using power spectra}
\vskip 0.2cm

Flux histories have been determined for 4U~1626--67 and GX~1+4 by
computing 1--5\,d power spectra of the x-ray light curves in DISCLA
channel 1 and using the amplitude of the dominant harmonic, in both
cases the fundamental, to estimate the pulsed amplitude.  The flux is
determined by convolving either a power-law or exponential model with
the detector response matrix, assuming a power-law index or
e--folding energy.  This technique assumes a constant spectral shape and
pulse profile, and can be used with non-sinusoidal profiles by
computing a correction factor between the amplitude of the dominant harmonic
and the pulsed flux.

\vskip 0.2cm
\centerline{Frequencies and Fluxes from PSR data}
\vskip 0.2cm

PSR data were used in the analysis of GRO J1744--28 because of its short
0.467 s pulse period. As well as data collected explicitly for GRO
J1744-28, all applicable PSR data that was collected in a single-sweep mode
(with no folding on board) were used. On the ground the 20-40 keV  rates
from these data were divided into intervals of $\sim 200$ s. For each
interval an empirical background model quadratic in time was fit to the
rates and subtracted. Then the rates in each interval were fit with at 6th
order Fourier pulse model. These (harmonically represented) profiles were 
then combined over $\sim$ 0.5 d intervals using a phase model, obtained
by bootstraping. The frequencies shown in section 4 were obtained by
fitting pulse phase measurements based on these profiles. To obtain fluxes,
a conversion factor from pulsed count rate to pulsed flux was calculated
with the detector response matrices by assuming a spectral form
$dN/dE = A E^{-2} \exp(-E/E_{fold}$ with $E_{fold} = 15$ keV.

%%%%%%%%%%%%%%%%%%%%%%%%%%%%%%%%%%%%%%%%%%%%%%%%%%%%%%%%%%%%%%%%%%%%%%%%%%%%%%%

%\input nocite.tex

% Nocites to force entries from table of pulsars
% and table or orbital parameters into reference list 

% Table of pulsars

\nocite{Finger96a}
\nocite{CWilson97}
\nocite{Deeter91}
\nocite{Levine93}
\nocite{Israel95}
\nocite{Ogelman96}
\nocite{Ilovaisky82}
\nocite{Makishima84}
\nocite{Skinner81}
\nocite{Stella85}
\nocite{Kelley83a}
\nocite{Wilson94d}
\nocite{Murakami87}
\nocite{Hellier94}
\nocite{Hughes94}
\nocite{Schmidtke95}
\nocite{Schwentker94}
\nocite{Seward86}
\nocite{Iwasawa92}
\nocite{Israel94}
\nocite{Koyama91b}
\nocite{Koyama91a}
\nocite{Koyama90a}
\nocite{Koyama90b}
\nocite{Tawara89}
\nocite{Levine91}
\nocite{Makishima87}
\nocite{Coe94_J1008}
\nocite{Colleen95}
\nocite{CWilson96}
\nocite{Haberl96}

%Table of orbital parameters

\nocite{Finger96a}
\nocite{Deeter91}
\nocite{Wilson94a}
\nocite{Levine91}
\nocite{Makishima87}
\nocite{Corbet93}
\nocite{Rubin94}
\nocite{Levine93}
\nocite{Finger93}
\nocite{Makishima84}
\nocite{Cook87b}
\nocite{Deeter87}
\nocite{Chakrabarty93}
\nocite{Koh97}
\nocite{Rappaport78}
\nocite{Cominsky94}
\nocite{Kelley83a}
\nocite{Stella85}
\nocite{Finger1417}
m\nocite{Stollberg94}
\nocite{Finger94e}

%%%%%%%%%%%%%%%%%%%%%%%%%%%%%%%%%%%%%%%%%%%%%%%%%%%%%%%%%%%%%%%%%%%%%%%%%%%%%%%

\newpage
{\baselineskip=10pt
\small
%%\nopagebreak
%%\bibliographystyle{apj_noskip}
\bibliography{ref}
}

\end{document}